\documentstyle[twocolumn,twoside]{article}
\newcommand{\AAA}{{\cal A \hspace*{-1.7ex} A}}
\newcommand{\BB}{{\cal B \hspace*{-1.7ex} B}}

\newcommand{\CC}{{\cal C \hspace*{-1.2ex} C}}
\newcommand{\dotnabla}{\nabla \hspace*{-1.26ex} \raisebox{0.4ex}{$\cdot \,$}}
\newcommand{\dotnablascr}{\nabla \hspace*{-1ex} \raisebox{0.1ex}{$\cdot$}}

\newcommand{\FF}{{\cal F \hspace*{-2ex} F}}
\newcommand{\FS}{{\sf F \hspace*{-1.1ex} F}}
\newcommand{\Lagr}{{\cal L \hspace*{-1.45ex} L}}
\newcommand{\MM}{{\cal M \hspace*{-2.7ex} M}}
\newcommand{\NN}{{\cal N \hspace*{-2.1ex} N}}
\newcommand{\plaision}{\rule[1.4ex]{1.4ex}{0.15ex} \hspace*{-1.4ex}
            \rule[0.17ex]{0.25ex}{1.3ex} \hspace*{-0.3ex} \Box}
\newcommand{\RR}{{\cal R \hspace*{-2.1ex} R}}
\newcommand{\RS}{{\sf R \hspace*{-1.3ex} R}}
\newcommand{\smallvec}[1]{\vec{\scriptstyle \rm #1}}
\newcommand{\smallcomp}[1]{{\scriptstyle \rm #1}}
\newcommand{\SS}{{\cal S \hspace*{-1.7ex} S}}

\newcommand{\VV}{{\cal V \hspace*{-1.5ex} V}}
\title{\Large \bf Five-Dimensional Tangent Vectors in Space-Time \\
\large \bf VI. Bivector Derivative and Its Application}
\author{Alexander Krasulin \\ \it Institute for Nuclear Research of the
Russian Academy of Sciences \\ krasulin@post.com}
\date{\normalsize \bf Abstract \\ \mbox{ } \\ \begin{minipage}{460pt}
\begin{center} \begin{minipage}{400pt}
\normalsize
In this concluding part of the series I first consider the bivector
derivative for four-vector and four-tensor fields in the case of arbitrary
Riemannian geometry. I then define this derivative for five-vector and
five-tensor fields, examine the bivector analogs of the Riemann tensor,
and introduce the notion of the commutator for the fields of five-vector
bivectors. After that I examine a more general case of five-vector
affine connection, introduce the five-vector analog of the curvature
tensor, discuss the canonical stress-energy and angular momentum tensors
corresponding to the five-vector covariant derivative, and consider
a possible five-vector generalization of the Einstein and Kibble--Sciama
equations. In conclusion, I introduce the notion of the bivector derivative
for the fields of nonspacetime vectors and tensors, consider the corresponding 
gauge fields and their properties, and derive a possible generalization of Maxwell's equation.
\end{minipage} \end{center} \vspace{2ex} \begin{flushright}
\it To Valeri Dvoeglazov
\end{flushright} \end{minipage}}

\begin{document}

\maketitle

\begin{flushleft}
A. \it Bivector derivative in curved space-time
\end{flushleft}
In part III I have introduced the bivector derivative for scalar,
four-vector and four-tensor fields in flat space-time. Now I would like to
define this derivative in the general case. As one can see from the formulae
obtained in section D of part III, in the case of flat space-time the
bivector derivative of the indicated fields is determined only by the metric,
and since with respect to its metric properties any sufficiently smooth
space-time manifold is locally flat, the bivector derivative in the general
case can be defined by postulating that in local Lorentz coordinates it has
the same form at any space-time geometry. For scalar fields this means that
the bivector derivative of an arbitrary function $f$ is given by formula
(32) of part III, where ${\bf e}_{A}$ can be any active regular basis at the
considered point. For four-vector fields the above assertion means that the
bivector derivative of the basis fields ${\bf E}_{\alpha}$ corresponding to
any system of local Lorentz coordinates at the considered point is given by
formula (33) of part III, where ${\bf e}_{A}$ is now the active regular
five-vector basis associated with ${\bf E}_{\alpha}$. Furthermore, one should
assume that in the general case, too, the bivector derivative has the
properties expressed by equations (37) of part III, which will enable one to
define the derivative ${\sf D}_{\cal A} {\bf W}$ for any four-vector field
${\bf W}$ and any five-vector bivector $\AAA$, and that the bivector
derivative of the contraction and tensor product obeys the Leibniz rule,
which will enable one to define the action of operator $\sf D$ on all other
four-tensor fields.

As in the case of flat space-time, for any set of four-vector basis fields
${\bf E}_{\alpha}$ and any set of five-vector basis fields ${\bf e}_{A}$ one
can define the bivector connection coefficients according to formula (40) of
part III. In the particular case where ${\bf E}_{\alpha}$ is a basis that
corresponds to some system of local Lorentz coordinates at the considered
point and ${\bf e}_{A}$ is the associated active regular five-vector basis,
the bivector connection coefficients $\Gamma^{\mu}_{\; \nu AB}$ at that point
are given by formulae (41) of part III.

The bivector derivative for the considered type of fields in the general case
can also be defined without referring to local Lorentz coordinates. Instead,
one can postulate that as in the case of flat space-time, it is expressed
according to formulae (38) and (39) of part III for scalar fields in terms of
the directional derivative and for four-vector fields in terms of the
torsion-free $g$-conserving ordinary covariant derivative $\dotnabla$, which
is uniquely determined by space-time metric, and of the local operator
$\widehat{\bf M}$, whose argument is a four-vector bivector, on which
$\widehat{\bf M}$ depends linearly. Let us recall that in an arbitrary
four-vector basis ${\bf E}_{\alpha}$, operator $\widehat{\bf M}$ has the
following components:
\begin{equation}
(\widehat{\bf M}_{{\bf E}_{\alpha} \wedge {\bf E}_{\beta}})^{\mu}_{\; \nu}
= \delta^{\mu}_{\, \beta} \, g_{\alpha \nu} - \delta^{\mu}_{\, \alpha} \,
g_{\beta \nu} \equiv (M_{\alpha \beta})^{\mu}_{\, \nu}.
\end{equation}

The bivector derivative of scalar, four-vector and four-tensor fields defined
above possesses one important property: at any five-vector affine connection
$\plaision$ with respect to which the metric tensor $g$ is covariantly
constant, the five-vector covariant derivative of any of these fields is
expressed linearly in terms of its bivector derivatives. More precisely
this property can be formulated as follows: at any given five-vector affine
connection that conserves the metric tensor, at each space-time point there
exists such a linear map $\sigma$ from the tangent space of five-vectors to
the tangent space of five-vector bivectors that for any five-vector $\bf u$
at that point
\begin{equation}
\plaision_{\bf u} \, {\cal G} = {\sf D}_{\sigma({\bf u})} \, {\cal G}
\end{equation}
for any field $\cal G$ from the considered class of fields. Let us now
prove this statement.

Consider an arbitrary point $Q$ and introduce in its vicinity some system
of local Lorentz coordinates with the origin at $Q$. Let ${\bf E}_{\alpha}$
be the four-vector basis corresponding to these coordinates and
${\bf e}_{A}$ be the associated active regular five-vector basis. Let
$\Upsilon^{\alpha}_{\; \, \beta A}$ denote the connection coefficients
for the basis ${\bf E}_{\alpha}$:
\begin{equation}
\plaision_{A} {\bf E}_{\alpha} = {\bf E}_{\beta}
\Upsilon^{\beta}_{\; \, \alpha A},
\end{equation}
where, as usual, $\plaision_{A} \equiv \plaision_{{\bf e}_{A}}$. It is
evident that the upper and the first lower indices of $\Upsilon^{\alpha}_{\;
\, \beta A}$ are four-vector, and its second lower index is five-vector.

The supposed covariant constancy of $g$ imposes certain constraints on the
coefficients $\Upsilon^{\alpha}_{\; \, \beta A}$. Indeed, owing to the
property of $\plaision$ expressed by relation (18) of part V, from the
fact that $g$ is covariantly constant as a five-tensor follows that it
is covariantly constant as a four-tensor, which in its turn means that
\begin{displaymath}
\plaision g = 0,
\end{displaymath}
where $g$ is considered a four-tensor. Writing down this equation in
components relative to the basis ${\bf E}_{\alpha}$ and considering that the
latter is associated with a system of local Lorentz coordinates, one obtains
that at $Q$
\begin{equation}
g_{\alpha \omega} \Upsilon^{\omega}_{\; \, \beta A} +
g_{\omega \beta} \Upsilon^{\omega}_{\; \, \alpha A} = 0.
\end{equation}
Let us now introduce the quantities
\begin{equation}
S^{\alpha \beta}_{\hspace{2ex} A} \equiv - \, g^{\beta \omega} \,
\Upsilon^{\alpha}_{\; \, \omega A}.
\end{equation}
   From equation (4) it follows that $S^{\alpha \beta}_{\; \; \; A}$ are
antisymmetric with respect to their upper indices. Furthermore, it is
easy to check that with transition to any other system of local Lorentz
coordinates with the origin at the same point, the quantities
$S^{\alpha \beta}_{\; \; \; A}$ transform as components of a five-vector
1-form whose values are four-vector bivectors. Consequently, the 1-form
constructed according to the formula
\begin{displaymath}
\widetilde{\bf S} \equiv S^{|\alpha \beta|}_{\hspace*{2.5ex} A} \,
{\bf E}_{\alpha} \wedge {\bf E}_{\beta} \otimes \widetilde{\bf o}^{A},
\end{displaymath}
where $\widetilde{\bf o}^{A}$ is the basis of five-vector 1-forms dual to
the basis ${\bf e}_{A}$, will be the same at any choice of the local Lorentz
coordinates. From definition (5) it follows that
\begin{displaymath}
\Upsilon^{\alpha}_{\; \beta A} = - g_{\tau \beta}
S^{\alpha \tau}_{\; \; \; A} = - \delta^{\alpha}_{\, \sigma} g_{\tau \beta}
S^{\sigma \tau}_{\; \; \; A} = (M_{\sigma \tau})^{\alpha}_{\, \beta} \,
S^{|\sigma \tau|}_{\hspace*{2.5ex} A}.
\end{displaymath}
Consequently, by virtue of equations (1) and (3), at $Q$ one has
\begin{equation}
\plaision_{A} {\bf E}_{\alpha} = {\bf E}_{\beta} \, (M_{\sigma \tau})
^{\beta}_{\, \alpha} \, S^{|\sigma \tau|}_{\hspace*{2.5ex} A}
= \widehat{\bf M}_{\widetilde{\bf S}({\bf e}_{A})} {\bf E}_{\alpha},
\end{equation}
where $\widetilde{\bf S}({\bf e}_{A})$ denotes the four-vector bivector
obtained by contracting the 1-form $\widetilde{\bf S}$ with the basis
five-vector ${\bf e}_{A}$.

Since the basis ${\bf E}_{\alpha}$ is associated with a system of local
Lorentz coordinates at $Q$, at that point one has
\begin{displaymath}
\dotnabla_{A} {\bf E}_{\alpha} = 0,
\end{displaymath}
and consequently, equation (6) can be rewritten as
\begin{equation}
\plaision_{A} {\bf E}_{\alpha} = \dotnabla_{A} {\bf E}_{\alpha}
+ \widehat{\bf M}_{\widetilde{\bf S}({\bf e}_{A})} {\bf E}_{\alpha}.
\end{equation}
Due to the linear dependence of operators $\plaision$, $\dotnabla$, and
$\widehat{\bf M}$ on their arguments, from the latter formula it follows that
\begin{displaymath}
\plaision_{\bf u} {\bf E}_{\alpha} = \dotnabla_{\bf u} {\bf E}_{\alpha}
+ \widehat{\bf M}_{\widetilde{\bf S}({\bf u})} {\bf E}_{\alpha}
\end{displaymath}
for any five-vector $\bf u$. Furthermore, owing to the properties of the
derivative $\dotnabla$ expressed by equations (2b) and (2c) of part V; to
similar properties of the operator $\plaision$ expressed by equations (11b)
and (11c) of part V; and to the linearity and locality of $\widehat{\bf M}$,
from the latter equation one obtains that
\begin{equation}
\plaision_{\bf u} {\bf W} = \dotnabla_{\bf u} {\bf W}
+ \widehat{\bf M}_{\widetilde{\bf S}({\bf u})} {\bf W}
\end{equation}
for any four-vector field $\bf W$. Finally, by making use of the
relation between the bivector derivative and operators $\dotnabla$
and $\widehat{\bf M}$, expressed by equations (39) of part III, one
can present equation (8) as
\begin{equation}
\plaision_{\bf u} {\bf W} = {\sf D}_{{\cal A} ({\bf u})} {\bf W}
\end{equation}
where $\AAA ({\bf u})$ denotes the five-vector bivector whose
$\cal E$-component corresponds to the four-vector that corresponds to
$\bf u$, and whose $\cal Z$-component corresponds to the four-vector bivector
$\widetilde{\bf S}({\bf u})$. Owing to the linearity of the latter two
correspondences, the bivector $\AAA ({\bf u})$ can be presented as a
contraction of $\bf u$ with some five-vector 1-form $\widetilde{\bf s}$
whose values are five-vector bivectors. It is easy to see that in an
arbitrary active regular basis this 1-form has the following components:
\begin{equation}
s^{\alpha 5}_{\hspace{2ex} A} = - s^{5 \alpha}_{\hspace{2ex} A} =
\delta^{\alpha}_{\, A} \; \mbox{ and } \; s^{\alpha \beta}_{\hspace{2ex} A}
= S^{\alpha \beta}_{\; \; \; A},
\end{equation}
where $S^{\alpha \beta}_{\; \; \; A}$ are the components of the 1-form
$\widetilde{\bf S}$ in the associated four-vector basis. By using
$\widetilde{\bf s}$ one can rewrite equation (9) as
\begin{displaymath}
\plaision_{\bf u} {\bf W} = {\sf D}_{< \widetilde{\bf s}, {\bf u} >} {\bf W},
\end{displaymath}
which, if one puts $\sigma({\bf u}) \equiv \; < \widetilde{\bf s},{\bf u} >$,
coincides with equation (2) for four-vector fields. From formula (15) of part
V and formula (38) of part III it follows that at such $\sigma({\bf u})$
equation (2) will also hold for arbitrary scalar fields. Finally, since the
action of both $\plaision$ and $\sf D$ on the contraction and tensor product
obeys the Leibniz rule, equation (2) with $\sigma({\bf u})$ selected this
way will hold for all other four-tensor fields as well.

In the particular case where the fifth component of the five-vector covariant
derivative for four-vector fields is zero, one can establish a simple
relation between the 1-form $\widetilde {\bf S}$ and the four-vector torsion
tensor. Since the latter is a {\em four}-tensor, it is more convenient to
consider the covariant derivative a linear function of a four-vector rather
than of a five-vector, which is possible since in this case $\plaision$ is
equivalent to an ordinary covariant derivative. Furthermore, to present the
formulae involving torsion in a more familiar form, instead of $\plaision$
I will write $\nabla$. Instead of $\Upsilon^{\alpha}_{\; \, \beta A}$
one will then have ordinary four-vector connection coefficients
$\Gamma^{\alpha}_{\; \beta \mu}$, and instead of formula (5),
\begin{equation}
S^{\alpha \beta}_{\; \; \; \mu} \equiv - \, g^{\beta \omega} \,
\Gamma^{\alpha}_{\; \omega \mu},
\end{equation}
so $S^{\alpha \beta}_{\; \; \; \mu}$ will now be the components of a
four-vector 1-form rather than of a five-vector one.

Let us now recall the definition of four-vector torsion:
\begin{equation}
\nabla_{\bf U} {\bf V} - \nabla_{\bf V} {\bf U} - [{\bf U,V}] \equiv
- 2 {\bf T(U,V)},
\end{equation}
where $\bf U$ and $\bf V$ are any two four-vector fields and the factor
$-2$ is introduced for convenience. Since $\nabla$ depends on its argument
linearly, the quantity $\bf T(U,V)$, which is a certain four-vector field,
will depend linearly on $\bf U$ and $\bf V$, and consequently can be
presented as a contraction of some four-vector 2-form, $\widetilde{\bf T}$,
whose values are four-vectors, with the bivector $\bf U \wedge V$. Usually,
the components of $\widetilde{\bf T}$ are defined as follows:
\begin{equation}
\widetilde{\bf T} = T_{|\mu \nu|}^{\hspace{2.5ex} \alpha} \, {\bf E}_{\alpha}
\otimes \widetilde{\bf O}^{\mu} \wedge \widetilde{\bf O}^{\mu},
\end{equation}
where $\widetilde{\bf O}^{\mu}$ is the basis of four-vector 1-forms dual to
the basis ${\bf E}_{\alpha}$, and it is a simple matter to show that in
any coordinate basis they can be expressed in terms of the corresponding
connection coefficients in the following familiar way:
\begin{equation}
T_{\mu \nu}^{\; \; \; \alpha} = \Gamma^{\alpha}_{\; [\mu \nu]}.
\end{equation}
Comparing this formula for the case where ${\bf E}_{\alpha}$ is associated
with a system of local Lorentz coordinates, with formula (11), one finds
that
\begin{equation}
T_{\mu \nu}^{\; \; \; \alpha} = - \, S^{\alpha}_{\; [\mu \nu]},
\end{equation}
where $S^{\alpha}_{\; \mu \nu} = g_{\mu \omega} \, S^{\alpha \omega}_{\; \;
\; \nu}$. Since both $\widetilde{\bf S}$ and $\widetilde{\bf T}$ are tensors,
the latter equation will hold in {\em any} four-vector basis. Making use
of the antisymmetry of $\widetilde{\bf S}$ with respect to its first two
indices, in the usual way one can derive the formula opposite to formula
(15), which expresses the components of $\widetilde{\bf S}$ in terms of
those of $\widetilde{\bf T}$:
\begin{displaymath}
S^{\alpha \beta}_{\; \; \; \mu} = g^{\alpha \sigma} g^{\beta \tau} \,
(T_{\sigma \tau \mu} - T_{\tau \mu \sigma} - T_{\mu \sigma \tau}),
\end{displaymath}
where $T_{\sigma \tau \mu}=T_{\sigma \tau}^{\; \; \; \omega}g_{\omega \mu}$.
Therefore, the four-vector-valued 2-form $\widetilde{\bf T}$ and the
bivector-valued 1-form $\widetilde{\bf S}$ contain exactly the same
information. Later on we will see that a similar relation exists between
the five-vector 1-form $\widetilde{\bf s}$ introduced above and the
five-vector torsion tensor.

In conclusion, let me note that the four-vector 2-form $\widetilde{\bf S}$
actually coincides with the so-called contorsion tensor, $\widehat{\bf K}$,
which can be defined as an operator whose action on an arbitrary four-vector
field $\bf W$ is given by the formula
\begin{displaymath}
\widehat{\bf K}_{\bf U} {\bf W} \equiv
(\dotnabla_{\bf U} - \nabla_{\bf U}) {\bf W}.
\end{displaymath}
For practical reasons, the components of $\widehat{\bf K}$ relative to
some four-vector basis ${\bf E}_{\alpha}$ are defined as follows:
\begin{displaymath}
\widehat{\bf K}_{{\bf E}_{\mu}} {\bf E}_{\alpha} =
K_{\mu \alpha}^{\; \; \; \beta} \, {\bf E}_{\beta}.
\end{displaymath}
Comparing the latter two definitions with formula (8) adapted to the case
we are now considering, one finds that
\begin{displaymath}
K_{\mu \alpha}^{\; \; \; \beta} = g_{\alpha \omega} \,
S^{\beta \omega}_{\; \; \; \mu},
\end{displaymath}
so $\widetilde{\bf S}$ differs from $\widehat{\bf K}$ only in the arrangement
of its indices.

\vspace{3ex} \begin{flushleft}
B. \it Bivector derivative of five-vector \\
   \hspace*{2ex} and five-tensor fields
\end{flushleft}
In this section I wish to define the bivector derivative for five-vector
fields and for the fields of all other five-tensors. As before, let us begin
by considering flat space-time and after that generalize the formulae
obtained to the case of arbitrary Riemannian geometry by following the same
recipe that has been used in the previous section to define the bivector
derivative of scalar, four-vector and four-tensor fields in the general case.

As in all the cases considered earlier, the bivector derivative of
five-vector and five-tensor fields in flat space-time can be defined
according to formula (26) of part III, where $\cal G$ can now be an arbitrary
five-vector or five-tensor field and ${\bf \Pi}_{s} \{ {\cal G} \}$ denotes
the image of $\cal G$ relative to active Poincare transformations from some
one-parameter family, $\cal H$, that includes the identity transformation
(the latter corresponding to the value of the family parameter $s=0$). From
formula (31) of part I one then obtains that for an arbitrary $O$-basis
${\bf e}_{A}$,
\begin{equation}
{\sf D}_{{\bf e}_{\mu} \wedge {\bf e}_{5}} {\bf e}_{A} = {\bf 0}
\; \mbox{ and } \; {\sf D}_{{\bf e}_{\mu} \wedge {\bf e}_{\nu}} {\bf e}_{A}
= {\bf e}_{B} \, (M_{\mu \nu})^{B}_{\; A},
\end{equation}
where $(M_{KL})^{A}_{\; B} \equiv \delta^{A}_{\, L} \, g_{KB} -
\delta^{A}_{\, K} \, g_{LB}$. For an arbitrary $P$-basis ${\bf p}_{A}$
one will have
\begin{displaymath}
{\sf D}_{{\bf p}_{K} \wedge {\bf p}_{L}} {\bf p}_{A} =
{\bf p}_{B} \, (M_{KL})^{B}_{\; A}.
\end{displaymath}

Though such a definition of the bivector derivative for five-vector fields
is quite permissible, it is not difficult to see that in that case the
relation between $\plaision$ and $\sf D$ expressed by equation (2)
{\em cannot exist}. Indeed, the five-vector covariant derivative of
an arbitrary field from $\FF_{\! \cal Z}$ has in general a nonzero
$\cal E$-component, whereas the bivector derivative of any such field
defined as described above will always be a field from $\FF_{\! \cal Z}$,
as is readily seen from formulae (16). Since in the further analysis the
mentioned relation between $\plaision$ and $\sf D$ will play an essential
role, let us try to define the bivector derivative for five-vector fields
in a different way: so that relation (2) could hold in this case as well.

To understand how this should be done, let us define for an arbitrary set
of five-vector basis fields ${\bf e}_{A}$ in flat space-time the bivector
connection coefficients according to the formula similar to equation (40)
of part III:
\begin{equation}
{\sf D}_{KL} {\bf e}_{A} = {\bf e}_{B} G^{B}_{\; AKL}.
\end{equation}
Since we wish that equation (2) hold and since the derivative $\plaision$
has property (18) of part V, one should require that
\begin{equation}
{\bf v} \equiv {\bf w} \, ({\rm mod} \; R) \Longrightarrow
{\sf D}_{\cal A} {\bf v} \equiv {\sf D}_{\cal A} {\bf w} \, ({\rm mod} \; R)
\end{equation}
for any bivector field $\AAA$, and that the derivative ${\sf D}_{\cal A}
{\bf W}$ of any four-vector field $\bf W$ be the equivalence class of all
the derivatives of the form ${\sf D}_{\cal A} {\bf w}$ with $\bf w
\in W$. From these requirements it follows that in any standard basis
\begin{equation}
G^{\alpha}_{\; 5KL} = 0
\end{equation}
and
\begin{displaymath}
G^{\alpha}_{\; \beta KL} = \Gamma^{\alpha}_{\; \beta KL},
\end{displaymath}
where $\Gamma^{\alpha}_{\; \beta KL}$ are the bivector connection
coefficients corresponding to the associated four-vector basis. Thus,
according to formulae (41) of part III, for an $O$-basis one has
\begin{equation}
G^{\alpha}_{\; \beta \mu 5} = 0 \; \mbox{ and } \;
G^{\alpha}_{\; \beta \mu \nu} = (M_{\mu \nu})^{\alpha}_{\; \beta}.
\end{equation}

Of the yet undetermined bivector connection coefficients for five-vector
fields, the quantities $G^{5}_{\; B \mu 5}$ can be found by considering a
particular case where the five-vector affine connection is such that there
exists a local symmetry similar to the one which has been discussed in
section 3 of part II for $\nabla$ (see section D of this paper). In
that case, for any Lorentz four-vector basis the connection coefficients
$\Upsilon^{\alpha}_{\; \, \beta A}$ introduced in the previous section
are identically zero, and from formulae (5) and (10) one finds that
\begin{equation}
\plaision_{\mu} \, {\cal G} = {\sf D}_{\mu 5} \, {\cal G}
\end{equation}
for any scalar, four-vector or four-tensor field $\cal G$. Requiring that
this relation between the covariant and bivector derivatives hold in the
case of five-vector fields as well, for an arbitrary $O$-basis one obtains
\begin{equation}
G^{5}_{\; \beta \mu 5} = - \, g_{\beta \mu}
\; \mbox{ and } \; G^{5}_{\; 5 \mu 5} = 0.
\end{equation}

Finally, let me observe that the latter two equations, the first equation
in (20), and equation (19) for $(KL) = (\mu 5)$ can be combined into a
single formula:
\begin{displaymath}
G^{A}_{\; B \mu 5} = - \, (M_{\mu 5})^{A}_{\; B}.
\end{displaymath}
Likewise, the second equation in (20) and formula (19) for $(KL) =
(\mu \nu)$ can be combined into
\begin{displaymath}
G^{\alpha}_{\; B \mu \nu} = (M_{\mu \nu})^{\alpha}_{\; B}.
\end{displaymath}
If one now supposes that for an $O$-basis the connection coefficients
$G^{5}_{\; B \mu \nu}$ are also proportional to $(M_{\mu \nu})^{5}_{\; B}$,
one will have
\begin{equation}
G^{5}_{\; 5 \mu \nu} = G^{5}_{\; \beta \mu \nu} = 0.
\end{equation}
A more serious argument in favour of the latter equations is the following.
It is reasonable to think that as in the case of four-vector and four-tensor
fields, the bivector derivative of five-vector fields is determined only by
the metric, and since with respect to its metric properties flat space-time
is {\em homogeneous} and {\em isotropic}, the bivector connection
coefficients $G^{A}_{\; B KL}$ should have the same form in any Lorentz
coordinate system. Reasoning as in section 3 of part I, one can find
the following general form of the bivector connection coefficients for
an arbitrary $O$-basis, which satisfy condition (18):
\begin{displaymath}
G^{A}_{\; B \mu 5} \propto (M_{\mu 5})^{A}_{\; B} \; \mbox{ and } \;
G^{A}_{\; B \mu \nu} = (M_{\mu \nu})^{A}_{\; B}.
\end{displaymath}
Fixing the proportionality factor in the first relation from equation (21),
one finally gets
\begin{equation}
G^{A}_{\; B \mu 5} = - \, (M_{\mu 5})^{A}_{\; B} \; \mbox{ and } \;
G^{A}_{\; B \mu \nu} = (M_{\mu \nu})^{A}_{\; B},
\end{equation}
which coincides with formulae (19), (20), (22), and (23). One should
observe that the sign in the right-hand side of the first equation in
(24) will not change if in all the formulae one replaces ${\bf e}_{5}$
with $-{\bf e}_{5}$, for such a replacement will change the sign of
$G^{5}_{\; \beta \mu}$ and the sign in the right-hand side of equation
(21), but will not change the sign of $G^{5}_{\; \beta \mu 5}$.

Let me also note that the obtained connection coefficients $G^{A}_{\; BKL}$
for an $O$-basis, regarded as matrices with respect to the indices $A$ and
$B$, satisfy the commutation relations for the generators of the Poincare
group, which in the matrix form can be expressed as
\begin{displaymath} \begin{array}{l}
[ G_{KL}, G_{MN} ] = g_{KM} G_{LN} - g_{LM} G_{KN} \\ \hspace{17ex}
 - \; g_{KN} G_{LM} + g_{LN} G_{KM},
\end{array} \end{displaymath}
where $(G_{KL})^{A}_{\; B} \equiv G^{A}_{\; BKL}$. The same commutation
relations are satisfied by the bivector connection coefficients
corresponding to derivative (16), only in that case the matrices
$G_{\mu 5}$ that correspond to the generators of translations are all zero.

Formulae (17) and (24) determine the bivector derivative for sets of
five-vector fields that make up an $O$-basis. To define the derivative
${\sf D}_{\cal A} {\bf w}$ for any five-vector field $\bf w$ and any field of
five-vector bivectors $\AAA$, one should take that in the case of five-vector
fields, too, the operator $\sf D$ has the properties expressed by equations
(37) of part III, in which the four-vector fields should now be replaced with
the five-vector ones. Furthermore, to define the bivector derivative for the
fields of all other five-tensors, one should suppose that the action of the
operator $\sf D$ on the contraction and tensor product obeys the Leibniz
rule. One should observe that in this case properties (37) of part III and
the Leibniz rule are {\em postulated}, whereas in the case of four-vector
fields these properties of $\sf D$ follow from definition (26) of part III.

For the bivector derivative of five-vector fields there exists a
representation similar to formula (39) of part III. Namely, from equations
(17) and (24) it follows that the operator ${\sf D}_{\cal A}$ can be
presented as a sum of two operators: $(i)$ the operator $\dotnabla$ of the
ordinary covariant derivative that corresponds to the connection considered
in section 3 of part II, whose argument will be the five-vector from
$\cal Z$ that corresponds to the $\cal E$-component of $\AAA$, and $(ii)$
the local linear operator $\widehat{\bf M}$, whose components in an
arbitrary standard five-vector basis equal $(M_{KL})^{A}_{\, B}$ and
whose argument will be the $\cal Z$-component of $\AAA$. Thus, for an
arbitrary five-vector field $\bf u$ one has
\begin{equation}
{\sf D}_{\cal A} {\bf u} = \dotnabla_{\bf a} {\bf u}
 + \widehat{\bf M}_{\cal A^{Z}} {\bf u},
\end{equation}
where $\bf a$ denotes the five-vector from $\cal Z$ that corresponds to
$\AAA^{\cal E}$.

By using the formulae obtained above, it is not difficult to define the
bivector derivative for five-vector and five-tensor fields in the case of
arbitrary Riemannian geometry of space-time. As in the case of four-vector
fields, it is sufficient to postulate that formulae (17) and (24) are
valid in any system of local Lorentz coordinates at a given point and that
the bivector derivative of five-vector fields in the general case has the
properties similar to those expressed by equations (37) of part III.
Alternatively, one can postulate equation (25), which expresses $\sf D$
in terms of $\dotnabla$ and $\widehat{\bf M}$. By using either of these
definitions, it is not difficult to calculate the bivector connection
coefficients for an arbitrary set of basis five-vector fields ${\bf e}_{A}$
that make up an active regular basis at every point. For such a set of
fields, formulae (19), (22), and (23) will be valid without any changes,
while instead of formula (20) one will have
\begin{equation}
G^{\alpha}_{\; \beta \mu 5} = G^{\alpha}_{\; \beta \mu} \; \mbox{ and } \;
G^{\alpha}_{\; \beta \mu \nu} = (M_{\mu \nu})^{\alpha}_{\; \beta},
\end{equation}
where $G^{\alpha}_{\; \beta \mu}$ are the connection coefficients for the
basis ${\bf e}_{A}$, associated with the derivative $\dotnabla$.

\vspace{3ex} \begin{flushleft}
C. \it Bivector analogs of the Riemann tensor \\ \hspace{2ex}
       and the commutator of bivector fields
\end{flushleft}
As one can see from the material presented in the previous sections of this
paper and in section D of part III, in many respects the bivector
derivative is similar to the covariant derivative and usually can be handled
in very much the same way. Let us now develop the analogy between these two
derivatives further and consider the bivector analog of the Riemann tensor.
It seems reasonable to suppose that the components of the latter in any
regular basis associated with some system of local Lorentz coordinates at
a given point can be obtained by considering the operator
\begin{equation}
\widehat{\RS}_{KLMN} \equiv {\sf D}_{KL}{\sf D}_{MN}-{\sf D}_{MN}{\sf D}_{KL}
\end{equation}
acting on four-vector fields. It is a simple matter to show that
$\widehat{\RS}_{KLMN}$ is a local operator, so if ${\bf E}_{\alpha}$ is
the associated four-vector basis, one can write
\begin{equation}
\widehat{\RS}_{KLMN}{\bf E}_{\alpha} = {\bf E}_{\beta}
{\sf R}^{\beta}_{\; \alpha \, KLMN}.
\end{equation}
By definition, the coefficients ${\sf R}^{\alpha}_{\; \beta \, KLMN}$ have
the following symmetry properties:
\begin{displaymath}
{\sf R}^{\alpha}_{\; \beta \, KLMN} = - \,
{\sf R}^{\alpha}_{\; \beta \, LKMN} = - \,
{\sf R}^{\alpha}_{\; \beta \, KLNM}
\end{displaymath} and \begin{displaymath}
{\sf R}^{\alpha}_{\; \beta \, KLMN} = - \,
{\sf R}^{\alpha}_{\; \beta \, MNKL},
\end{displaymath}
and it is evident that with transition to another system of local Lorentz
coordinates with the origin at the same point, they transform as components
of a four-tensor with respect to the indices $\alpha$ and $\beta$ and as
components of a five-tensor with respect to the indices $K$, $L$, $M$, and
$N$. Consequently, the quantity
\begin{equation}
\RS \equiv {\scriptstyle \frac{1}{4}} \, {\bf E}_{\alpha} \otimes
\widetilde{\bf O}^{\beta} \, {\sf R}^{\alpha}_{\; \beta \, KLMN}
\, (\widetilde{\bf o}^{K} \wedge \widetilde{\bf o}^{L}) \otimes
(\widetilde{\bf o}^{M} \wedge \widetilde{\bf o}^{N})
\end{equation}
will be the same at any choice of the local Lorentz coordinates, and it
would seem that it is it that one should take to be the analog of the
Riemann tensor for the bivector derivative.

By using definitions (27) and (28), one can express the components of
$\RS$ in terms of the bivector connection coefficients:
\begin{equation} \begin{array}{l}
{\sf R}^{\alpha}_{\; \beta \, KLMN} = {\sf D}_{KL} \Gamma^{\alpha}_{\;
\beta MN} - {\sf D}_{MN} \Gamma^{\alpha}_{\; \beta KL} \\ \hspace{5ex}
+ \; \Gamma^{\alpha}_{\; \omega KL} \Gamma^{\omega}_{\; \beta MN} -
\Gamma^{\alpha}_{\; \omega MN} \Gamma^{\omega}_{\; \beta KL}. \rule{0ex}{3ex}
\end{array} \end{equation}
   From this formula it follows that in an active regular basis associated
with some system of local Lorentz coordinates at the considered point
\begin{displaymath} \begin{array}{l}
{\sf R}^{\alpha}_{\; \beta \, \kappa 5 \mu 5} = \partial_{\kappa}
\Gamma^{\alpha}_{\; \beta \mu 5} - \partial_{\mu}
\Gamma^{\alpha}_{\; \beta \kappa 5} \\ \hspace{15ex} \rule{0ex}{2.5ex} + \;
\Gamma^{\alpha}_{\; \omega \kappa 5} \Gamma^{\omega}_{\; \beta \mu 5} -
\Gamma^{\alpha}_{\; \omega \mu 5} \Gamma^{\omega}_{\; \beta \kappa 5} \\
\hspace{8.5ex} \rule{0ex}{2.5ex} =  \partial_{\kappa}
\Gamma^{\alpha}_{\; \beta \mu} - \partial_{\mu}
\Gamma^{\alpha}_{\; \beta \kappa} \\ \hspace{23ex} \rule{0ex}{2.5ex} + \;
\Gamma^{\alpha}_{\; \omega \kappa} \Gamma^{\omega}_{\; \beta \mu} -
\Gamma^{\alpha}_{\; \omega \mu} \Gamma^{\omega}_{\; \beta \kappa},
\end{array} \end{displaymath}
where $\Gamma^{\alpha}_{\; \beta \mu}$ are the four-vector connection
coefficients associated with the derivative $\dotnabla$, so
\begin{equation}
{\sf R}^{\alpha}_{\; \beta \, \kappa 5 \mu 5}=R^{\alpha}_{\; \beta\kappa\mu},
\end{equation}
where $R^{\alpha}_{\; \beta \kappa \mu}$ are the components of the Riemann
tensor in the associated four-vector basis. In a similar manner one finds
that
\begin{equation}
{\sf R}^{\alpha}_{\; \beta \, \kappa 5 \mu \nu} =
{\sf R}^{\alpha}_{\; \beta \, \mu \nu \kappa 5} = 0
\end{equation}
and
\begin{eqnarray}
{\sf R}^{\alpha}_{\; \beta \, \kappa \lambda \mu \nu} & \hspace{-1ex} =
& \hspace{-1ex} (M_{\kappa \lambda})^{\alpha}_{\; \omega}
(M_{\mu \nu})^{\omega}_{\; \beta} - (M_{\mu \nu})^{\alpha}_{\; \omega}
(M_{\kappa \lambda})^{\omega}_{\; \beta} \nonumber \\ \rule{0ex}{2.5ex}
& \hspace{-1ex}  = &  \hspace{-1ex} g_{\kappa \mu}
(M_{\lambda \nu})^{\alpha}_{\; \beta} - g_{\lambda \mu}
(M_{\kappa \nu})^{\alpha}_{\; \beta} \\ & & \rule{0ex}{2.5ex}
- g_{\kappa \nu} (M_{\lambda \mu})^{\alpha}_{\; \beta}
+ g_{\lambda \nu} (M_{\kappa \mu})^{\alpha}_{\; \beta}, \nonumber
\end{eqnarray}
and it is obvious that formulae (31)--(33) will be valid in {\em any}
active regular basis.

Let us now try to define $\RS$ without explicitly referring to coordinates.
To this end, let us consider the operator
\begin{equation}
[ \, {\sf D}_{\cal A}, {\sf D}_{\cal B} \, ] \equiv
{\sf D}_{\cal A} {\sf D}_{\cal B} - {\sf D}_{\cal B} {\sf D}_{\cal A},
\end{equation}
where $\AAA$ and $\BB$ are two arbitrary fields of five-vector bivectors.
In an active regular basis associated with some system of local Lorentz
coordinates $x^{\alpha}$ at the considered point one has
\begin{displaymath} \begin{array}{l}
[ \, {\sf D}_{\cal A}, {\sf D}_{\cal B} \, ] = {\cal A}^{|KL|}
{\cal B}^{|MN|} \, [ \, {\sf D}_{KL}, {\sf D}_{MN} \, ] \\
\hspace{14ex} \rule{0ex}{3ex} + \; ({\sf D}_{\cal A} {\cal B}^{|KL|} -
{\sf D}_{\cal B} {\cal A}^{|KL|}) \, {\sf D}_{KL}.
\end{array} \end{displaymath}
The first term in the right-hand side is simply the contraction of $\RS$
with the values of the fields $\AAA$ and $\BB$, and the second term is
a bivector derivative operator whose argument is the value of a bivector
field constructed from the fields $\AAA$ and $\BB$ in such a way that its
structure resembles that of the commutator of two four-vector fields. As
we will see below, the quantity
\begin{equation}
({\sf D}_{\cal A} {\cal B}^{|KL|} - {\sf D}_{\cal B} {\cal A}^{|KL|}) \;
{\bf e}_{K} \wedge {\bf e}_{L},
\end{equation}
where ${\bf e}_{A}$ is the considered five-vector basis, does not depend
on the choice of the corresponding local Lorentz coordinates, so if one
takes it to be the commutator of the fields $\AAA$ and $\BB$ and denotes
it as $[ \, \AAA,\BB \, ]$, one can give $\RS$ the following coordinate-free
definition:
\begin{equation}
< \RS, \AAA \otimes \BB > \; = {\sf D}_{\cal A} {\sf D}_{\cal B} -
{\sf D}_{\cal B} {\sf D}_{\cal A} - {\sf D}_{\cal [ \, A,B \, ]},
\end{equation}
provided one can give a coordinate-free definition to $[ \, \AAA,\BB \, ]$.

Before we turn to this latter problem, let us find the expression for the
commutator of $\AAA$ and $\BB$ in an active regular basis associated with an
{\em arbitrary} coordinate system, $x^{\prime \alpha}$. Straightforward
calculations give
\begin{equation}
[ \, \AAA, \BB \, ] = ({\sf D}_{\cal A} {\cal B}^{\prime \, |KL|}
- {\sf D}_{\cal B} {\cal A}^{\prime \, |KL|}) \;
{\bf e}'_{K} \wedge {\bf e}'_{L} + \Delta'
\end{equation}
and
\begin{equation} \begin{array}{l}
\Delta' \equiv ({\cal A}^{\prime \, \sigma 5}{\cal B}^{\prime \, \mu \tau} -
{\cal B}^{\prime \, \sigma 5}{\cal A}^{\prime \, \mu \tau}) \\ \hspace{12ex}
\times \; (L^{-1})^{\nu}_{\, \omega} (\partial^{\, \prime}_{\sigma}
L^{\omega}_{\, \tau}) \cdot {\bf e}'_{\mu} \wedge {\bf e}'_{\nu},
\end{array} \end{equation}
where $\partial^{\, \prime}_{\sigma} = \partial / \partial x^{\prime \,
\sigma}$ and $L^{\alpha}_{\, \beta} = \partial x^{\alpha} / \partial
x^{\prime \, \beta}$. From these expressions we see that the suggested
definition of the commutator is indeed invariant in the sense that quantity
(35) is the same at any choice of the local Lorentz coordinates at the
considered point. At the same time, the above formulae show that in contrast
to the case of four-vector and five-vector fields, the expression for the
commutator of bivector fields in an arbitrary coordinate system acquires
an additional term proportional to the derivatives $\partial^{2} x^{\alpha}
/ \partial x^{\prime \mu} \partial x^{\prime \nu}$. To understand the origin
of this term and to obtain a coordinate-free expression for the commutator of
bivector fields, let us again consider operator (34) and evaluate it using
formulae (39) of part III, which express $\sf D$ in terms of $\dotnabla$ and
$\widehat{\bf M}$. One will have
\begin{equation} \begin{array}{rl}
[ \, {\sf D}_{\cal A}, {\sf D}_{\cal B} \, ] & \! = \,
[ \, \dotnabla_{\bf A'} + \widehat{\bf M}_{\bf A''}, \dotnabla_{\bf B'}
+ \widehat{\bf M}_{\bf B''} \, ] \\ & \! = \, [ \, \dotnabla_{\bf A'},
\dotnabla_{\bf B'} \, ] + [ \, \dotnabla_{\bf A'}, \widehat{\bf M}_{\bf B''}]
\\ & - \; [ \, \dotnabla_{\bf B'}, \widehat{\bf M}_{\bf A''}]
+ [ \, \widehat{\bf M}_{\bf A''}, \widehat{\bf M}_{\bf B''} \, ]
\end{array} \end{equation}
where $\bf A'$ and $\bf B'$ are the four-vector fields that correspond to
the $\cal E$-components of $\AAA$ and $\BB$ and $\bf A''$ and $\bf B''$ are
the four-vector bivector fields that correspond to their $\cal Z$-components.
By adding and subtracting the derivative $\dotnabla_{\bf [ \, A',B' \, ]}$,
one can present the right-hand side of the latter formula as a sum of three
terms:
\begin{eqnarray}
[ \, {\sf D}_{\cal A}, {\sf D}_{\cal B} \, ] & \hspace*{-5ex}
= \; (\dotnabla_{\bf A'} \dotnabla_{\bf B'} - \dotnabla_{\bf B'}
\dotnabla_{\bf A'} - \dotnabla_{\bf [ \, A',B' \, ]}) \nonumber \\
& \hspace*{-11ex} \left. \begin{array}{l} + \; (\widehat{\bf M}_{\bf A''}
\widehat{\bf M}_{\bf B''} - \widehat{\bf M}_{\bf B''}
\widehat{\bf M}_{\bf A''}) \\ + \; (\dotnabla_{\bf A'}
\widehat{\bf M}_{\bf B''} - \widehat{\bf M}_{\bf B''} \dotnabla_{\bf A'})
\rule{0ex}{3ex} \end{array} \right. \\ & \hspace*{-0.5ex} - \;
(\dotnabla_{\bf B'} \widehat{\bf M}_{\bf A''} - \widehat{\bf M}_{\bf A''}
\dotnabla_{\bf B'}) + \dotnabla_{\bf [ \, A',B' \, ]}. \nonumber
\end{eqnarray}
The first term is apparently the value of the Riemann tensor (corresponding
to derivative $\dotnabla$) on the bivector $\bf A' \wedge B'$, and by virtue
of equation (31), it equals $< \RS, \AAA^{\cal E} \otimes \BB^{\cal E} >$.
Likewise, the second term can be shown to equal $< \RS, \AAA^{\cal Z} \otimes
\BB^{\cal Z} >$. So, considering that according to equations (32), $< \RS,
\AAA^{\cal E} \otimes \BB^{\cal Z} > \; = \; < \RS, \AAA^{\cal Z} \otimes
\BB^{\cal E} > \; = 0$, we see that the sum of the first two terms in (40)
is exactly the contraction of $\RS$ with $\AAA$ and $\BB$.

The sum of all other terms in the right-hand side of formula (40) should
therefore be considered as arising from the commutator of $\AAA$ and $\BB$.
Simple calculations show that it equals
\begin{equation}
\dotnabla_{\bf [ \, A',B' \, ]} + \widehat{\bf M}_{
( \dotnablascr_{\bf A'} {\bf B''} - \dotnablascr_{\bf B'} {\bf A''})}.
\end{equation}
Thus, the commutator of $\AAA$ and $\BB$ should be such a bivector field
that its $\cal E$-component would correspond to the four-vector field
$\bf [\, A',B' \,]$ and its $\cal Z$-component would correspond to the
field of four-vector bivectors $(\dotnabla_{\bf A'}{\bf B''} -
\dotnabla_{\bf B'}{\bf A''})$. In an active regular basis associated with
an arbitrary coordinate system one therefore has
\begin{displaymath} \begin{array}{rcl}
( \, [ \, \AAA, \BB \, ] \, )^{\mu 5} & = & {\cal A}^{\sigma 5}
\partial_{\sigma} {\cal B}^{\mu 5} - {\cal B}^{\sigma 5} \partial_{\sigma}
{\cal A}^{\mu 5} \\ & = & {\cal A}^{\sigma 5} {\sf D}_{\sigma 5}
{\cal B}^{\mu 5} - {\cal B}^{\sigma 5} {\sf D}_{\sigma 5} {\cal A}^{\mu 5} \\
& = & {\sf D}_{\cal A} {\cal B}^{\mu 5} - {\sf D}_{\cal B} {\cal A}^{\mu 5}
\end{array} \end{displaymath} and \begin{displaymath} \begin{array}{l}
( \, [ \, \AAA, \BB \, ] \, )^{\mu \nu} = {\cal A}^{\sigma 5}
( \partial_{\sigma} {\cal B}^{\mu \nu} + \Gamma^{\mu}_{\; \tau \sigma}
{\cal B}^{\tau \nu} + \Gamma^{\nu}_{\; \tau \sigma} {\cal B}^{\mu \tau} ) \\
\hspace*{12ex} - \; {\cal B}^{\sigma 5} ( \partial_{\sigma}
{\cal A}^{\mu \nu} + \Gamma^{\mu}_{\; \tau \sigma} {\cal A}^{\tau \nu}
+ \Gamma^{\nu}_{\; \tau \sigma} {\cal A}^{\mu \tau} ) \\ \hspace*{6ex} =
( {\sf D}_{\cal A} {\cal B}^{\mu \nu} - {\sf D}_{\cal B} {\cal A}^{\mu \nu})
\\ \hspace*{6ex} + \; ( {\cal A}^{\sigma 5} {\cal B}^{\omega \tau} -
{\cal B}^{\sigma 5}{\cal A}^{\omega \tau}) \, ( \Gamma^{\nu}_{\; \tau \sigma}
\delta^{\mu}_{\omega} - \Gamma^{\mu}_{\; \tau \sigma} \delta^{\nu}_{\omega}),
\end{array} \end{displaymath}
where $\Gamma^{\mu}_{\; \tau \sigma}$ are the corresponding four-vector
connection coefficients associated with the derivative $\dotnabla$. Hence,
\begin{equation}
[ \, \AAA, \BB \, ] = ({\sf D}_{\cal A} {\cal B}^{|KL|} - {\sf D}_{\cal B}
{\cal A}^{|KL|}) \, {\bf e}_{K} \wedge {\bf e}_{L} + \Delta,
\end{equation}
where
\begin{equation}
\Delta = ( {\cal A}^{\sigma 5} {\cal B}^{\mu \tau} -
{\cal B}^{\sigma 5} {\cal A}^{\mu \tau} ) \cdot
\Gamma^{\nu}_{\; \tau \sigma} \cdot {\bf e}_{\mu} \wedge {\bf e}_{\nu}.
\end{equation}
Now, if the selected coordinate system is a local Lorentz one, the
connection coefficients $\Gamma^{\nu}_{\; \tau \sigma}$ at the considered
point are all zero, and formula (42) acquires the form of formula (35).
For an arbitrary coordinate system, the above connection coefficients can be
expressed in terms of the elements of the transformation matrix that relates
this coordinate system to a local Lorentz one. According to the standard
formula for tranformation of connection coefficients, one has
\begin{displaymath} \begin{array}{l}
\Gamma^{\prime \, \nu}_{\; \; \, \tau \sigma} = (L^{-1})^{\nu}_{\, \omega}
\Gamma^{\omega}_{\; \alpha \beta} L^{\alpha}_{\, \tau} L^{\beta}_{\, \sigma}
\\ \hspace*{4ex} + \; (L^{-1})^{\nu}_{\, \omega} (\partial_{\beta}
L^{\omega}_{\, \tau}) L^{\beta}_{\, \sigma} = (L^{-1})^{\nu}_{\, \omega}
(\partial^{\, \prime}_{\sigma} L^{\omega}_{\, \tau}),
\end{array} \end{displaymath}
where $\partial^{\, \prime}_{\sigma} = \partial / \partial x^{\prime \,
\sigma}$ and $L^{\alpha}_{\, \beta} = \partial x^{\alpha}/\partial x^{\prime
\, \beta}$, so formulae (42) and (43) with primes coincide with formulae
(37) and (38).

As one can see from expression (41) or from formulae (42) and (43), in
contrast to the commutators of four-vector and five-vector fields, which
are determined only by the differential structure of the manifold, for
determining the commutators of bivector fields one also needs to know the
Riemannian geometry of space-time. This dependece on the metric manifests
itself in several ways, one of which has to do with the existence of a set
of basis bivector fields for which all the commutators are identically zero
within a given region of space-time. It turns out that such a set exists if
and only if within this region space-time is flat. Let us prove this
statement.

If space-time is flat within a given region of space-time, then in the
latter one can introduce a system of global Lorentz coordinates, and from
formulae (42) and (43) it follows that for any regular basis ${\bf e}_{A}$
associated with these coordinates one will have $[ \, {\bf e}_{A} \wedge
{\bf e}_{B}, {\bf e}_{C} \wedge {\bf e}_{D} \, ] = {\bf 0}$.

Let us now suppose that in some finite region of space-time there exists
a set of ten basis bivector fields, $\, \AAA_{1}, \AAA_{2}, \dots , \AAA_{10}
\,$, such that everywhere in this region $[ \, \AAA_{i}, \AAA_{j} \, ] =
{\bf 0}$ at all $i$ and $j$. The latter equations are apparently equivalent
to the equations
\begin{displaymath}
[ \, {\bf A}'_{i}, {\bf A}'_{j} \, ] = {\bf 0}  \; \mbox{ and } \;
\dotnabla_{{\bf A}'_{i}} {\bf A}''_{j} - \dotnabla_{{\bf A}'_{j}}
{\bf A}''_{i} = {\bf 0},
\end{displaymath}
where ${\bf A}'_{i}$ denotes the four-vector field that corresponds to
$\AAA^{\cal E}_{i}$ and ${\bf A}''_{i}$ denotes the field of four-vector
bivectors that corresponds to $\AAA^{\cal Z}_{i}$.

It is evident that from the set ${\bf A}'_{1}, {\bf A}'_{2}, \ldots ,
{\bf A}'_{10}$ one can always select four fields that make up a four-vector
basis. Without any loss in generality one can take that these are the first
four fields of the set. Let us now recall one useful theorem about
four-vector fields: if the fields ${\bf V}_{i}$ ($i=1,2,3,4$) make up
a basis and are such that everywhere within a certain region of space-time
$[ \, {\bf V}_{i}, {\bf V}_{j} \, ] = {\bf 0}$ for all $i$ and $j$, then any
field $\bf U$ such that $[ \, {\bf U}, {\bf V}_{i} \, ] = {\bf 0}$ for all
$i$ within the considered region is a linear combination of the fields
${\bf V}_{i}$ {\em with constant coefficients}. Applying this theorem to the
considered set of fields, one obtains that each of the fields ${\bf A}'_{5},
\ldots , {\bf A}'_{10}$ is a linear combination with constant coefficients
of the fields ${\bf A}'_{1}, \ldots , {\bf A}'_{4}$, and consequently the
$\cal E$-components of the fields $\AAA_{5}, \ldots , \AAA_{10}$ are linear
combinations with constant coefficients of the $\cal E$-components of
$\AAA_{1}, \ldots , \AAA_{4}$. This fact enables one to construct a new
set of commuting fields: $\BB_{i} = \AAA_{i}$ for $i = 1,2,3,4$ and
$\BB_{i} = \AAA_{i} - C_{i1} \AAA_{1} - \ldots - C_{i4} \AAA_{4}$ for
$i = 5, \ldots ,10$, where the coefficients $C_{ij}$ are selected in such
a way that the $\cal E$-components of $\BB_{5}, \dots , \BB_{10}$ be zero.
Since all $C_{ij}$ are constants, everywhere within the considered region
of space-time one will have $[ \, \BB_{i}, \BB_{j} \, ] = {\bf 0}$ at any
$i$ and $j$. It is evident that for $1 \leq i \leq 4$ and $5 \leq j \leq 10$
the latter equation is equivalent to the following four-tensor equation:
\begin{displaymath}
\dotnabla_{{\bf B}'_{i}} {\bf B}''_{j} = {\bf 0},
\end{displaymath}
where ${\bf B}'_{i}$ and ${\bf B}''_{i}$ are related to $\BB_{i}$ the same
way as ${\bf A}'_{i}$ and ${\bf A}''_{i}$ are related to $\AAA_{i}$. This
latter equation means that in the considered region of space-time there
exist six linearly independent and covariantly constant fields of
four-vector bivectors, which is only possible if everywhere in this
region space-time is flat.

Let us now obtain an expression for the commutator of bivector fields that
would not involve coordinates. To this end let us observe that since the
derivative $\dotnabla$ is torsion-free, one has $\bf [\, A',B' \,] =
\dotnabla_{\bf A'}{\bf B'}-\dotnabla_{\bf B'}{\bf A'}$. Consequently, the
bivector field $[ \, \AAA, \BB \, ]$ corresponds to the following pair of
four-tensor fields:
\begin{displaymath}
( \; \dotnabla_{\bf A'}{\bf B'} - \dotnabla_{\bf B'}{\bf A'} \, , \,
\dotnabla_{\bf A'}{\bf B''} - \dotnabla_{\bf B'}{\bf A''} \; ),
\end{displaymath}
and by using formula (25) of the previous section one can easily obtain that
\begin{equation} \begin{array}{l}
[ \, \AAA, \BB \, ] = ({\sf D}_{{\cal A}^{\cal E}} \BB^{\cal E} -
{\sf D}_{{\cal B}^{\cal E}} \AAA^{\cal E}) \\ \hspace*{14ex} + \;
({\sf D}_{{\cal A}^{\cal E}} \BB^{\cal Z} - {\sf D}_{{\cal B}^{\cal E}}
\AAA^{\cal Z})^{\cal Z}.
\end{array} \end{equation}
The expression in the right-hand side is obviously a part of the quantity
${\sf D}_{\cal A} \BB - {\sf D}_{\cal B} \AAA$, and compared to the latter
it lacks the following three terms:
\begin{displaymath} \begin{array}{l}
({\sf D}_{{\cal A}^{\cal E}} \BB^{\cal Z} - {\sf D}_{{\cal B}^{\cal E}}
\AAA^{\cal Z})^{\cal E}, \; \; ({\sf D}_{{\cal A}^{\cal Z}} \BB^{\cal E} -
{\sf D}_{{\cal B}^{\cal Z}} \AAA^{\cal E}) \, , \\ \hspace*{10ex}
\mbox{and} \;
({\sf D}_{{\cal A}^{\cal Z}} \BB^{\cal Z} -
{\sf D}_{{\cal B}^{\cal Z}} \AAA^{\cal Z}) \, . \rule{0ex}{3ex}
\end{array} \end{displaymath}
Writing out explicitly the expressions for these terms in components, one
finds that the sum of the first two terms {\em is identically zero}, so
instead of (44) one can write
\begin{displaymath}
[ \, \AAA, \BB \, ] = ({\sf D}_{\cal A} \BB^{\cal E} -
{\sf D}_{\cal B} \AAA^{\cal E}) + ({\sf D}_{{\cal A}^{\cal E}}
\BB^{\cal Z} - {\sf D}_{{\cal B}^{\cal E}} \AAA^{\cal Z})
\end{displaymath}
or
\begin{equation}
[ \, \AAA, \BB \, ] = ({\sf D}_{\cal A} \BB - {\sf D}_{\cal B} \AAA) -
({\sf D}_{{\cal A}^{\cal Z}} \BB^{\cal Z} - {\sf D}_{{\cal B}^{\cal Z}}
\AAA^{\cal Z}).
\end{equation}

Looking at the latter formula, one may think that in the above analysis we
did something wrong and that the ``right'' expression for the commutator
should be
\begin{equation}
[ \, \AAA, \BB \, ] = {\sf D}_{\cal A} \BB - {\sf D}_{\cal B} \AAA.
\end{equation}
To see if this is possible, let us analyze more closely how we have arrived
at formula (45).

It is evident that our definition of the bivector commutator is a consequence
of our definition of the tensor $\RS$ and of the assumption that the two
quantities are related by equation (36). It is also evident that our
definition of $\RS$ implicitly includes the assumption that the commutators
of basis bivector fields corresponding to a regular five-vector basis
associated with a system of local Lorentz coordinates are all zero at the
origin of the latter. Since this latter assumption is based solely on the
analogy with the case of four-vector fields, which may very well be wrong,
one should regard it as merely a reasonable initial hypothesis the
correctness of which is to be confirmed or disproved by further analysis. We
now see that judging by the formulae obtained above, it may seem more natural
to define the commutator of two bivector fields according to formula (46)
rather than to formula (45). Compared to the former, the latter lacks the
term
\begin{displaymath} \begin{array}{rcl}
({\sf D}_{{\cal A}^{\cal Z}} \BB^{\cal Z} - {\sf D}_{{\cal B}^{\cal Z}}
\AAA^{\cal Z}) & = & \widehat{\bf M}_{{\cal A}^{\cal Z}} \BB^{\cal Z} -
\widehat{\bf M}_{{\cal B}^{\cal Z}} \AAA^{\cal Z} \\ & = &
( 2 g_{\alpha \beta} \, {\cal A}^{\alpha \mu} {\cal B}^{\beta \nu}) \,
{\bf e}_{\mu} \wedge {\bf e}_{\nu}, \rule{0ex}{3ex}
\end{array} \end{displaymath}
which is simply a contraction of $\AAA$ and $\BB$ with a Lorentz-invariant
five-tensor, so its addition to the commutator will not change anything
in essence. Such a redefinition of the commutator implies that from the
left-hand side of formula (36) one should now subtract the term
\begin{displaymath}
{\sf D}_{({\sf D}_{{\cal A}^{\cal Z}} {\cal B}^{\cal Z} -
{\sf D}_{{\cal B}^{\cal Z}} {\cal A}^{\cal Z})} = 2 \, (
\widehat{\bf M}_{{\cal A}^{\cal Z}} \widehat{\bf M}_{{\cal B}^{\cal Z}} -
\widehat{\bf M}_{{\cal B}^{\cal Z}} \widehat{\bf M}_{{\cal A}^{\cal Z}}),
\end{displaymath}
which is also a contraction of $\AAA$ and $\BB$ with a Lorentz-invariant
five-tensor, so it will not lead to any essential changes in $\RS$ either.
If one does define the commutator for bivector fields this way, then
instead of formula (27) one will have
\begin{equation}
\widehat{\RS}_{KLMN} \equiv {\sf D}_{KL} {\sf D}_{MN} -{\sf D}_{MN}
{\sf D}_{KL} - {\sf D}_{[KL,MN]},
\end{equation}
where ${\sf D}_{[KL,MN]} \equiv {\sf D}_{[{\bf e}_{K} \wedge {\bf e}_{L},
{\bf e}_{M} \wedge {\bf e}_{N}]}$, but definitions (28) and (29) will be
the same. Since for any active regular basis associated with a system of
local Lorentz coordinates, at the origin of the latter one still has
\begin{equation}
[ \, {\bf e}_{\kappa} \wedge {\bf e}_{5}, {\bf e}_{\mu} \wedge
{\bf e}_{5} \, ] = [ \, {\bf e}_{\kappa} \wedge {\bf e}_{5},
{\bf e}_{\mu} \wedge {\bf e}_{\nu} \, ] = 0,
\end{equation}
formulae (31) and (32) for the components of $\RS$ will still be valid.
However, since for the same basis one will now have
\begin{equation} \begin{array}{l}
\hspace*{-1ex} [ \, {\bf e}_{\kappa} \wedge {\bf e}_{\lambda}, {\bf e}_{\mu}
\wedge {\bf e}_{\nu} \, ] \\ \hspace*{7ex} = 2 \, \{ \, g_{\kappa \mu} \,
{\bf e}_{\lambda} \wedge {\bf e}_{\nu} - g_{\lambda \mu} \, {\bf e}_{\kappa}
\wedge {\bf e}_{\nu} \\ \hspace*{14ex} - \; g_{\kappa \nu} \,
{\bf e}_{\lambda} \wedge {\bf e}_{\mu} + g_{\lambda \nu} \, {\bf e}_{\kappa}
\wedge {\bf e}_{\mu} \, \} ,
\end{array} \end{equation}
the expression for the components ${\sf R}^{\alpha}_{\; \beta \, \kappa
\lambda \mu \nu}$ will acquire an additional term equal to
\begin{displaymath} \begin{array}{l}
- \, 2 \, \{ \, g_{\kappa \mu} (M_{\lambda \nu})^{\alpha}_{\; \beta} -
g_{\lambda \mu} (M_{\kappa \nu})^{\alpha}_{\; \beta} \\ \hspace*{10ex}
- \; g_{\kappa \nu} (M_{\lambda \mu})^{\alpha}_{\; \beta} +
g_{\lambda \nu} (M_{\kappa \mu})^{\alpha}_{\; \beta} \, \} ,
\end{array} \end{displaymath}
so the right-hand side of formula (33) will simply change its sign.

The latter formula suggests still another way of choosing the commutator for
bivector fields, namely,
\begin{equation}
[ \, \AAA, \BB \, ] = ({\sf D}_{\cal A} \BB - {\sf D}_{\cal B} \AAA) -
{\scriptstyle \frac{1}{2}} ({\sf D}_{{\cal A}^{\cal Z}} \BB^{\cal Z} -
{\sf D}_{{\cal B}^{\cal Z}} \AAA^{\cal Z}).
\end{equation}
Compared to the right-hand side of equation (46) the right-hand side of the
latter equation lacks the term $g_{\alpha \beta} \, {\cal A}^{\alpha \mu}
{\cal B}^{\beta \nu} \, {\bf e}_{\mu} \wedge {\bf e}_{\nu}$. Definitions (28),
(29), and (47) will still be valid, and so will formulae (31), (32) and (48).
However, instead of equation (49) one will now have
\begin{equation} \begin{array}{l}
\hspace*{-1ex} [ \, {\bf e}_{\kappa} \wedge {\bf e}_{\lambda}, {\bf e}_{\mu}
\wedge {\bf e}_{\nu} \, ] \\ \hspace*{7ex} = g_{\kappa \mu} \,
{\bf e}_{\lambda} \wedge {\bf e}_{\nu} - g_{\lambda \mu} \, {\bf e}_{\kappa}
\wedge {\bf e}_{\nu} \\ \hspace*{14ex} - \; g_{\kappa \nu} \,
{\bf e}_{\lambda} \wedge {\bf e}_{\mu} + g_{\lambda \nu} \, {\bf e}_{\kappa}
\wedge {\bf e}_{\mu},
\end{array} \end{equation}
so instead of formula (33) one will simply have
\begin{equation}
{\sf R}^{\alpha}_{\; \beta \, \kappa \lambda \mu \nu} = 0.
\end{equation}
In the rest of this paper it will be assumed that the commutator for bivector
fields is defined according to formula (50).

\vspace{3ex}

It is evident that $\RS$ is not the only analog of the Riemann tensor that
one can think of. There are at least two more analogs of the latter that may
be of interest. Both of them can be obtained by considering operator (47),
but now one should let the latter act, in the one case, on five-vector fields
and, in the other case, on the fields of five-vector bivectors. As before, it
is easy to prove that on both types of fields $\widehat{\RS}_{KLMN}$ acts as
a local operator. This enables one to define two sets of quantities similar
to ${\sf R}^{\alpha}_{\; \beta \, KLMN}$ and then construct out of them
two five-tensors, which I will denote as $\RS^{(5)}$ and $\RS^{(10)}$.
Accordingly, the tensor $\RS$ itself can now be denoted as $\RS^{(4)}$
and the notation $\RS$ reserved for those cases where one needs to write
a formula valid for all three tensors.

Let us now calculate the components of $\RS^{(5)}$ and $\RS^{(10)}$. For
simplicity, in the following formulae I will omit the lables $\rule{0ex}
{1ex}^{(5)}$ and $\rule{0ex}{1ex}^{(10)}$, since the upper indices of the
components will unambiguously indicate which tensor they belong to. Let us
again consider an arbitrary active regular basis associated with some system
of local Lorentz coordinates. By virtue of equations (48) and (51), one has
\begin{displaymath}
{\sf D}_{[\kappa 5, \mu 5]} = {\sf D}_{[\kappa 5, \mu \nu]} =
{\sf D}_{[\kappa \lambda, \mu 5]} = 0
\end{displaymath}
and
\begin{equation} \begin{array}{l}
{\sf D}_{[\kappa \lambda , \mu \nu]} =  g_{\kappa \mu}
{\sf D}_{\lambda \nu} - g_{\lambda \mu} {\sf D}_{\kappa \nu} \\
\hspace*{19ex} - \; g_{\kappa \nu} {\sf D}_{\lambda \mu} +
g_{\lambda \nu} {\sf D}_{\kappa \mu}.
\end{array} \end{equation}
Consequently,
\begin{displaymath} \begin{array}{rcl}
{\sf R}^{A}_{\; B \, \kappa 5 \mu 5} & = & \partial_{\kappa}
G^{A}_{\; B \mu 5} - \partial_{\mu} G^{A}_{\; B \kappa 5} \\ & &
\hspace*{4ex} + \; G^{A}_{\; C \kappa 5} G^{C}_{\; B \mu 5} -
G^{A}_{\; C \mu 5} G^{C}_{\; B \kappa 5} \\ & = & \partial_{\kappa}
G^{A}_{\; B \mu} - \partial_{\mu} G^{A}_{\; B\kappa} \\ & &
\hspace*{4ex} + \; G^{A}_{\; C \kappa} G^{C}_{\; B \mu} -
G^{A}_{\; C \mu} G^{C}_{\; B \kappa},
\end{array} \end{displaymath}
where $G^{A}_{\; B \mu}$ are the five-vector connection coefficients
associated with the derivative $\dotnabla$. By using formulae (47), (48),
and (50) of part II, one finds that
\begin{equation}
{\sf R}^{\alpha}_{\; 5 \, \kappa 5 \mu 5} = {\sf R}^{5}_{\; 5 \, \kappa 5
\mu 5} = 0 \; \mbox{ and } \; {\sf R}^{\alpha}_{\; \beta \, \kappa 5 \mu 5}
= R^{\alpha}_{\; \beta\kappa\mu},
\end{equation}
where $R^{\alpha}_{\; \beta \kappa \mu}$ are the components of the Riemann
tensor in the associated four-vector basis. In a similar manner, since the
derivative $\dotnabla$ is torsion-free, one finds that
\begin{equation} \begin{array}{rl}
{\sf R}^{5}_{\; \beta \, \kappa 5 \mu 5} \hspace*{-2ex} & =
g_{\beta \kappa ; \mu} - g_{\beta \mu ; \kappa} + g_{\beta \omega} \,
( G^{\omega}_{\; \kappa \mu} - G^{\omega}_{\; \mu \kappa} ) \\ & =
2 \, g_{\beta \omega} \Gamma^{\omega}_{\; [\kappa \mu]} = 0.
\end{array} \end{equation}
By using formulae (24), for the ``mixed'' components one obtains
\begin{displaymath} \begin{array}{rl}
{\sf R}^{A}_{\; B \, \kappa 5 \mu \nu} \hspace*{-2ex} & = \partial_{\kappa}
G^{A}_{\; B \mu \nu} + G^{A}_{\; C \kappa 5} G^{C}_{\; B \mu \nu} -
G^{A}_{\; C \mu \nu} G^{C}_{\; B \kappa 5} \\ & = \partial_{\kappa}
(M_{\mu \nu})^{A}_{\; B} - (M_{\kappa 5})^{A}_{\; C}
(M_{\mu \nu})^{C}_{\; B} \rule{0ex}{3ex} \\ & \hspace*{22ex} + \;
(M_{\mu \nu})^{A}_{\; C} (M_{\kappa 5})^{C}_{\; B}, \rule{0ex}{3ex}
\end{array} \end{displaymath}
and since the selected basis is associated with a local Lorentz coordinate
system, one has
\begin{displaymath} \begin{array}{rl}
{\sf R}^{A}_{\; B \, \kappa 5 \mu \nu} \hspace*{-2ex} & = - \, [ \,
M_{\kappa 5}, M_{\mu \nu} \, ]^{A}_{\, B} \\ & = - \, g_{\kappa \mu}
(M_{5 \nu})^{A}_{\; B} + g_{\kappa \nu} (M_{5 \mu})^{A}_{\; B}
\rule{0ex}{3ex} \\ & = \; \delta^{A}_{\; 5} \, ( g_{\kappa \mu}
g_{\nu B} - g_{\kappa \nu} g_{\mu B} ). \rule{0ex}{3ex}
\end{array} \end{displaymath}
Hence,
\begin{displaymath} \begin{array}{l}
{\sf R}^{5}_{\; 5 \, \kappa 5 \mu \nu} =
{\sf R}^{\alpha}_{\; 5 \, \kappa 5 \mu \nu} =
{\sf R}^{\alpha}_{\; \beta \, \kappa 5 \mu \nu} = 0 \\
{\sf R}^{5}_{\; \beta \, \kappa 5 \mu \nu} =
g_{\kappa \mu} g_{\nu \beta} - g_{\kappa \nu} g_{\mu \beta}.
\end{array} \end{displaymath}
Finally, using formulae (24) again and considering (53), one obtains that
\begin{displaymath}
{\sf R}^{5}_{\; 5 \, \kappa \lambda \mu \nu} =
{\sf R}^{\alpha}_{\; 5 \, \kappa \lambda \mu \nu} =
{\sf R}^{5}_{\; \beta \, \kappa \lambda \mu \nu} =
{\sf R}^{\alpha}_{\; \beta \, \kappa \lambda \mu \nu} = 0.
\end{displaymath}

The components of the tensor $\RS^{(10)}$ are defined according to the
formula
\begin{displaymath}
\widehat{\RS}_{KLMN} \, ({\bf e}_{C} \wedge {\bf e}_{D}) =
{\bf e}_{A} \wedge {\bf e}_{B} \, {\sf R}^{|AB|}_{\hspace*{3ex} CD \, KLMN},
\end{displaymath}
and can be easily calculated by noting that for any two five-vector fields
$\bf u$ and $\bf v$,
\begin{displaymath}
\widehat{\RS}_{KLMN} \, ({\bf u} \wedge {\bf v}) = (\widehat{\RS}_{KLMN} \,
{\bf u}) \wedge {\bf v} + {\bf u} \wedge (\widehat{\RS}_{KLMN} \, {\bf v}),
\end{displaymath}
so in any five-vector basis,
\begin{equation} \begin{array}{l}
\hspace*{-2ex} {\sf R}^{AB}_{\hspace*{2ex} CD \, KLMN} = \delta^{A}_{\; C}
{\sf R}^{B}_{\; D \, KLMN} - \delta^{B}_{\; C} {\sf R}^{A}_{\; D \, KLMN} \\
\hspace*{10ex}
- \; \delta^{A}_{\; D} {\sf R}^{B}_{\; C \, KLMN}
+ \delta^{B}_{\; D} {\sf R}^{A}_{\; C \, KLMN}. \rule{0ex}{3ex}
\end{array} \hspace*{-1ex} \end{equation}

\vspace{3ex}

Tensors $\RS^{(4)}$, $\RS^{(5)}$, and $\RS^{(10)}$ have two important
properties, one of which is algebraic and follows from the fact that
Poincare transformations conserve the scalar product $g$, and the other
is differential and is of the kind possessed by all analogs of the Riemann
tensor.

The first property can be derived either by using the expressions for the
components of these tensors and recalling that the Riemann tensor is
antisymmetric in its first two indices, or directly---by writing out the
equation that expresses the fact that Poincare transformations conserve the
scalar product $g$ and differentiating it twice. Either way, for the
components of $\RS^{(4)}$ one obtains the equation
\begin{equation}
g_{\alpha \omega} {\sf R}^{\omega}_{\; \beta \, KLMN} +
g_{\beta \omega} {\sf R}^{\omega}_{\; \alpha \, KLMN} = 0,
\end{equation}
and similar equations for the components of $\RS^{(5)}$ and $\RS^{(10)}$.

Before we turn to the second property, let us see if the commutator of
bivector fields satisfies the Jacobi identity. To this end let us consider
the sum
\begin{displaymath}
[ \, \AAA, [ \, \BB, \CC \, ]] + [ \, \BB, [ \, \CC, \AAA \, ]] +
[ \, \CC, [ \, \AAA, \BB \, ]],
\end{displaymath}
where $\AAA$, $\BB$, and $\CC$ are arbitrary fields of five-vector
bivectors. By using definition (50), one can easily find that it equals
\begin{displaymath} \begin{array}{l}
{\bf e}_{K} \wedge {\bf e}_{L} \cdot R^{|KL|}_{\hspace{3.5ex} AB \, CDEF} \\
\rule{0ex}{4ex} \hspace{2ex} \times ( \, {\cal A}^{|AB|} \, {\cal B}^{|CD|}
\, {\cal C}^{|EF|} \; + \mbox{ cyclic permutations} \, ).
\end{array} \end{displaymath}
Direct calculation shows that the only term not identically zero in the
latter expression is
\begin{equation} \begin{array}{l}
\hspace*{1ex} {\bf e}_{\kappa} \wedge {\bf e}_{\lambda} \cdot
R^{| \kappa \lambda |}_{\hspace{3ex} \alpha \beta \, \mu 5 \nu 5} \cdot
( \, {\cal A}^{| \alpha \beta |} {\cal B}^{\mu 5} {\cal C}^{\nu 5} \\
\rule{0ex}{3ex} \hspace{10ex} + \; {\cal B}^{| \alpha \beta |}
{\cal C}^{\mu 5} {\cal A}^{\nu 5} + {\cal C}^{| \alpha \beta |}
{\cal A}^{\mu 5} {\cal B}^{\nu 5} \, ) \\ = \; {\bf e}_{\kappa} \wedge
{\bf e}_{\beta} \cdot R^{\kappa}_{\; \alpha \mu \nu} \cdot ( \,
{\cal A}^{\alpha \beta} {\cal B}^{\mu 5} {\cal C}^{\nu 5} \rule{0ex}{3ex} \\
\rule{0ex}{3ex} \hspace{10ex} + \; {\cal B}^{\alpha \beta} {\cal C}^{\mu 5}
{\cal A}^{\nu 5}+{\cal C}^{\alpha \beta}{\cal A}^{\mu 5}{\cal B}^{\nu 5} \,).
\end{array} \end{equation}
Denoting this quantity as $- {\bf \Delta}(\AAA ,\BB ,\CC )$, one obtains
the following equation that replaces the Jacobi identity in the case of
commutators of bivectors fields:
\begin{equation} \begin{array}{l}
[ \, \AAA, [ \, \BB, \CC \, ]] + [ \, \BB, [ \, \CC, \AAA \, ]] \\
\rule{0ex}{3ex} \hspace{5ex} + \; [ \, \CC, [ \, \AAA, \BB \, ]] +
{\bf \Delta}(\AAA ,\BB ,\CC ) = {\bf 0}.
\end{array} \end{equation}
The last term in the left-hand side of this equation is identically zero
only if space-time is flat, so this is another example of how the properties
of bivector commutators depend on the Riemannian geometry of space-time.

Let us finally discuss the analog of the Bianchi identity for tensors $\RS$.
Let us first derive this identity in a general form applicable to any
derivative that depends linearly on its argument. The latter can be a vector
or a tensor of any kind provided that for the fields of such vectors
or tensors one can define the notion of a commutator. In the following
derivation I will denote this derivative as $D$ and the vectors or tensors
that can be its argument as $\sf A$, $\sf B$, and $\sf C$. By virtue of the
Jacobi identity for ordinary commutators of operators $D$, one has
\begin{displaymath} \begin{array}{rcl}
0 & = & {[} \, D_{\sf A}, {[} \, D_{\sf B}, D_{\sf C} \, {]]} \; +
\mbox{ cyclic permutations} \\ & = & {[} \, D_{\sf A}, ( \, D_{\sf B}
D_{\sf C} - D_{\sf C} D_{\sf B} - D_{\sf {[} \, B,C \, {]}} \, ) \, {]}
\rule{0ex}{3ex} \\ & & + \; {[} \, D_{\sf A}, D_{\sf {[} \, B,C \, {]}} \,
{]} \; + \mbox{ cyclic permutations} \\ & = & {[} \, D_{\sf A}, ( \,
D_{\sf B} D_{\sf C} - D_{\sf C} D_{\sf B} - D_{\sf {[} \, B,C  \,{]}} \, )
\, {]} \rule{0ex}{3ex} \\ & & + \; ( \, D_{\sf A} D_{\sf {[}  \,B,C \, {]}}
- D_{\sf {[} \, B,C \, {]}} D_{\sf A} -  D_{\sf {[} \, A, \, {[} \, B,C \,
{]]}} \, ) \\ & & + \; \, D_{\sf {[} \, A, \, {[} \, B, \, C{]]}} \; +
\mbox{ cyclic permutations}.
\end{array} \end{displaymath}
Introducing the notation $\Re ({\sf A,B}) \equiv D_{\sf A} D_{\sf B} -
D_{\sf B} D_{\sf A} - D_{\sf {[}A,B{]}} \,$, one can rewrite the latter
equation as
\begin{equation} \begin{array}{l}
0 \; = \; {[} \, D_{\sf A}, \Re ({\sf B,C}) \, {]} \; + \; \Re
({\sf A, \, {[} \, B,C \, {]}} \, ) \\ \rule{0ex}{3ex} \hspace{2ex} + \;
D_{\sf {[} \, A, \, {[} \, B, C \, {]]}} \; + \mbox{ cyclic permutations }.
\end{array} \end{equation}
In a standard way one can show that $\Re ({\sf A,B})$ is a local operator
and therefore can be regarded as a tensor of rank $(1,1)$ over the space of
the vectors or tensors upon which act the operators $D$. One can then show
that $[ \, D_{\sf A}, \Re ({\sf B,C}) \, ] = D_{\sf A} \Re ({\sf B,C})$,
where in the right-hand side the operator $D_{\sf A}$ acts on $\Re
({\sf B,C})$ as on a tensor. Finally, since $\Re ({\sf A,B})$ is a linear
function of its arguments and is antisymmetric in them, one can present
it as a contraction of a certain tensor $\RR$ independent of $\sf A$ and
$\sf B$ with the antisymmetrized tensor product of $\sf A$ and $\sf B$.
Denoting this latter product as $\sf A \wedge B$ and substituting all this
into equation (60), one obtains the identity
\begin{equation} \begin{array}{l}
D_{\sf A} < \RR ,{\sf B \wedge C} > + \; D_{\sf B} < \RR ,{\sf C \wedge A} >
\\ \rule{0ex}{2ex} + \; D_{\sf C} < \RR ,{\sf A \wedge B} > - < \RR ,{[} \,
{\sf A,B} \, {]} \wedge {\sf C} > \\ \rule{0ex}{2ex} \, - < \RR , {[} \,
{\sf B,C} \, {]} \wedge {\sf A} > - < \RR , {[} \, {\sf C,A} \, {]} \wedge
{\sf B} > \\ \rule{0ex}{2ex} + \; D_{\sf \, ( \, {[} \, A, \, {[} \, B, C \,
{]]} \; + \; {[} \, B, \, {[} \, C, A \, {]]} \; + \; {[} \, C, \, {[} \,
A, B \, {]]} \, )} \; = \; 0.
\end{array} \end{equation}

When $D$ is an ordinary covariant derivative, $\RR$ is the Riemann tensor.
The last term in the left-hand side of equation (61) is then identically
zero, since the commutators of four-vector fields satisfy the Jacobi identity,
and on comparing the remaining terms with the right-hand side of equation
(45b) of part V, which defines the exterior derivative of nonscalar-valued
four-vector 2-forms, one obtains the usual Bianchi identity: $\bf d R = 0$.

When $D$ is a bivector derivative, $\RR$ is one of the tensors $\RS$, and
identity (61) acquires the form
\begin{equation} \begin{array}{l}
{\sf D}_{\cal A} < \RS ,{\BB \wedge \CC} > + \; {\sf D}_{\cal B} < \RS ,
{\CC \wedge \AAA} > \\ \; + \; {\sf D}_{\cal C} < \RS ,{\AAA \wedge \BB} >
- < \RS ,{[} \, {\AAA , \BB} \, {]} \wedge {\CC} > \\ \; - < \RS , {[} \,
{\BB, \CC} \, {]} \wedge {\AAA} > - < \RS , {[} \, {\CC, \AAA} \, {]}
\wedge {\BB} > \\ \hspace{15ex} = \; {\sf D}_{{\bf \Delta}({\cal A,B,C})}.
\end{array} \end{equation}
Since both sides of this equation are linear and antisymmetric in $\AAA$,
$\BB$, and $\CC$, each of them can be presented as a contraction of a
certain five-tensor with the antisymmetrized tensor product of these three
bivector fields. Considering that the left-hand side of equation (62) in
its structure resembles the left-hand of equation (45b) of part V, one may
suppose that the tensor corresponding to it is the analog of the exterior
derivative for $\RS$, and one may denote it as ${\bf d}^{\scriptstyle \sf
(D)} \RS$. The right-hand side of equation (62) equals
\begin{equation} \begin{array}{l}
\widehat{\bf M}_{\sigma \tau} \cdot R^{\sigma}_{\; \alpha \mu \nu} \,
( \, {\cal A}^{\alpha \tau} {\cal B}^{\mu 5} {\cal C}^{\nu 5} \\
\hspace{8ex} + \; {\cal B}^{\alpha \tau} {\cal C}^{\mu 5} {\cal A}^{\nu 5}
+ {\cal C}^{\alpha \tau}{\cal A}^{\mu 5}{\cal B}^{\nu 5} \,),
\end{array} \end{equation}
and one may denote the tensor corresponding to it as $\widehat{\bf M}_
{\bf \Delta}$. Since $\AAA$, $\BB$, and $\CC$ are arbitrary bivector
fields, one obtains the equation
\begin{equation}
{\bf d}^{\scriptstyle \sf (D)} \RS = \widehat{\bf M}_{\bf \Delta},
\end{equation}
which should be regarded as the analog of the Bianchi identity for tensors
$\RS$.

\vspace{3ex} \begin{flushleft}
D. \it A more general case of five-vector \\ \hspace{2ex} affine connection
\end{flushleft}
So far I have considered only one particular case of the connection for
five-vector fields, which corresponds to an ordinary covariant derivative
and where there exists a local symmetry described in section 3 of part II.
Let us now see what the requirement of the same local symmetry will give
in the case of the five-vector covariant derivative.

Repeating the reasoning presented in the cited section of part II, one
finds that in any regular basis associated with a system of local Lorentz
coordinates at the considered point, the only connection coefficients
that do not have to be zero are $H^{\alpha}_{\; \beta 5}$,
$H^{5}_{\; \beta \mu}$, and $H^{5}_{\; 55}$ and that one should have
\begin{equation}
H^{\alpha}_{\; \beta 5} \propto \delta^{\alpha}_{\, \beta}, \; \;
H^{5}_{\; \beta \mu} \propto g_{\beta \mu} \; \mbox{ and } \;
H^{5}_{\; 55} = {\rm const}.
\end{equation}
Requiring also that with respect to this connection the metric tensor be
covariantly constant, one finds that the quantities $H_{\alpha \beta 5}
\equiv g_{\alpha \omega} H^{\omega}_{\; \beta 5}$ should be antisymmetric
in $\alpha$ and $\beta$, which is compatible with the first relation in
(65) only if $H^{\alpha}_{\; \beta 5} = 0$. Selecting the length of the
fifth basis five-vector so that the proportionality factor between
$H^{5}_{\; \beta \mu}$ and $g_{\beta \mu}$ be minus unity, one finally
obtains that
\begin{equation}
H^{5}_{\; \beta \mu} = - \, g_{\beta \mu} \; \mbox{ and } \;
H^{5}_{\; 55} = \omega,
\end{equation}
where $\omega$ is a constant of dimension $(interval)^{-2}$, and all other
connection coefficients at the considered point for such a basis are zero.

Let us now reformulate the results obtained in terms of parallel transport.
By virtue of condition (18) of part V, for the connection coefficients
corresponding to the four-vector basis associated with the same coordinate
system one should have
\begin{equation}
\Upsilon^{\alpha}_{\; \beta A} = H^{\alpha}_{\; \beta A},
\end{equation}
which is the analog of equation (48) of part II. From the results
we have just obtained it then follows that at the considered point all
$\Upsilon^{\alpha}_{\; \beta A}$ are zero, which means that for four-vector
fields the derivative $\plaision$ coincides with derivative $\dotnabla$, and
the parallel transport of four-vectors corresponding to $\plaision$ (defined
in accordance with the interpretation of $\plaision$ discussed in section B
of part V) coincides with their transport defined by the metric. This
result can be obtained more directly by observing that according to what has
been said after equations (66), the $\cal Z$-components of the transported
five-vectors in the considered case are the same as in the case of the
connection examined in part II. Let us now see what happens with the
$\cal E$-components.

It is evident that at $\omega = 0$ the derivative $\plaision$ coincides with
the derivative $\dotnabla$ for five-vector fields as well. Consequently, in
this case a five-vector from $\cal E$ is transported into a five-vector from
$\cal E$ of the same length (relative to $\bf 1$), and the change in the
fifth component of any five-vector $\bf u$ in an active regular basis at each
infinitesimal step of the transport equals the scalar product $g$ of $\bf u$
with the infinitesimal tangent five-vector that characterizes the element of
the transport path covered at that step. At $\omega \neq 0$ the length of
the five-vectors from $\cal E$ is not conserved when the transport is made
along timelike or spacelike curves. Consequently, the change in the
$\cal E$-component of any five-vector $\bf u$ at each infinitesimal step
of the transport will in general be a sum of a quantity proportional to the
mentioned scalar product of $\bf u$ with the infinitesimal tangent vector
and of a quantity equal to the change in the initial $\cal E$-component of
$\bf u$ at that step.

Since at any $\omega$ the derivative $\plaision$ for four-vector fields
coincides with the derivative $\dotnabla$, for such a connection four-vector
torsion is identically zero. In view of this, one may wish to examine
some other case of five-vector affine connection, where the constraints
on the connection coefficients would not be so stringent and four-vector
torsion could be nonvanishing.

As it has been shown in section A, at any five-vector connection with respect
to which the metric tensor is covariantly constant there exists a relation
between the derivatives $\plaision$ and $\sf D$ of any scalar, four-vector
or four-tensor field, expressed by equation (2), where $\sigma({\bf u})$ is
a contraction of $\bf u$ with a certain bivector-valued five-vector 1-form
$\widetilde{\bf s}$ whose components $s^{\alpha 5}_{\hspace{2ex} A}$ in any
regular basis are fixed and components $s^{\alpha \beta}_{\hspace{2ex} A}$
are uniquely determined by four-vector torsion. The inverse theorem is also
valid: if for a given five-vector connection $\plaision$ there exists such
a linear map $\sigma : \, \FF \rightarrow \FF \wedge \FF$ that equation (2)
holds for any scalar, four-vector or four-tensor field, then relative to this
connection the metric tensor is covariantly constant. Consequently, requiring
$g$ to be covariantly constant is equivalent to requiring equation (2) to
hold at some $\sigma$ for all four-vector fields, since, as it has been shown
in section A, the latter condition is sufficient for equation (2) to hold
for all scalar and all four-tensor fields as well.

Since derivatives $\plaision$ and $\sf D$ both preserve the correspondence
between four- and five-vectors, from the validity of equation (2) for
four-vector fields it follows that for any five-vector field $\bf w$
\begin{equation}
[ \, \plaision_{\bf u} \, {\bf w} ]^{\cal Z} =
[ {\sf D}_{\sigma({\bf u})} \, {\bf w} ]^{\cal Z}
\end{equation}
at the same $\sigma$. Considering this, it seems natural to examine the case
where for any $\bf w$
\begin{equation}
\plaision_{\bf u} \, {\bf w} = {\sf D}_{\sigma({\bf u})} \, {\bf w}.
\end{equation}
Since equation (68) holds without any additional assumptions, requirement
(69) imposes constraints only on those five-vector connection coefficients
that determine the $\cal E$-component of the covariant derivative. From
equation (69) it follows that in any five-vector basis
\begin{displaymath}
H^{A}_{\; BC} = G^{A}_{\; BKL} \, s^{|KL|}_{\hspace*{3ex} C},
\end{displaymath}
and by using formulae (19), (22), (23), and (26) of section B, for the
five-vector connection coefficients in an arbitrary active regular basis one
finds that
\begin{equation} \begin{array}{l}
H^{A}_{\; B \mu} = G^{A}_{\; B \mu} + (M_{\sigma \tau})^{A}_{\; B} \,
s^{|\sigma \tau|}_{\hspace*{3ex} \mu} \\ H^{A}_{\; B5} = (M_{\sigma \tau})
^{A}_{\; B} \, s^{|\sigma \tau|}_{\hspace*{3ex} 5} \, , \rule{0ex}{3ex}
\end{array} \end{equation}
where $G^{A}_{\; B \mu}$ are ordinary connection coefficients for the
considered basis, associated with the derivative $\dotnabla$. Writing out
the latter formulae in detail, one obtains
\begin{equation} \begin{array}{l}
H^{\alpha}_{\; \beta \mu} = G^{\alpha}_{\; \beta \mu} -
s^{\alpha}_{\; \beta \mu}, \; \; H^{\alpha}_{\; 5 \mu} = 0 \\
H^{\alpha}_{\; \beta 5} = - s^{\alpha}_{\; \beta 5}, \; \;
H^{\alpha}_{\; 55} = 0 \rule{0ex}{3ex} \vspace*{-1ex}
\end{array} \end{equation} and \begin{equation}
H^{5}_{\; \beta \mu} = - \, g_{\beta \mu}, \; \; H^{5}_{\; \beta 5}
 = H^{5}_{\; 5 \mu} = H^{5}_{\; 55} = 0,
\end{equation}
where $s^{\alpha}_{\; \beta C} \equiv g_{\beta \omega}
s^{\alpha \omega}_{\hspace{2ex} C}$. Thus, all the connection coefficients
except for $H^{\alpha}_{\; \beta C}$ in this case are the same as in the
case of the locally symmetric connection considered earlier, at $\omega = 0$.

   From the formulae obtained it follows that four-vectors transported along
a given curve according to the transport rules associated with $\plaision$
can turn (all together) arbitrarily with respect to the same four-vectors
but transported according to the rules fixed by the metric. Furthermore,
one can see that compared to ordinary parallel transport, four-vectors can
experience an additional rotation that does not depend on the direction of
the transport and whose magnitude at each infinitesimal step of the latter
is proportinal to $|ds|$, where $ds$ is the infinitesimal interval covered at
that step. Finally, from the formulae obtained it follows that the length of
the five-vectors from $\cal E$ does not change during the transport and that
at each infinitesimal step of the latter the fifth component of a transported
five-vector acquires an increment equal to its scalar product $g$ with the
infinitesimal tangent five-vector that characterizes the element of the
transport path covered at that step.

To understand more clearly how general is the case we have just considered,
let us compare it with the case where the parallel transport of five-vectors
is constrained only by the following three requirements: $(i)$ it should
preserve the correspondence between four- and five-vectors; $(ii)$ it should
conserve the scalar product $g$; and $(iii)$ it should conserve the length
of the five-vectors from $\cal E$. (The latter requirement seems quite
reasonable if one considers that by virtue of the first condition, the
five-vectors from $\cal E$ are transported into five-vectors from $\cal E$,
and that for the latter there exists a natural measure determined only by
the differential structure of the manifold. In addition, as it has been shown
in section G of part IV, the conservation of the length of five-vectors from
$\cal E$ is a necessary condition of the five-vector Levi-Civita tensor being
covariantly constant.)

Let us consider an arbitrary point $Q$, introduce in its neighbourhood a
system of local Lorentz coordinates, and construct the corresponding active
regular basis ${\bf e}_{A}$. As it has been shown earlier, the above three
conditions on five-vector parallel transport lead to the following
constraints on the five-vector connection coefficients corresponding to
the basis ${\bf e}_{A}$:
\begin{displaymath}
H^{\alpha}_{\; 5C}(Q) = H^{5}_{\; 5C}(Q) = 0
\end{displaymath} and \begin{displaymath}
g_{\alpha \omega} H^{\omega}_{\; \beta C}(Q) + g_{\omega \beta}
H^{\omega}_{\; \alpha C}(Q) = 0.
\end{displaymath}
A comparison of these relations with equations (6) and (7) of part
III, which determine the components of the five-tensor $\SS$ introduced in
that paper, makes it apparent that under the considered constraints the
connection coefficients at $Q$ can be presented as
\begin{equation}
H^{A}_{\; BC} = (M_{\sigma \tau})^{A}_{\; B} \, r^{|\sigma \tau|}_
{\hspace*{3ex} C} + (M_{5 \tau})^{A}_{\; B} \, r^{5 \tau}_{\; \; \; C},
\end{equation}
where $r^{A \beta}_{\;\;\; C} = - \, g^{\beta \omega} H^{A}_{\;\; \omega C}$.
If, as in the case of the tensor $\RR$ considered in part III, one puts
\begin{displaymath}
r^{\alpha 5}_{\; \; \; C} = - \, r^{5 \alpha}_{\; \; \; C},
\end{displaymath}
the quantities $r^{AB}_{\; \; \; C}$ will become antisymmetric in $A$ and
$B$, and one will be able to rewrite formula (73) as
\begin{equation}
H^{A}_{\; BC} = (M_{KL})^{A}_{\; B} \, r^{|KL|}_{\hspace*{3.5ex} C},
\end{equation}
which is similar to equation (13) of part III. It is apparent that
the quantities $r^{AB}_{\; \; \; C}$ are analogs of the quantities
$S^{\alpha \beta}_{\; \; \; C}$ introduced in section A, and one can easily
prove that the five-vector 1-form constructed according to the formula
\begin{displaymath}
\widetilde{\bf r} \; \equiv \; r^{|AB|}_{\hspace*{3.5ex} C} \,
{\bf e}_{A} \wedge {\bf e}_{B} \otimes \widetilde{\bf o}^{C}
\end{displaymath}
will be the same at any choice of the local Lorentz coordinates.

The fact that with respect to the indices $A$ and $B$ the connection
coefficients $H^{A}_{\; BC}$ for the considered basis have the same form
as the parameters of an infinitesimal Poincare transformation is certainly
not a coincidence. Indeed, at such a choice of the five-vector basis fields,
the first four basis five-vectors at the point $Q'$ with coordinates
$x^{\alpha} (Q') = x^{\alpha}(Q) + dx^{\alpha}$ are orthonomal to the
{\em second} order in $dx^{\alpha}$, so under the considered constraints
one will have
\begin{displaymath}
[{\bf e}_{A}(Q')]^{{\rm transported \; to} \; Q}={\bf e}_{B}(Q)C^{B}_{\; A},
\end{displaymath}
where $C^{5}_{\; 5} = 1$, $C^{\alpha}_{\; 5} = 0$, and $C^{\alpha}_{\; \beta}
\in$ SO(3,1) to the second order in $dx^{\alpha}$. Consequently, for such a
basis the connection coefficients $H^{A}_{\; BC}(Q)$ should have the same
form with respect to the indices $A$ and $B$ as the components of the
tensor $\SS$ constructed from parameters of an infinitesimal Poincare
transformation.

It is evident that in any regular basis $r^{\alpha \beta}_ {\; \; \; C} =
S^{\alpha \beta}_{\; \; \; C}$, where $S^{\alpha \beta}_{\; \; \; C}$
are the components of the 1-form $\widetilde{\bf S}$ in the associated
four-vector basis, so these components of $\widetilde{\bf r}$ determine how
the $\cal Z$-components of transported five-vectors turn relative to the
$\cal Z$-components of the same five-vectors but transported according to
the rules fixed by the metric. The components $r^{5 \alpha}_{\; \; \; C} =
- r^{\alpha 5}_{\; \; \; C}$ determine the change in the $\cal E$-component
of a transported five-vector, and since they can be arbitrary, this component
can change arbitrarily, and its variation in general will not be correlated
in any way with the Riemannian geometry of space-time.

Let us now determine how the 1-form $\widetilde{\bf r}$ is related to
five-vector torsion. The latter can be defined according to a formula
similar to formula (12). Namely, for any five-vector fields $\bf u$ and
$\bf v$ one puts
\begin{equation}
\plaision_{\bf u} {\bf v} - \plaision_{\bf v} {\bf u} - [{\bf u,v}]
\equiv - 2 \, {\bf t(u,v)}.
\end{equation}
As in the case of the ordinary covariant derivative, it is a simple matter
to prove that $\bf t(u,v)$ is a linear function of $\bf u$ and $\bf v$ and
therefore can be presented as a contraction of $\bf u \wedge v$ with a
certain five-vector 2-form $\widetilde{\bf t}$, which I will call the
{\em five-vector torsion tensor}. Defining the components of the latter
according to the formula
\begin{displaymath}
\widetilde{\bf t} = t_{|KL|}^{\hspace*{3ex} A} \, {\bf e}_{A}
\otimes \widetilde{\bf o}^{K} \wedge \widetilde{\bf o}^{L},
\end{displaymath}
one can show that in any five-vector basis for which all the commutators
are zero one has
\begin{displaymath}
t_{KL}^{\hspace*{2.5ex} A} = H^{A}_{\; [KL]},
\end{displaymath}
which is the analog of formula (14). Substituting expression (74) into the
latter equation, one obtains
\begin{displaymath}
t_{KL}^{\hspace*{2.5ex} A} = - \, r^{A}_{\; [KL]},
\end{displaymath}
which is the analog of equation (15). In detail, the latter means that
\begin{displaymath}
t_{\mu \nu}^{\; \; \; A} = - \, r^{A}_{\; [\mu \nu]} \; \mbox{ and } \;
t_{\mu 5}^{\; \; \; A} = - {\scriptstyle \frac{1}{2}} \, r^{A}_{\; \mu 5}.
\end{displaymath}
A comparison of the first of these equations at $A = \alpha$ with equation
(15) makes it apparent that in any regular basis
\begin{displaymath}
r^{\alpha \beta}_{\; \; \; \mu} = g^{\alpha \sigma} g^{\beta \tau} \,
(t_{\sigma \tau \mu} - t_{\tau \mu \sigma} - t_{\mu \sigma \tau}),
\end{displaymath}
where $t_{\sigma \tau \mu} \equiv t_{\sigma \tau}^{\hspace*{2ex} \omega}
g_{\omega \mu}$, so the components of the five-vector torsion tensor
determine all the components of the 1-form $\widetilde{\bf r}$ except for
the symmetric part of $r^{5}_{\; \mu \nu}$.

By comparing formulae (70) for an active regular basis associated with some
system of local Lorentz coordinates at the considered point with formula
(74), one finds that in the case of the five-vector affine connection
satisfying requirement (69) one has
\begin{displaymath}
r^{\alpha \beta}_{\; \; \; C} = s^{\alpha \beta}_{\hspace{2ex} C}
\; \mbox{ and } \; r^{5 \beta}_{\; \; \; C} = \delta^{\beta}_{\, C}
= - \, s^{5 \beta}_{\hspace{2ex} C},
\end{displaymath}
so
\begin{equation}
t_{\alpha \beta}^{\hspace{2ex} \mu} = - \, s^{\mu}_{\; [\alpha \beta]},
\; \; t_{\alpha 5}^{\hspace{2ex} \mu} = - {\scriptstyle \frac{1}{2}} \,
s^{\mu}_{\; \alpha 5}, \; \; t_{\alpha C}^{\hspace{2.5ex} 5} = 0.
\end{equation}
Therefore, in this paricular case the components $t_{AB}^{\hspace{2.5ex}\mu}$
can be arbitrary and the components $t_{AB}^{\hspace*{2.5ex} 5}$ are
identically zero.

\vspace{3ex} \begin{flushleft}
E. \it Five-vector curvature tensor
\end{flushleft}
In this section I will consider the five-vector analog of the curvature
tensor for the derivative $\plaision$ and will discuss some of its basic
properties and one important physical application. This tensor can be defined
in exactly the same way as in section E of part V I have defined the
five-vector analogs of the field strength tensor: as a five-vector 2-form,
$\bf R$, whose values are tensors of rank $(1,1)$ over $V_{5}$ and which is
such that for any five-vector field $\bf w$ regarded as a five-vector-valued
0-form
\begin{displaymath}
\bf ddw = R(w),
\end{displaymath}
where the notation $\bf R(w)$ means that the value of $\bf R$ acts on the
value of $\bf w$ as a linear operator. By using the analog of formula (45a)
of part V for five-vector forms and derivative $\plaision$, one can find that
\begin{equation}
< {\bf R \, , u \wedge v} > \; = \, \plaision_{\bf u} \plaision_{\bf v} -
\plaision_{\bf v} \plaision_{\bf u} - \plaision_{\bf [u,v]},
\end{equation}
where the expression in the left-hand side is simply the value of the 2-form
$\bf R$ on the bivector $\bf u \wedge v$. From this formula one can easily
obtain a familiar expression for the components of $\bf R$ in a five-vector
basis for which all the commutators are zero, in terms of the corresponding
five-vector connection coefficients:
\begin{displaymath} \begin{array}{l}
R^{A}_{\; BCD} = \partial_{C} H^{A}_{\; BD} - \partial_{D} H^{A}_{\; BC} \\
\rule{0ex}{3ex} \hspace{15ex} + \; H^{A}_{\; KC} H^{K}_{\; BD} -
H^{A}_{\; KD} H^{K}_{\; BC}.
\end{array} \end{displaymath}
If the basis is a standard one, then by virtue of equations (20) of part V
and (67) one will have
\begin{displaymath} \begin{array}{l}
R^{\alpha}_{\; 5 CD} = \partial_{C} H^{\alpha}_{\; 5D} - \partial_{D}
H^{\alpha}_{\; 5C} + H^{\alpha}_{\; \omega C} H^{\omega}_{\; 5D} \\
\rule{0ex}{3ex} \hspace{5ex} + \; H^{\alpha}_{\; 5C} H^{5}_{\; 5D}
- H^{\alpha}_{\; \omega D} H^{\omega}_{\; 5C} - H^{\alpha}_{\; 5D}
H^{5}_{\; 5C} \; = \; 0
\end{array} \end{displaymath}
and
\begin{displaymath} \begin{array}{rcl}
R^{\alpha}_{\; \beta CD} & \hspace{-1.5ex} = & \hspace{-1.5ex} \partial_{C}
H^{\alpha}_{\; \beta D} - \partial_{D} H^{\alpha}_{\; \beta C} \\ & &
\hspace{4ex} + \; H^{\alpha}_{\; \omega C} H^{\omega}_{\; \beta D} -
H^{\alpha}_{\; \omega D} H^{\omega}_{\; \beta C} \\ & \hspace{-1.5ex} =
& \hspace{-1.5ex} \partial_{C} \Upsilon^{\alpha}_{\; \beta D} -
\partial_{D} \Upsilon^{\alpha}_{\; \beta C} \\ & & \hspace{4ex} +
\Upsilon^{\alpha}_{\; \omega C} \Upsilon^{\omega}_{\; \beta D} -
\Upsilon^{\alpha}_{\; \omega D} \Upsilon^{\omega}_{\; \beta C},
\end{array} \end{displaymath}
which means that at any five-vector connection that satisfies requirement
(18) of part V, in any standard basis the components $R^{\alpha}_{\; 5 CD}$
are identically zero and the components $R^{\alpha}_{\; \beta CD}$ are
completely determined by the connection coefficients for four-vector fields.

Let us now calculate the components of $\bf R$ for the connection that
satisfies condition (69). By using formulae (71) and (72) one finds
that in any active regular basis
\begin{equation} \begin{array}{l}
R^{\alpha}_{\; 5CD} = R^{5}_{\; 5CD} = 0, \; \; R^{\alpha}_{\; \beta \mu \nu}
= R^{{\scriptscriptstyle (\nabla)} \, \alpha}_{\hspace{3.5ex} \beta \mu \nu},
\\ R^{5}_{\; \beta CD} = - \, 2 g_{\beta \omega} s^{\omega}_{\; \; {[}CD{]}}
= 2 t_{CD \beta}, \\ R^{\alpha}_{\; \beta \mu 5} = - \, \{ \, \partial_{\mu}
s^{\alpha}_{\; \beta 5} + H^{\alpha}_{\; \omega \mu} s^{\omega}_{\; \beta 5}
+ s^{\alpha}_{\; \omega 5} H^{\omega}_{\; \beta \mu} \, \} ,
\end{array} \end{equation}
where $R^{{\scriptscriptstyle (\nabla)} \, \alpha}_{\hspace{3.5ex} \beta \mu
\nu}$ are the components of the Riemann tensor corresponding to the ordinary
covariant derivative $\nabla$ related to $\plaision$ according to equation
(12) of part V, in the associated four-vector basis. From the fact that $g$ is
covariantly constant it follows that
\begin{displaymath}
g_{\alpha \omega} R^{\omega}_{\; \beta CD} +
g_{\beta \omega} R^{\omega}_{\; \alpha CD} = 0,
\end{displaymath}
and comparing this equation and the first double equation in (78) with
equations (6) and (7) of part III, we see that one can associate with $\bf R$,
in the way now familiar to us, a certain five-vector 2-form, $\bf K$, whose
values are five-vector bivectors and whose components are related to those
of $\bf R$ as
\begin{displaymath}
K^{A \beta}_{\hspace{2ex} CD} = - \, K^{\beta A}_{\hspace{2ex} CD} =
g^{\beta \omega} R^{A}_{\; \; \omega CD},
\end{displaymath}
where $g^{\beta \omega}$ is the inverse of the $4 \times 4$ matrix $g_{\beta
\omega}$. From formulae (78) one finds that in any active regular basis
\begin{equation} \begin{array}{l}
K^{\alpha 5}_{\hspace{2ex} CD} = - \, K^{5 \alpha}_{\hspace{2ex} CD}
= 2 s^{\alpha}_{\; \; {[}CD{]}} = - \, 2 t_{CD}^{\hspace{3ex} \alpha}, \\
K^{\alpha \beta}_{\hspace{2ex} \mu 5} = - \, \{ \, \partial_{\mu}
s^{\alpha \beta}_{\hspace{1.5ex} 5} + H^{\alpha}_{\; \omega \mu}
s^{\omega \beta}_{\hspace{2ex} 5} + H^{\beta}_{\; \omega \mu}
s^{\alpha \omega}_{\hspace{1.5ex} 5} \, \} , \rule{0ex}{3ex} \hspace{-1ex}
\\ K^{\alpha \beta}_{\hspace{1.5ex} \mu \nu} = g^{\beta \omega}
R^{{\scriptscriptstyle (\nabla)} \, \alpha}_{\hspace{3ex} \omega \mu \nu}.
\rule{0ex}{3ex} \end{array} \end{equation}
Since in the rest of this section I will no longer deal with the components
of the five-tensor $\bf R$ itself, in the following I will omit the
superscript $\rule{0ex}{0.5ex}^{\scriptscriptstyle (\nabla)}$ in the
notations for the components of the Riemann tensor corresponding to $\nabla$
and of all other four-tensors constructed out of it.

Let us now try to construct the five-vector analog of the Einstein tensor.
Since the latter is related to the four-vector Riemann tensor as
\begin{displaymath}
G^{\mu}_{\; \alpha} = {\scriptstyle \frac{1}{4}} \; \epsilon_{\alpha \lambda
\rho \omega} R^{\rho \omega}_{\hspace{1.5ex} \sigma \tau} \,
\epsilon^{\sigma \tau \lambda \mu},
\end{displaymath}
let us first consider the tensor
\begin{equation}
Y^{AB}_{\hspace{2.5ex} CD} \; = \; {\scriptstyle \frac{1}{4}} \;
{\rm sign \, \xi} \cdot \epsilon_{CDXRQ} \, K^{RQ}_{\hspace{2ex} ST}
\, \epsilon^{STXAB}.
\end{equation}
By using the formulae presented in section G of part IV one can easily
find that
\begin{equation} \hspace*{-1.3ex}  \begin{array}{l}
Y^{\mu 5}_{\hspace{2ex} \alpha 5} = {\scriptstyle \frac{1}{4}} \;
{\rm sign \, \xi} \cdot \epsilon_{\alpha 5 \lambda \rho \omega} \,
K^{\rho \omega}_{\hspace{1.5ex} \sigma \tau} \, \epsilon^{\sigma \tau
\lambda \mu 5} \\ \hspace*{2ex} = K^{\sigma \mu}_{\hspace{1.5ex}
\sigma \alpha} - {\scriptstyle \frac{1}{2}} \, \delta^{\mu}_{\, \alpha}
K^{\sigma \tau}_{\hspace{1.5ex} \sigma \tau} = R^{\mu}_{\; \alpha} -
{\scriptstyle \frac{1}{2}} \, \delta^{\mu}_{\, \alpha} R =
G^{\mu}_{\; \alpha}, \rule{0ex}{3ex} \hspace*{-1ex}
\end{array} \end{equation}
where $R$ is the scalar curvature and $R^{\mu}_{\; \alpha}$ and
$G^{\mu}_{\; \alpha}$ are the components of the Ricci tensor and Einstein
tensor in the associated four-tensor basis, all corresponding to the
derivative $\nabla$. In a similar manner one finds that
\begin{equation} \begin{array}{rcl}
Y^{\mu 5}_{\hspace{2ex} \alpha\beta} & = & {\scriptstyle \frac{1}{2}} \;
{\rm sign \, \xi} \cdot \epsilon_{\alpha \beta \lambda \rho 5} \, K^{\rho 5}
_{\hspace{1.5ex} \sigma \tau} \, \epsilon^{\sigma \tau \lambda \mu 5} \\
& = & - \, K^{\mu 5}_{\hspace{1.5ex} \alpha \beta} + \delta^{\mu}_{\, \beta}
K^{\sigma 5}_{\hspace{1.5ex} \alpha \sigma} - \delta^{\mu}_{\, \alpha}
K^{\sigma 5}_{\hspace{1.5ex} \beta \sigma} \rule{0ex}{3ex} \\ & = & 2 \,
\{ \, t_{\alpha \beta}^{\hspace{2ex} \mu} + \delta^{\mu}_{\, \alpha} \,
t_{\beta \sigma}^{\hspace{2ex} \sigma} - \delta^{\mu}_{\, \beta} \,
t_{\alpha \sigma}^{\hspace{2ex} \sigma} \, \} \rule{0ex}{3ex} \\ & \! = & \!
2 \, \{ \, T_{\alpha \beta}^{\hspace{1.5ex} \mu} + \delta^{\mu}_{\, \alpha}
\, T_{\beta \sigma}^{\hspace{1.5ex} \sigma} - \delta^{\mu}_{\, \beta} \,
T_{\alpha \sigma}^{\hspace{1.5ex} \sigma} \, \} , \rule{0ex}{3ex}
\end{array} \end{equation}
where $T_{\alpha \beta}^{\hspace{1.5ex} \mu}$ are the components of the
four-vector torsion tensor defined by equations (12) and (13), and the
combination in the curly brackets in the right-hand side is known as the
modified torsion tensor.

Formulae (81) and (82) have an interesting bearing on physics. In
accordance with what has been said in part V, the five-vector connection
$\plaision$ can be regarded as a composite structure consisting of the
ordinary affine connection $\nabla$ and of another structure, which in the
case we are now considering is fixed by a field of four-vector bivectors
whose components in any four-vector basis coincide with the components
$s^{\alpha \beta}_{\hspace{2ex} 5}$ of the 1-form $\widetilde{\bf s}$ in
the associated active regular five-vector basis. Now, if one supposes
that despite the fact that $\nabla$ is not necessarily torsion-free, the
four-vector Einstein tensor corresponding to it is still related to the
canonical stress--energy tensor by the Einstein equation, one will have
\begin{equation}
Y^{\mu 5}_{\hspace{2ex} \alpha 5} = G^{\mu}_{\; \alpha} = k \hspace{0.1ex}
\Theta^{\mu}_{\, \alpha} = - \, k \hspace{0.1ex}{\cal M}^{\mu}_{\, \alpha 5},
\end{equation}
where $k$ is Newton's gravitational constant times $8 \pi c^{-4}$ and $\MM$
is the stress--enegry--angular momentum tensor introduced in part I. For
the moment let us pay no attention to the fact that the four-vector index
$\mu$ in the right-hand side of the latter equation corresponds to the
antisymmetrized pair of five-vector indices $\mu 5$ in its left-hand
side, and let us just concentrate on the lower indices. If one supposes
that the same relation as above exists between the quantities
$Y^{\mu 5}_{\hspace{1.5ex} AB}$ and ${\cal M}^{\mu}_{\, AB}$ at all
other values of the indices $A$ and $B$, in addition to (83) one will
have the equation
\begin{equation}
Y^{\mu 5}_{\hspace{2ex} \alpha \beta} =
- k \hspace{0.1ex} {\cal M}^{\mu}_{\, \alpha \beta},
\end{equation}
and considering that in any regular basis ${\cal M}^{\mu}_{\alpha \beta}$
coincide with the components of the spin angular momentum, one can rewrite
equation (84) as
\begin{equation}
T_{\alpha \beta}^{\hspace{1.5ex} \mu} + \delta^{\mu}_{\, \alpha} \,
T_{\beta \sigma}^{\hspace{1.5ex} \sigma} - \delta^{\mu}_{\, \beta} \,
T_{\alpha \sigma}^{\hspace{1.5ex} \sigma} = - \, {\scriptstyle \frac{1}{2}}
k \hspace{0.1ex} \Sigma^{\mu}_{\, \alpha \beta},
\end{equation}
which is exactly the Kibble--Sciama equation that relates four-vector
torsion to spin.\footnote{\rule{0ex}{4ex}Some authors hide the factor
$-{\scriptstyle \frac{1}{2}}$ by defining the four-vector torsion tensor
with a different sign and by choosing a different normalization for the spin
angular momentum. The simplest way to compare the definitions of these
quantities adopted in a particular paper with ours is to evaluate the
proportionality factor between $\Sigma^{\mu}_{\, \alpha \beta ; \mu} - 2
\, T_{\mu \omega}^{\hspace{2ex} \omega} \Sigma^{\mu}_{\, \alpha \beta}$ and
$g_{\beta \mu} \Theta^{\mu}_{\, \alpha} - g_{\alpha \mu} \Theta^{\mu}_{\,
\beta}$ (in our case it is unity) and the proportionality factor between
$T_{\alpha \beta}^{\hspace{1.5ex} \mu}$ and $\Gamma^{\alpha}_{\; [\mu \nu]}$
(in our case the latter is unity, too, provided the definition of the
four-vector connection coefficients is the same as ours). The sign and
normalization of the stress-energy tensor is fixed by the condition that
$\Theta^{0}_{\, 0}$ be the energy density of matter.}

To understand how one can eliminate the discrepancy between the upper indices
in the left- and right-hand sides of equations (83) and (84), one should
examine more closely the tensor $\MM$, or rather, its analog in the case
where the Lagrangian density depends on the {\em five}-vector covariant
derivatives of the fields.

\vspace{3ex} \begin{flushleft}
F. \it Stress--energy--angular momentum five-tensor
\end{flushleft}
Let us consider a situation where one has $n$ matter fields, $\smallvec{U}_
{\ell}$, whose values can be vectors or tensors of any nature (the index
$\ell$ runs 1 through $n$ and lables the fields, not their components)
and where the Lagrangian density $\Lagr$ that describes these fields is a
function of the values of the fields themselves and of their five-vector
covariant derivatives. For our purposes it will be sufficient to examine
a simplified situation where there are no gauge fields. As in ordinary
theory, from the requirement of local isotropy and homogeneity of space-time
one can derive certain relations, from which, by using the equations of
motion for the considered fields, one can then derive equations that can
be interpreted as a conservation law for a certain tensor quantity whose
components in the limit of flat space-time coincide with the five-vector
analogs of the Noether currents associated with the symmetry under global
Poincare transformations. The formulation of the local isotropy and
homogeneity requirement is quite apparent and I will not present it. The
relations that follow from it are
\begin{equation}
0 \; = \; \sum_{\ell} \, \{ \, \frac{\partial {\bf L}}{\partial
\smallvec{U}_{\ell}} \, {\sf D}_{\mu \nu} \smallvec{U}_{\ell} +
\frac{\partial {\bf L}}{\partial (\hspace{0.2ex} \plaision
\smallvec{U}_{\ell})} \, {\sf D}_{\mu \nu} (\hspace{0.2ex} \plaision
\smallvec{U}_{\ell}) \, \}
\end{equation} and \begin{equation} \begin{array}{l}
\hspace*{-1ex} \displaystyle 0 \; = \; \sum_{\ell} \, \{ \,
\frac{\partial {\bf L}}{\partial \smallvec{U}_{\ell}} \, {\sf D}_{\mu 5}
\smallvec{U}_{\ell} \\ \displaystyle  \hspace*{12ex} + \;
\frac{\partial {\bf L}}{\partial (\hspace{0.2ex} \plaision
\smallvec{U}_{\ell})} \, {\sf D}_{\mu 5} (\hspace{0.2ex} \plaision
\smallvec{U}_{\ell}) \, \} - \frac{d {\bf L}}{d x^{\mu}} ,
\end{array} \end{equation}
where the quantities $\plaision \smallvec{U}_{\ell}$, when differentiated, are
regarded as (nonscalar-valued) five-vector 1-forms and $\sf D$ in this case,
and this is quite essential, is the bivector derivative whose action on
five-vector fields is defined according to equation (16) and not as it has
been described in the rest of section B. It is evident that equations (86)
and (87) can be presented as a single equation:
\begin{displaymath}
0 = {\sf D}_{AB} {\bf L} -  \sum_{\ell} \, \{ \,
\frac{\partial {\bf L}}{\partial \smallvec{U}_{\ell}} \, {\sf D}_{AB}
\smallvec{U}_{\ell} + \frac{\partial {\bf L}}{\partial (\hspace{0.2ex}
\plaision \smallvec{U}_{\ell})} \, {\sf D}_{AB} (\hspace{0.2ex}
\plaision \smallvec{U}_{\ell}) \, \} ,
\end{displaymath}
the right-hand side of which is nothing but the bivector derivative of the
Lagrangian $\bf L$ regarded as a scalar field whose value at each point is
a function of the values of $n$ fields $\smallvec{U}_{\ell}$ and $n$ fields
$\plaision \smallvec{U}_{\ell}$.

The equations of motion for the fields $\smallvec{U}_{\ell}$ in this case
have the form
\begin{equation}
\frac{\partial {\bf L}}{\partial \smallvec{U}_{\ell}} \, = \,
\plaision_{\hat{A}} \{ \frac{\partial (e {\bf L})}{\partial (\hspace{0.2ex}
\plaision_{\hat{A}} \smallvec{U}_{\ell})} \} ,
\end{equation}
where $e$ denotes the square root of minus the determinant of the metric
tensor and the hat over $A$ means that when one evaluates the derivative
of $\partial (e {\bf L}) / \partial (\hspace{0.2ex} \plaision_{A}
\smallvec{U}_{\ell})$, this index should be treated as an {\em external}
one. By using these equations of motion, from equation (86) one obtains that
\begin{equation} \begin{array}{l}
\hspace*{-2ex} (\partial_{\alpha} + e^{-1}\partial_{\alpha} e) \,
{\cal M}^{\alpha}_{\, \mu \nu} \! - \! {\cal M}^{A}_{\, \omega \nu}
H^{\omega}_{\; \; \mu A} \! - \! {\cal M}^{A}_{\, \mu \omega}
H^{\omega}_{\; \; \nu A} \\ \displaystyle \hspace*{6ex} \rule{0ex}{4ex}
= \; \sum_{\ell} \, \frac{\partial {\bf L}}{\partial (\hspace{0.2ex}
\plaision_{\alpha} \smallvec{U}_{\ell})} \, ( g_{\alpha \nu} \plaision_{\mu}
- g_{\alpha \mu} \plaision_{\nu}) \, \smallvec{U}_{\ell} ,
\end{array} \hspace*{-1ex} \end{equation}
where
\begin{equation}
{\cal M}^{A}_{\, \mu \nu} \; = \; - \, \sum_{\ell} \, \frac{\partial {\bf L}}
{\partial (\hspace{0.2ex} \plaision_{A} \smallvec{U}_{\ell})} \,
{\sf D}_{\mu \nu} \smallvec{U}_{\ell}.
\end{equation}
It is apparent that the latter quantities are direct analogs of the
components of the total spin angular momentum. The right-hand side of
equation (89) can be presented as
\begin{displaymath} \begin{array}{l}
\displaystyle \{ \, \delta^{\alpha}_{\mu} {\bf L} - \sum_{\ell}
\frac{\partial {\bf L}} {\partial (\hspace{0.2ex} \plaision_{\alpha}
\smallvec{U}_{\ell})} \, \plaision_{\mu} \smallvec{U}_{\ell} \, \}
\cdot H^{5}_{\; \nu \alpha} \\ \displaystyle \hspace*{5ex} - \; \{ \,
\delta^{\alpha}_{\nu} {\bf L} - \sum_{\ell} \frac{\partial {\bf L}}
{\partial (\hspace{0.2ex}\plaision_{\alpha} \smallvec{U}_{\ell})} \,
\plaision_{\nu} \smallvec{U}_{\ell} \, \} \cdot H^{5}_{\; \mu \alpha},
\end{array} \end{displaymath}
whence it is seen that if one takes
\begin{equation}
{\cal M}^{A}_{\, \mu 5} = - \, {\cal M}^{A}_{\, 5 \mu} =
\delta^{A}_{\mu} {\bf L} - \sum_{\ell} \frac{\partial {\bf L}}
{\partial (\hspace{0.2ex} \plaision_{A} \smallvec{U}_{\ell})} \,
\plaision_{\mu} \smallvec{U}_{\ell} \, ,
\end{equation}
   from equation (89) one will get
\begin{equation} \hspace*{-1ex} \begin{array}{rcl}
0 & \hspace*{-1ex} = & \hspace*{-1ex} (\partial_{\alpha} + e^{-1}
\partial_{\alpha} e) \, {\cal M}^{\alpha}_{\, \mu \nu} \\ & & \rule{0ex}{3ex}
- \; {\cal M}^{A}_{\, K \nu} H^{K}_{\; \; \mu A} - {\cal M}^{A}_{\, \mu K}
H^{K}_{\; \; \nu A} \\ & \hspace*{-1ex} = & \hspace*{-1ex} \rule{0ex}{3ex}
{\cal M}^{A}_{\, \mu \nu \, ; A} + (H^{K}_{\; \; KA} - H^{K}_{\; \; AK}) \,
{\cal M}^{A}_{\, \mu \nu} \\ & \hspace*{-1ex} = & \hspace*{-1ex}
\rule{0ex}{3ex} {\cal M}^{A}_{\, \mu \nu \, ; A} - 2 \,
t_{AK}^{\hspace{2.5ex} K} \, {\cal M}^{A}_{\, \mu \nu} \; = \;
( \, \stackrel{\ast}{\plaision}_{A} \! \MM )^{A}_{\, \mu \nu} ,
\end{array} \end{equation}
where the semicolon denotes the covariant differentiation associated with
$\plaision$ and the operator $\, \stackrel{\ast}{\plaision}_{A} \; \equiv \;
\plaision_{A} - 2 \, t_{AK}^{\hspace{2.5ex} K}$ is the direct generalization
of the corresponding four-vector operator $\stackrel{\ast}{\nabla}_{\alpha}
\; \equiv \; \nabla_{\alpha} - 2 \, T_{\alpha \omega}^{\hspace{2ex} \omega}$.
By analogy with the usual terminology, the expression in the right-hand side
of (92) will be called the {\em modified} divergence.

Definition (91) may seem somewhat surprizing. Indeed, considering
definition (90), one would expect that
\begin{equation}
{\cal M}^{A}_{\, \mu 5} = - \, {\cal M}^{A}_{\, 5 \mu} =
\delta^{A}_{\mu} {\bf L} - \sum_{\ell} \, \frac{\partial {\bf L}}
{\partial (\hspace{0.2ex} \plaision_{A} \smallvec{U}_{\ell})} \,
{\sf D}_{\mu 5} \smallvec{U}_{\ell} \, .
\end{equation}
In this connection let me observe that owing to the invariance of the
$\widetilde{\cal E}$--$\widetilde{\cal Z}$ decomposition for five-vector
2-forms, the components ${\cal M}^{A}_{\, \mu \nu}$ and ${\cal M}^{A}_{\,
\mu 5}$ are absolutely independent from each other. In particular, one can
equally well take the latter to be given by formula (91) or by formula
(93). However, depending on how the components ${\cal M}^{A}_{\, \mu 5}$
are selected, the $\widetilde{\cal Z}$-component of the modified divergence
of $\MM$ will have different values, and equation (89) tells us that
${\cal M}^{A}_{\, \mu 5}$ can be chosen in such a way that this
$\widetilde{\cal Z}$-component would be zero.

The expression for the $\cal E$-component of the mentioned divergence can be
found from relation (87) by using once more the equations of motion (88).
Simple calculations give
\begin{equation}
( \, \stackrel{\ast}{\plaision}_{A} \! \MM )^{A}_{\, \mu 5} =
{\cal M}^{A}_{\, \sigma \tau} K^{| \sigma \tau |}_{\hspace{2.5ex} \mu A}
+ 2 \, {\cal M}^{A}_{\, \sigma 5} s^{\sigma}_{\; [ \mu A ]},
\end{equation}
which, by using formulae (79), can be cast into the following form:
\begin{equation}
( \, \stackrel{\ast}{\plaision}_{A} \! \MM )^{A}_{\, \mu 5} =
{\cal M}^{A}_{\, ST} K^{|ST|}_{\hspace{2.5ex} \mu A}.
\end{equation}

Let us now observe that in the latter equation and in formula (91) the
five-vector index $\mu$ in the right-hand side corresponds to the pair
of antisymmetrized indices $\mu 5$ in the left-hand side. To eliminate
this discrepancy let us first introduce the quantities $\delta^{A}_{BC}$
defined as
\begin{equation}
\delta^{A}_{\mu 5} = - \, \delta^{A}_{5 \mu} = \delta^{A}_{\mu} \;
\mbox{ and } \; \delta^{A}_{\mu \nu} = - \, \delta^{A}_{\nu \mu} = 0,
\end{equation}
where the symbol $\delta^{A}_{\mu}$ with only one lower index is defined
in the usual way. It is easy to see that the quantities $\delta^{A}_{BC}$
are components of the Lorentz-invariant five-tensor which when acting on
five-vector bivectors as an operator, transforms each of them into the
five-vector from $\cal Z$ that corresponds to the $\cal E$-component of
this bivector.

Secondly, let us define a new type of bivector derivative, whose operator
will be denoted as $\overline{\sf D}$. By definition, let us take that in
any active regular basis
\begin{equation}
\overline{\sf D}_{\mu \nu} = {\sf D}_{\mu \nu} \; \; \mbox{ but } \; \;
\overline{\sf D}_{\mu 5} = \plaision_{\mu} \, .
\end{equation}
The same can also be expressed as follows:
\begin{displaymath}
\overline{\sf D}_{\cal A} = \nabla_{\bf a} + \widetilde{\bf M}_{\cal A^{Z}},
\end{displaymath}
where $\bf a$, as in formula (25), denotes the five-vector from $\cal Z$
that corresponds to the $\cal E$-component of $\AAA$. It is evident that at
zero torsion the derivatives $\sf D$ and $\overline{\sf D}$ coincide. To
distinguish these two kinds of bivector derivative one from the other,
one can call the first of them {\em metric} and the second one {\em affine}.
Let me also mention that the connection coefficients for $\overline{\sf D}$,
which I will denote as $\overline{G}^{\, A}_{\, BKL}$ and which can be
defined according to a formula similar to equation (17), are related to
the connection coefficients for the derivatives $\plaision$ and $\sf D$
in the following way:
\begin{displaymath}
\overline{G}^{\, A}_{\, B \mu 5} = H^{A}_{\, B \mu} \; \mbox { and } \;
\overline{G}^{\, A}_{\, B \mu \nu} = G^{A}_{\, B \mu \nu} \, .
\end{displaymath}

By using the new notations one can present formulae (90) and (91) as a
single equation:
\begin{equation}
{\cal M}^{A}_{\, CD} \; = \; \delta^{A}_{CD} \, {\bf L} -  \sum_{\ell}
\frac{\partial {\bf L}}{\partial (\hspace{0.2ex} \plaision_{A}
\smallvec{U}_{\ell})} \, \overline{\sf D}_{CD} \smallvec{U}_{\ell} \, ,
\end{equation}
where now there is complete correspondence between the five-vector indices
in the right- and left-hand sides of the equation. To achieve the same in
equation (95), to the expression in the right-hand side of the latter
one should assign an additional index 5, doing this in such a way that
the expression as a whole would be antisymmetric with respect to the
transposition $5 \leftrightarrow \mu$. It is evident that this additional
index should be assigned to $K^{ST}_{\hspace{2ex} \mu A}$, and consequently
equation (95) will acquire the form
\begin{equation}
( \, \stackrel{\ast}{\plaision}_{A} \! \MM )^{A}_{\, \mu 5} =
{\cal M}^{A}_{\, ST} K^{|ST|}_{\hspace{3ex} \mu 5 \, A} \, ,
\end{equation}
where one should put
\begin{equation}
K^{ST}_{\hspace{2ex} \mu 5 \, A} = - \,
K^{ST}_{\hspace{2ex} 5 \mu \, A} = K^{ST}_{\hspace{2ex} \mu A} \, .
\end{equation}
If in addition to this, one should wish that the equation
$( \, \stackrel{\ast}{\plaision}_{A} \! \MM )^{A}_{\, \mu \nu} = 0$
could be presented in a similar form, one should require also that
\begin{equation}
K^{ST}_{\hspace{2ex} \mu \nu \, A} = 0 \, .
\end{equation}

It turns out that the tensor with such components can be defined in a
manner similar to how one defines the curvature tensor, only instead of
corresponding to the commutator of two identical derivatives, it will
correspond to the commutator of the derivatives $\, \plaision_{\bf u} \,$
and $\, \overline{\sf D}_{\cal A} \,$, acting on four-vector fields. In
order to obtain an operator that depends linearly on $\bf u$ and $\AAA$, from
the mentioned commutator one should subtract a certain nontrivial
combination of the derivatives $\, \plaision \,$ and $\, \overline{\sf D}
\, $, whose purpose is similar to that of the last term in the right-hand
side of formula (77) and which is selected in such a way that the
components of the resulting analog of the curvature tensor have the
desired form. Omitting the details, let me only present the final formula:
\begin{displaymath} \begin{array}{l}
[ \, \plaision_{\bf u} , \overline{\sf D}_{\cal A} \, ] \,
- \, \overline{\sf D}_{( \Box_{\bf u} {\cal A})} \, + \,
\nabla_{(\overline{\sf D}_{\cal A} {\bf u})} \\
\rule{0ex}{3ex} \hspace*{15ex} = \; {\cal A}^{|BC|} \, u^{E}
K^{|ST|}_{\hspace{3ex} BC \, E} \cdot \overline{\sf D}_{ST} \, ,
\end{array} \end{displaymath}
where the quantities in the right- and left-hand sides are regarded as
operators acting on four-vector fields.

By using the components of this new tensor, one can present equation (99)
and the similar conservation law for the $\widetilde{\cal Z}$-component of
$\MM$ as a single equation:
\begin{equation}
( \, \stackrel{\ast}{\plaision}_{A} \! \MM )^{A}_{\, BC} =
{\cal M}^{A}_{\, ST} K^{|ST|}_{\hspace{3ex} BC \, A} \, .
\end{equation}
Furthermore, this tensor enables one to eliminate the discrepancy between
the upper indices in equations (83) and (84). Indeed, for that one should
take the analog of the Einstein tensor to be
\begin{equation}
Y^{A}_{BC} \; = \; {\scriptstyle \frac{1}{8}} \; {\rm sign \, \xi} \cdot
\epsilon_{BCXPQ} \, K^{PQ}_{\hspace{2.5ex} RST} \, \epsilon^{RSTXA}.
\end{equation}
By using formulae (100) and (101), one easily finds that
\begin{displaymath}
Y^{\alpha}_{\mu 5} = - \, G^{\alpha}_{\; \mu} \;
\mbox{ and } \; Y^{\alpha}_{\mu \nu} = - \, 2 \,
T^{\, {\scriptscriptstyle \rm (mod)} \, \alpha}_{\; \mu \nu} \, ,
\end{displaymath}
where $T^{\, {\scriptscriptstyle \rm (mod)} \, \alpha}_{\; \mu \nu}$ are
the components of the modified torsion tensor mentioned in section E:
\begin{displaymath}
T^{\, {\scriptscriptstyle \rm (mod)} \, \alpha}_{\; \mu \nu} \equiv
T_{\mu \nu}^{\hspace{1.5ex} \alpha} + \delta^{\alpha}_{\, \mu} \,
T_{\nu \sigma}^{\hspace{1.5ex} \sigma} - \delta^{\alpha}_{\, \nu} \,
T_{\mu \sigma}^{\hspace{1.5ex} \sigma} \, .
\end{displaymath}
Consequently, the Einstein and Kibble--Sciama equations can be presented as
follows:
\begin{equation}
Y^{\alpha}_{BC} = k \hspace{0.1ex} {\cal M}^{\alpha}_{\, DC} \, .
\end{equation}
Since neither $G^{\alpha}_{\; \mu}$ nor $T^{\, {\scriptscriptstyle \rm (mod)}
\, \alpha}_{\; \mu \nu}$ depend on $s^{\alpha \beta}_{\hspace{2ex} 5}$ and
since
\begin{equation}
Y^{5}_{\mu 5} = Y^{5}_{\mu \nu} = 0 ,
\end{equation}
   from these equations one can say nothing about the components of
five-vecor torsion corresponding to $\plaision_{5}$. What case equations
(104) and (105) correspond to will be said in the next section.

\vspace{3ex} \begin{flushleft}
G. \it Five-vector generalization of the Einstein \\ \hspace*{2ex}
       and Kibble--Sciama equations
\end{flushleft}
As is known, the Einstein and Kibble--Sciama equations can be obtained from
the action principle if the Lagrangian describing the geometry of space-time
is taken (in our notations) to be $(-1/2k)R$, where $R$ is the curvature
scalar constructed out of the four-vector curvature tensor, and the varied
parameters are the components $g_{\mu \nu}$ of the metric tensor and the
components $T_{\alpha \beta}^{\hspace{1.5ex}\mu}$ of the four-vector torsion
tensor. Let us suppose that the graviational equations in the case of
five-vector affine connection can be obtained in a similar way. By virtue
of equations (79) and owing to the antisymmetry of the quantities
$s^{\alpha \beta}_{\hspace{2ex} 5}$ in their upper indices, one has
\begin{displaymath} \begin{array}{rcl}
K^{AB}_{\hspace{2ex} AB} & = & 2 K^{\alpha 5}_{\hspace{2ex} \alpha 5} +
K^{\alpha \beta}_{\hspace{2ex} \alpha \beta} \\ & = & 4
s^{\alpha}_{\; \alpha 5} + g^{\beta \omega} R^{\alpha}_{\; \; \omega
\alpha \beta} = g^{\beta \omega} R_{\omega \beta} = R \, ,
\end{array} \end{displaymath}
and since the components $R^{\alpha}_{\; \; \beta \mu \nu}$ are independent
of $s^{\alpha \beta}_{\hspace{2ex} 5}$, to obtain a full system of
equations from the action principle in the case of five-vector affine
connection, to the Lagrangian $(-1/2k) R$ one should add some additional
term, which I will denote as ${\bf L}_{\rm add}$. Thus,
\begin{displaymath}
{\bf L}_{\rm geom} = (-1/2k) R + {\bf L}_{\rm add} \, .
\end{displaymath}

As varied parameters let us choose $g_{\mu \nu}$ and
$t_{\alpha \beta}^{\hspace{1.5ex} \mu} = T_{\alpha \beta}^{\hspace{1.5ex}
\mu}$, and also the six quantities $s^{\alpha \beta}_{\hspace{2ex} 5}$.
By direct calculation one finds that
\begin{displaymath} \begin{array}{l}
2 e^{-1} \, \delta ( e {\bf L}_{\rm geom}) \\ \hspace*{3ex} = \; \delta
g_{\mu \nu} \cdot \{ \, k^{-1} G^{\{ \mu \nu \}} - k^{-1} (\stackrel{\ast}
{\nabla}_{\omega} \! T^{\, \scriptscriptstyle \rm (mod)})^{\mu \omega \nu}\\
\hspace*{13ex} - \; k^{-1} (\stackrel{\ast}{\nabla}_{\omega} \!
T^{\, \scriptscriptstyle \rm (mod)} )^{\nu \omega \mu} + \, g^{\mu \nu}
{\bf L}_{\rm add} \\ \hspace*{28ex} \rule{0ex}{2.5ex} + 2 \; (\delta
{\bf L}_{\rm add} / \delta g_{\mu \nu} ) \; \} \\ \rule{0ex}{3ex}
\hspace*{2.5ex} - \; \delta T_{\mu \nu}^{\hspace{1.5ex} \alpha} \cdot \{ \,
2 k^{-1} g_{\alpha \omega} \, ( \, T^{\, {\scriptscriptstyle \rm (mod)} \,
\mu \omega \nu} - T^{\, {\scriptscriptstyle \rm (mod)} \, \nu \omega \mu} \\
\rule{0ex}{2.5ex} \hspace*{12ex} + \; T^{\, {\scriptscriptstyle \rm (mod)}
\, \mu \nu \omega} \, ) - \; 2 (\delta {\bf L}_{\rm add} / \delta
T_{\mu \nu}^{\hspace{1.5ex} \alpha} ) \; \} \; \\ \rule{0ex}{2.5ex}
\hspace*{2.5ex} + \; \delta s^{\alpha \beta}_{\hspace{2ex} 5} \cdot \{ \, 2
\,(\delta{\bf L}_{\rm add}/\delta s^{\alpha \beta}_{\hspace{2ex}5}) \; \}\, ,
\end{array} \end{displaymath}
where $G^{\mu \nu} = G^{\mu}_{\; \sigma} g^{\sigma \nu}$,
$\; T^{\, {\scriptscriptstyle \rm (mod)} \, \mu \omega \nu} = g^{\mu \sigma}
g^{\omega \tau} \, T^{\, {\scriptscriptstyle \rm (mod)} \, \nu}_{\; \sigma
\tau}$, and the derivative $\stackrel{\ast}{\nabla}_{\omega}$ acts on
$T^{\, \scriptscriptstyle \rm (mod)}$ as on a four-tensor. Varying with
respect to the same parameters the part of the Lagrangian density that
describes matter, one obtains
\begin{displaymath} \hspace*{-1ex} \begin{array}{rcl}
2e^{-1} \, \delta ( e {\bf L}_{\rm matter}) \hspace*{-1.5ex} & =
\hspace*{-1.5ex} & - \; \delta g_{\mu \nu} \cdot \{ \, \Theta^{\{ \mu \nu \}}
+ \frac{1}{2} (\stackrel{\ast}{\nabla}_{\omega} \! \Sigma )^{\, \mu \omega
\nu} \\ && + \; {\scriptstyle \frac{1}{2}} (\stackrel{\ast} {\nabla}_{\omega}
\! \Sigma )^{\, \nu \omega \mu} + g_{\sigma \tau}s^{\mu \sigma}_{\hspace{2ex}
5} {\cal M}^{\nu \tau 5} \} \\ & & - \; \rule{0ex}{2.5ex} \delta
T_{\mu \nu}^{\hspace{1.5ex} \alpha} \cdot \{ \, g_{\alpha \omega} \, ( \,
\Sigma^{\, \mu \omega \nu} - \Sigma^{\, \nu \omega \mu} \\
& & \hspace*{2ex} + \; \rule{0ex}{2.5ex} \Sigma^{\, \mu \nu \omega} \, )
\, \} - \delta s^{\alpha \beta}_{\hspace{2ex}5} \cdot \{ \,
{\cal M}^{\, 5}_{\, \alpha \beta} \} ,
\end{array} \end{displaymath}
where $\Theta^{\mu \nu} = {\cal M}^{\mu}_{\, 5 \sigma} \, g^{\sigma \nu}$,
$\Sigma^{\, \mu \nu \alpha} = g^{\mu \sigma} g^{\nu \tau}
{\cal M}^{\alpha}_{\, \sigma \tau}$, ${\cal M}^{\mu \nu 5} =
g^{\mu \sigma} g^{\nu \tau} {\cal M}^{\, 5}_{\, \sigma \tau}$, and
$\stackrel{\ast}{\nabla}_{\omega}$ acts on $\Sigma$ as on a four-tensor.
Requiring the variation of the total action to vanish, one obtains
\begin{equation} \begin{array}{l}
G^{\{ \mu \nu \}} - (\stackrel{\ast}{\nabla}_{\omega} \!
T^{\, \scriptscriptstyle \rm (mod)} )^{\mu \omega \nu} - (\stackrel{\ast}
{\nabla}_{\omega} \! T^{\, \scriptscriptstyle \rm (mod)})^{\nu \omega \mu} \\
\hspace{8ex} + \; k \hspace{0.1ex} g^{\mu \nu} {\bf L}_{\rm add} + 2k
\hspace{0.1ex} (\delta {\bf L}_{\rm add} / \delta g_{\mu \nu} )
\rule{0ex}{2.5ex} \\ \hspace{4ex} = \; \; k \hspace{0.1ex}
\Theta^{\{ \mu \nu \}} - k \hspace{0.1ex} g_{\sigma \tau}
s^{\sigma \{ \mu}_{\hspace{2.5ex} 5} {\cal M} \rule{0ex}{1.8ex}^{\, \nu \}
\tau 5} \rule{0ex}{3ex} \\ \hspace{8ex} + \; \frac{1}{2} k \hspace{0.1ex}
( \stackrel{\ast}{\nabla}_{\omega} \! \Sigma \, )^{\, \mu \omega \nu} +
\frac{1}{2} k \hspace{0.1ex} (\stackrel{\ast}{\nabla}_{\omega}
\! \Sigma \, )^{\, \nu \omega \mu} \, ,
\end{array} \end{equation}
then
\begin{equation} \begin{array}{l}
( T^{\, {\scriptscriptstyle \rm (mod)} \, \mu \lambda \nu} -
T^{\, {\scriptscriptstyle \rm (mod)} \, \nu \lambda \mu} +
T^{\, {\scriptscriptstyle \rm (mod)} \, \mu \nu \lambda} ) \\ \hspace{18ex}
- \; k \hspace{0.1ex} g^{\lambda \omega} ( \delta {\bf L}_{\rm add} /
\delta T_{\mu \nu}^{\hspace{1.5ex} \omega} ) \\ \rule{0ex}{3ex} \hspace{6ex}
= - \, \frac{1}{2} k \hspace{0.1ex} ( \Sigma^{\, \mu \lambda \nu} -
\Sigma^{\, \nu \lambda \mu} + \Sigma^{\, \mu \nu \lambda} ) \, ,
\end{array} \end{equation}
and finally
\begin{equation}
\delta {\bf L}_{\rm add} / \delta s^{\alpha \beta}_{\hspace{2ex} 5} \; = \;
{\scriptstyle \frac{1}{2}} \, {\cal M}^{\, 5}_{\, \alpha \beta} \, .
\end{equation}

One should notice that none of the components ${\cal M}^{\, 5}_{\mu 5}$
act as a source. Moreover, none of them have any effect on the conservation
law for $\MM$, since in the right-hand side of equation (102) they appear
only in the term
\begin{displaymath}
{\cal M}^{5}_{\, \sigma 5} K^{\sigma 5}_{\hspace{2.5ex} \mu 5} =
2 \, {\cal M}^{5}_{\, \sigma 5} s^{\sigma}_{\; [ \mu 5 ]} =
{\cal M}^{5}_{\, \sigma 5} s^{\sigma}_{\; \mu 5} \, ,
\end{displaymath}
and in the left-hand side, only in the term
\begin{displaymath} \begin{array}{rcl}
( \, \stackrel{\ast}{\plaision}_{5} \! \MM )^{5}_{\, \mu 5} & = &
{\cal M}^{5}_{\, \mu 5 \, ;5} - 2 \, t_{5K}^{\hspace{2.5ex} K} \,
{\cal M}^{5}_{\, \mu 5} \\ & = & {\cal M}^{5}_{\, \mu 5 \, ; 5} +
(H^{K}_{\; \; K 5} - H^{K}_{\; \; 5 K}) \, {\cal M}^{5}_{\, \mu 5} \\
& = & - \, {\cal M}^{5}_{\, \sigma 5} H^{\sigma}_{\; \; \mu 5} \; = \;
{\cal M}^{5}_{\, \sigma 5} s^{\sigma}_{\; \mu 5} \, .
\end{array} \end{displaymath}
Consequently, their contributions cancel out. In view of this, in addition
to the ``canonical'' tensor $\MM$, whose components are given by formula
(98), one can introduce the ``dynamical'' tensor $\MM$, which will
differ from the former only in that its ${\cal M}^{\, 5}_{\mu 5}$
components will be identically zero.

Let us now try to select ${\bf L}_{\rm add}$ in such a way that the
field equations resulting from equations (106)--(108) in which the
role of the source is played by ${\cal M}^{\, \alpha}_{\mu 5}$ and
${\cal M}^{\, \alpha}_{\mu \nu}$ would differ as little as possible from
the Einstein and Kibble--Sciama equations, respectively. In the latter
case this can be achieved quite easily: one has only to require that
${\bf L}_{\rm add}$ {\em be independent of} $T_{\mu \nu}^{\hspace{1.5ex}
\alpha}$. Equation (107) will then give
\begin{displaymath}
T^{\, {\scriptscriptstyle \rm (mod)} \, \alpha \beta \mu} = - \,
{\scriptstyle \frac{1}{2}} k \hspace{0.1ex} \Sigma^{\, \alpha \beta \mu} ,
\end{displaymath}
which is equivalent to equation (85). Substituting this value of
$T^{\, \scriptscriptstyle \rm (mod)}$ into equation (106), one obtains
\begin{equation} \begin{array}{l}
G^{\{ \mu \nu \}} + k \hspace{0.1ex} g^{\mu \nu} {\bf L}_{\rm add}
+ 2k \hspace{0.1ex} (\delta {\bf L}_{\rm add} / \delta g_{\mu \nu} ) \\
\hspace*{12ex} = \; k \hspace{0.1ex} \Theta^{\{ \mu \nu \}}
- k \hspace{0.1ex} g_{\sigma \tau} s^{\sigma \{ \mu}_{\hspace{2.5ex} 5}
{\cal M} \rule{0ex}{1.8ex}^{\, \nu \} \tau 5}. \rule{0ex}{3ex}
\end{array} \end{equation}
It is impossible in general to get rid of the second term in the left-hand
side of this equation, and as one will see below, there is no need to. One
can, however, try to select ${\bf L}_{\rm add}$ in such a way that the
last term in the left-hand side would calcel out with the last term in
the right-hand side. This requirement gives one the second condition on
${\bf L}_{\rm add}$:
\begin{equation}
\delta {\bf L}_{\rm add} / \delta g_{\mu \nu} \; = \; - \, {\scriptstyle
\frac{1}{2}} g_{\sigma \tau} s^{\sigma \{ \mu}_{\hspace{2.5ex} 5}
{\cal M} \rule{0ex}{1.8ex}^{\, \nu \} \tau 5} ,
\end{equation}
and equation (109) then acquires the form
\begin{displaymath}
G^{\{ \mu \nu \}} + k \hspace{0.1ex} g^{\mu \nu} {\bf L}_{\rm add}
\; = \; k \hspace{0.1ex} \Theta^{\{ \mu \nu \}} .
\end{displaymath}

As one can see, the symmetric parts of $G^{\mu \nu}$ and $k\Theta^{\mu \nu}$
are no longer equal to each other. However, one can try to choose
${\bf L}_{\rm add}$ in such a way that the {\em anti}symmetric parts of
these tensors would coincide:
\begin{equation}
G^{[ \mu \nu ]} \; = \; k \hspace{0.1ex} \Theta^{[ \mu \nu ]} .
\end{equation}
If one succeeds, then after adding the latter two equations one will obtain
\begin{equation}
G^{\mu \nu} + k \hspace{0.1ex} g^{\mu \nu} {\bf L}_{\rm add}
\; = \; k \hspace{0.1ex} \Theta^{\mu \nu} .
\end{equation}
To derive from requirement (111) a constraint on ${\bf L}_{\rm add}$, let us
recall the differential identity that relates the modified four-dimensional
divergence of $T^{\, \scriptscriptstyle \rm (mod)}$ to the antisymmetric
part of the Einstein tensor:
\begin{displaymath}
(\stackrel{\ast}{\nabla}_{\alpha} \! T^{\, \scriptscriptstyle \rm (mod)}
)^{\alpha}_{\mu \nu} \; = \; G_{[ \mu \nu ]} .
\end{displaymath}
Combining this identity with equation (111) and using (85) and (102),
one finds that
\begin{displaymath} \begin{array}{l}
0 = (\stackrel{\ast}{\nabla}_{\alpha} \! \Sigma )^{\alpha}_{\mu \nu} +
2 \hspace{0.1ex} \Theta_{[ \mu \nu ]} = (\partial_{\alpha} + e^{-1}
\partial_{\alpha} e) \, {\cal M}^{\alpha}_{\, \mu \nu} \\ \hspace*{0.5ex}
- \; {\cal M}^{\alpha}_{\, \omega \nu} \Gamma^{\omega}_{\; \; \mu \alpha}
- {\cal M}^{\alpha}_{\, \mu \omega} \Gamma^{\omega}_{\; \; \nu \alpha}
+ g_{\mu \alpha} \, {\cal M}^{\alpha}_{\, 5 \nu} - g_{\nu \alpha} \,
{\cal M}^{\alpha}_{\, 5 \mu} \\ \hspace*{2ex} = (\partial_{\alpha} + e^{-1}
\partial_{\alpha} e) \, {\cal M}^{\alpha}_{\, \mu \nu} \\ \hspace*{0.5ex}
- \; {\cal M}^{\alpha}_{\, \omega \nu} H^{\omega}_{\, \mu \alpha}
- {\cal M}^{\alpha}_{\, \mu \omega} H^{\omega}_{\, \nu \alpha}
- {\cal M}^{\alpha}_{\, 5 \nu} H^{5}_{\, \mu \alpha}
- {\cal M}^{\alpha}_{\, \mu 5} H^{5}_{\, \nu \alpha} \hspace*{-0.5ex} \\
\hspace*{2ex} = ( \, \stackrel{\ast}{\plaision}_{A} \! \MM )^{A}_{\, \mu \nu}
+ {\cal M}^{\; 5}_{\, K \nu} H^{K}_{\; \; \mu 5}
+ {\cal M}^{\; 5}_{\, \mu K} H^{K}_{\; \; \nu 5} \\ \hspace*{2ex}
= - {\cal M}^{\; 5}_{\, \omega \nu} s^{\omega}_{\; \; \mu 5}
- {\cal M}^{\; 5}_{\, \mu \omega} s^{\omega}_{\; \; \nu 5} \, ,
\end{array} \end{displaymath}
whence it follows that
\begin{equation}
{\cal M}^{\mu \; \; 5}_{\; \; \omega} s^{\omega}_{\; \; \nu \hspace{0ex} 5}
- s^{\mu}_{\; \; \omega \hspace{0ex} 5} \,
{\cal M}^{\omega \; \; 5}_{\; \; \nu} = 0 ,
\end{equation}
meaning that the quantities $s^{\mu}_{\; \; \nu \hspace{0ex} 5}$ and
${\cal M}^{\mu \; \; 5}_{\; \; \nu}$ regarded as matrices with respect
to the indices $\mu$ and $\nu$ should commute with each other. Together
with equation (108), the latter relation gives us one more constraint on
${\bf L}_{\rm add}$.

Let us finally recall that in the case of arbitrary four-vector torsion the
Einstein tensor satisfies the differential identity
\begin{displaymath}
(\stackrel{\ast}{\nabla}_{\alpha} \! G \, )^{\alpha}_{\mu} \; = \;
R^{\sigma \tau}_{\hspace{1.5ex} \mu \alpha} \,
T^{\, {\scriptscriptstyle \rm (mod)} \, \alpha}_{\; \sigma \tau} - 2 \,
T_{\mu \alpha}^{\hspace{1.5ex} \sigma} \, G^{\alpha}_{\; \sigma} \, .
\end{displaymath}
Combining the latter with equations (85) and (112) and using (94), one has
\begin{displaymath} \begin{array}{rcl}
0 & = & \stackrel{\ast}{\nabla}_{\alpha} \! (\Theta^{\alpha}_{\mu} -
\delta^{\alpha}_{\mu}{\bf L}_{\rm add}) + R^{\sigma \tau}_{\hspace{1.5ex} \mu
\alpha} \cdot {\scriptstyle \frac{1}{2}} \Sigma^{\alpha}_{\, \sigma \tau} \\
& & \hspace*{2ex} + \; 2 \, T_{\mu \alpha}^{\hspace{1.5ex} \sigma} \,
(\Theta^{\alpha}_{\sigma} - \delta^{\alpha}_{\sigma} {\bf L}_{\rm add}) \\
& = & (\partial_{\alpha} + e^{-1} \partial_{\alpha} e) \,
{\cal M}^{\alpha}_{\, 5 \mu} - {\cal M}^{\alpha}_{\, 5 \omega}
\Gamma^{\omega}_{\; \mu \alpha} - \partial_{\mu} {\bf L}_{\rm add} \\
& & \hspace*{2ex} + \; R^{|\sigma \tau|}_{\hspace{3ex} \mu \alpha} \,
{\cal M}^{\alpha}_{\, \sigma \tau} + 2 \, T_{\mu \alpha}^{\hspace{1.5ex}
\sigma} {\cal M}^{\alpha}_{\, 5 \sigma} \\ & & \hspace*{2ex} - \;
( e^{-1} \partial_{\mu} e - \Gamma^{\omega}_{\; \mu \omega} + 2 \,
T_{\mu \omega}^{\hspace{1.5ex} \omega}) \, {\bf L}_{\rm add} \\ & = & - \;
( \, \stackrel{\ast}{\plaision}_{\alpha} \! \MM )^{\alpha}_{\, \mu 5} +
R^{|\sigma \tau|}_{\hspace{3ex} \mu \alpha} \, {\cal M}^{\alpha}_{\,
\sigma \tau} \\ & & \hspace*{2ex} + \; 2 \, s^{\sigma}_{\; [ \mu \alpha ]}
{\cal M}^{\alpha}_{\, \sigma 5} - \partial_{\mu} {\bf L}_{\rm add} \\ & = &
- \; K^{| \sigma \tau |}_{\hspace{2.5ex} \mu A} {\cal M}^{A}_{\, \sigma \tau}
- 2 \, s^{\sigma}_{\; [ \mu \alpha ]} {\cal M}^{\alpha}_{\, \sigma 5} \\
& & \hspace*{2ex} + \; R^{|\sigma \tau|}_{\hspace{3ex} \mu \alpha} \,
{\cal M}^{\alpha}_{\, \sigma \tau} + 2 \, s^{\sigma}_{\; [ \mu \alpha ]}
{\cal M}^{\alpha}_{\, \sigma 5} - \partial_{\mu}{\bf L}_{\rm add} \\
& = & - \; \partial_{\mu}{\bf L}_{\rm add}
- K^{| \sigma \tau |}_{\hspace{2.5ex} \mu 5} {\cal M}^{5}_{\, \sigma \tau},
\end{array} \end{displaymath}
   from which, using formulae (79), one obtains the last condition on
${\bf L}_{\rm add}$:
\begin{equation} \begin{array}{l}
\partial_{\mu} {\bf L}_{\rm add} =  {\scriptstyle \frac{1}{2}} \,
\{ \, \partial_{\mu} s^{\sigma \tau}_{\hspace{1.5ex} 5} +
H^{\sigma}_{\; \omega \mu} s^{\omega \tau}_{\hspace{1.5ex} 5} \\
\hspace*{23ex} + \; H^{\tau}_{\; \omega \mu}
s^{\sigma \omega}_{\hspace{1.5ex} 5} \, \} \, {\cal M}^{5}_{\, \sigma \tau}.
\end{array} \end{equation}

The simplest way to satisfy requirement (113) is to take $s_{\sigma \tau 5}$
proportional to ${\cal M}^{5}_{\, \sigma \tau}$. As one can see from equation
(108), for that one should choose
\begin{equation}
{\bf L}_{\rm add} = a \cdot g_{\alpha \sigma} g_{\beta \tau} h^{55}
s^{\alpha \beta}_{\hspace{1.5ex} 5} \, s^{\sigma \tau}_{\hspace{1.5ex} 5},
\end{equation}
where $a$ is a certain constant and the factor $h^{55}$ has been introduced
so that the latter would not depend on the normalization of the fifth basis
vector. Accordingly, one has
\begin{equation}
2 a \, h^{55} s_{\sigma \tau 5} =
{\scriptstyle \frac{1}{2}} \, {\cal M}^{5}_{\, \sigma \tau}.
\end{equation}
It is a simple matter to check that at such ${\bf L}_{\rm add}$ conditions
(114) and (110) are also satisfied. Indeed, by differentiating (115) and
using the covariant constancy of $g$, one obtains that
\begin{displaymath} \begin{array}{l}
\partial_{\mu}{\bf L}_{\rm add} = \; 2a \, h^{55} s_{\sigma \tau 5}
\cdot s^{\sigma \tau}_{\hspace{1.5ex} 5 \, ; \, \mu} \\ \hspace*{7.8ex} = \;
{\scriptstyle \frac{1}{2}} \, \{ \, \partial_{\mu}
s^{\sigma \tau}_{\hspace{1.5ex} 5} + H^{\sigma}_{\; \omega \mu}
s^{\omega \tau}_{\hspace{1.5ex} 5} + H^{\tau}_{\; \omega \mu}
s^{\sigma \omega}_{\hspace{1.5ex} 5} \, \} {\cal M}^{5}_{\, \sigma \tau}.
\end{array} \end{displaymath}
Similarly, by varying (115) with respect to $g_{\mu \nu}$ and using (116),
one obtains
\begin{displaymath} \begin{array}{rcl}
\delta {\bf L}_{\rm add} & \! \! = & \! \! \delta g_{\mu \nu} \cdot 2 a \,
g_{\sigma \tau} h^{55} s^{\mu \sigma}_{\hspace{1.5ex} 5} \,
s^{\nu \tau}_{\hspace{1.5ex} 5} \\ & \! \! = & \! \! \delta g_{\mu \nu}
\cdot \{ \, {\scriptstyle \frac{1}{2}} \, g_{\sigma \tau}
s^{\mu \sigma}_{\hspace{1.5ex} 5} \, {\cal M}^{\nu \tau 5} \, \} \, ,
\end{array} \end{displaymath}
whence follows (110).

The dimension of the constant $a$ can be easily established from formula
(115). Since in the normalized regular basis $h^{55}$ is dimensionless and
the components $s^{\alpha \beta}_{\hspace{1.5ex} 5}$ have the same dimension
as $s^{\alpha \beta}_{\hspace{1.5ex} \mu}$, the expression following $a$ in
formula (115) should have the same dimension as $R$, so $a^{-1}$ should have
the same dimension as $k$. In view of this, let us put $a = (-1/2k) \,
\varrho$, where $\varrho$ is some unknown dimensionless constant, whose
value should be found experimentally. One will then have
\begin{equation}
{\bf L}_{\rm geom} = (-1/2k) \, ( R + \varrho \cdot g_{\alpha \sigma}
g_{\beta \tau} h^{55} s^{\alpha \beta}_{\hspace{1.5ex} 5} \,
s^{\sigma \tau}_{\hspace{1.5ex} 5} ) ,
\end{equation}
and the gravitational equations in the four-tensor notations will acquire
the following form:
\begin{equation} \begin{array}{l}
G_{\mu \nu} - g_{\mu \nu} \, {\scriptstyle \frac{1}{2}} \hspace{0.1ex}
\varepsilon \hspace{0.1ex} X_{\sigma \tau} X^{\sigma \tau} \; = \; k
\hspace{0.1ex} \Theta_{\mu \nu} \\ T_{\mu \nu}^{\hspace{1.5ex} \alpha}
+ \delta^{\alpha}_{\, \mu} \, T_{\nu \sigma}^{\hspace{1.5ex} \sigma} -
\delta^{\alpha}_{\, \nu} \, T_{\mu \sigma}^{\hspace{1.5ex} \sigma} \; = \; -
\, {\scriptstyle \frac{1}{2}} \hspace{0.1ex}k \, \Sigma^{\alpha}_{\, \mu \nu}
\rule{0ex}{3ex} \\ X_{\mu \nu} \; = \; - \, {\scriptstyle \frac{1}{2}}
\, \varepsilon^{-1} k \, \Xi_{\mu \nu} \, , \rule{0ex}{3ex}
\end{array} \end{equation}
where I have denoted $\; X^{\mu \nu} \equiv s^{\mu \nu}_{\hspace{1.5ex}5}
\cdot |h^{55}|^{1/2} \,$, $\; \Xi_{\mu \nu} \equiv {\cal M}^{5}_{\, \mu \nu}
\cdot |h^{55}|^{-1/2} \;$, and $\; \varepsilon \equiv \varrho \, {\rm sign}
\hspace{0.1ex} h^{55}$. Turning back to the five-tensor notations and using
the tensor $Y^{A}_{BC}$ introduced in the previous section, one obtains
\begin{equation} \begin{array}{rcl}
Y^{\alpha}_{\mu 5} + \delta^{\alpha}_{\, \mu} \, \varrho \, s^{|\sigma \tau|}
_{\hspace{3ex} 5} \, s_{\sigma \tau}^{\hspace{2ex} 5} &  = & k
\hspace{0.1ex} {\cal M}^{\alpha}_{\, \mu 5} \\ Y^{\alpha}_{\mu \nu} & = & k
\hspace{0.1ex}{\cal M}^{\alpha}_{\, \mu \nu} \\ - \; 2 \varrho \, s_{\mu \nu}
^{\hspace{2ex} 5} & = & k \hspace{0.1ex} {\cal M}^{5}_{\, \mu \nu} \, ,
\end{array} \end{equation}
where $s_{\mu \nu}^{\hspace{2ex} 5} \equiv s_{\mu \nu 5} \, h^{55}$.
We thus see that from our point of view equations (104) and (105)
correspond to the particular case where ${\cal M}^{5}_{\, \mu \nu}
= s_{\mu \nu}^{\hspace{2ex} 5} = 0$. To present equations (119) as
a single five-tensor equation, one should introduce still another
five-tensor, whose components are
\begin{displaymath}
Z^{\alpha}_{\mu 5} = \delta^{\alpha}_{\, \mu} \,
s^{|\sigma \tau|}_{\hspace{3ex} 5} \, s_{\sigma \tau}^{\hspace{2ex} 5}
\, , \; Z^{\alpha}_{\mu \nu} = Z^{5}_{\mu 5} = 0, \; Z^{5}_{\mu \nu}
= - \, 2 \hspace{0.1ex} s_{\mu \nu}^{\hspace{2ex} 5}.
\end{displaymath}
By using the latter, one can present equations (119) as
\begin{displaymath}
Y^{A}_{BC} + \varrho Z^{A}_{BC} \; = \; k \hspace{0.1ex}
{\cal M}^{A}_{\, BC} \, ,
\end{displaymath}
where it is assumed that ${\cal M}^{A}_{\, BC}$ are the components of the
``dynamical'' tensor $\MM$.

\vspace{3ex} \begin{flushleft}
H. \it Bivector derivative for the fields \\ \hspace*{2ex}
of nonspacetime vectors and tensors
\end{flushleft}
In section C of part V I have defined the five-vector covariant
derivative for the fields whose values are nonspacetime vectors or tensors,
thereby obtaining a certain five-vector generalization of the traditional
gauge field theory framework. It turns out that one can do the same with the
bivector derivative and obtain a more particular generalization where the
five-vector gauge fields introduced in part V are viewed as composite
quantities constructed from more elementary connection coefficients---from
those associated with the bivector derivative. This latter generalization
is obtained by postulating that for the fields of nonspacetime vectors and
tensors there exists a derivative whose argument is a five-vector bivector
and that for all such fields this derivative is related to their five-vector
covariant derivative according to equation (2), where $\sigma({\bf u})$ is
the same as it is for four-vector fields.

Let us first discuss the formal side of the matter. As in section C of
part V, let us consider a set $\VV$ of all sufficiently smooth fields
whose values are some $n$-dimensional nonspacetime vectors, which I will
denote as $\smallvec{U}$, $\smallvec{V}$, $\smallvec{W}$, etc. Defining the
bivector derivative for such fields is equivalent to specifying a map
\begin{displaymath}
{\sf D}: \; \FF \wedge \FF \times \VV \rightarrow \VV.
\end{displaymath}
The latter should satisfy the usual requirements: for any scalar functions
$f$ and $g$, any bivector fields $\AAA$ and $\BB$, and any fields
$\smallvec{U}$ and $\smallvec{V}$ from $\VV$,
\begin{flushright}
\hfill ${\sf D}_{(f{\cal A} + g{\cal B})} \smallvec{U} = f \cdot
{\sf D}_{\cal A} \smallvec{U} + g \cdot {\sf D}_{\cal B} \smallvec{U}$
\hfill {\rm (120a)} \\ \hfill ${\sf D}_{\cal A} (\smallvec{U} +
\smallvec{V}) = {\sf D}_{\cal A} \smallvec{U} + {\sf D}_{\cal A}
\smallvec{V}$ \hspace*{4.5ex} \rule{0ex}{3ex} \hfill {\rm (120b)} \\
\hfill ${\sf D}_{\cal A} (f \smallvec{U}) = {\sf D}_{\cal A} f \cdot
\smallvec{U} + f \cdot {\sf D}_{\cal A} \smallvec{U}$, \hspace*{0.5ex}
\rule{0ex}{3ex} \hfill {\rm (120c)}
\end{flushright} \setcounter{equation}{120}
where the bivector derivative of the scalar field $f$ is defined as in
section A. To define in a natural way the bivector derivative for all
other tensor fields over $\VV$, let us postulate that the action of
$\sf D$ on the contraction and tensor product obeys the Leibniz rule.

If $\smallvec{E}_{i}$ ($i = 1, \ldots, n$) is some set of basis fields from
$\VV$, one can define for it the connection coefficients associated with
the derivative $\sf D$ according to the formula
\begin{equation}
{\sf D}_{AB} \smallvec{E}_{i} = \smallvec{E}_{j} C^{j}_{\; i AB},
\end{equation}
where, as usual, ${\sf D}_{AB} \equiv {\sf D}_{{\bf e}_{A} \wedge
{\bf e}_{B}}$ and ${\bf e}_{A}$ is the selected five-vector basis. I will
call these connection coefficients the {\em bivector} gauge fields. Using
them, one can obtain the following expression for the components of the
bivector derivative of an arbitrary field $\smallvec{U} = \smallcomp{U}^{i}
\smallvec{E}_{i}$:
\begin{displaymath}
({\sf D}_{AB} \smallvec{U})^{i} = {\sf D}_{AB} \smallcomp{U}^{i} +
C^{i}_{\; j AB} \smallcomp{U}^{j} \equiv \smallcomp{U}^{i}_{\; : AB}.
\end{displaymath}
Therefore, in any active regular basis
\begin{displaymath}
\smallcomp{U}^{i}_{\; : \alpha 5} = \partial_{\alpha} \smallcomp{U}^{i} +
C^{i}_{\; j \alpha 5} \smallcomp{U}^{j} \; \; \mbox{ and } \; \;
\smallcomp{U}^{i}_{\; : \alpha \beta} = C^{i}_{\; j \alpha \beta}
\smallcomp{U}^{j}.
\end{displaymath}
Under the transformation ${\bf e}'_{A} = {\bf e}_{B} L^{B}_{\; A}$ of the
five-vector basis the bivector gauge fields transform simply as
\begin{displaymath}
C'^{\, i}_{\; \, jAB} = C^{i}_{\; jST} L^{S}_{\; A} L^{T}_{\; B}.
\end{displaymath}
Under the transformation $\smallvec{E} \rule{0ex}{1ex}^{\, \prime}_{i} =
\smallvec{E}_{j} \Lambda^{j}_{\, i}$ of the basis in $\VV$ these gauge
fields transform as
\begin{displaymath}
C'^{\, i}_{\; \, jAB} = (\Lambda^{-1})^{i}_{\, k} C^{k}_{\; lAB} \Lambda
^{l}_{\, j} + (\Lambda^{-1})^{i}_{\, k} {\sf D}_{AB} \Lambda^{k}_{\, j},
\end{displaymath}
so in any active regular basis one has
\begin{equation} \begin{array}{rcl}
C'^{\, i}_{\; \, j \alpha 5} & = & (\Lambda^{-1})^{i}_{\, k}
C^{k}_{\; l \alpha 5} \Lambda^{l}_{\, j} + (\Lambda^{-1})^{i}_{\, k}
\partial_{\alpha} \Lambda^{k}_{\, j} \\ C'^{\, i}_{\; \, j \alpha \beta}
& = & (\Lambda^{-1})^{i}_{\, k} C^{k}_{\; l \alpha \beta}
\Lambda^{l}_{\, j}. \rule{0ex}{3ex}
\end{array} \end{equation}
Thus, in such a basis the quantities $C^{i}_{\; j \alpha 5}$ transform as
ordinary gauge fields, while the quantities $C^{i}_{\; j \alpha \beta}$
transform as components of a tensor and cannot be nullified at a given
space-time point by an appropriate choice of the basis in $\VV$.

Let us now write down explicitly the relation between the derivatives
$\plaision$ and $\sf D$ for the considered type of fields. As it has been
said above, for any field $\smallvec{U}$ from $\VV$ one should have
\begin{equation}
\plaision_{\bf v} \smallvec{U} = {\sf D}_{\sigma({\bf v})} \smallvec{U}
\end{equation}
at any $\bf v$. For $\smallvec{U} = \smallvec{E}_{i}$ and ${\bf v} =
{\bf e}_{A}$ one has
\begin{displaymath} \begin{array}{l}
\smallvec{E}_{i} B^{i}_{\; j A} = \, \plaision_{A} \smallvec{E}_{j}
= {\sf D}_{\sigma({\bf e}_{A})} \smallvec{E}_{j} \\ \hspace{7.3ex} =
s^{|KL|}_{\hspace{3ex} A} {\sf D}_{KL} \smallvec{E}_{j} = \, \smallvec{E}_{i}
C^{i}_{\; j KL} s^{|KL|}_{\hspace{3ex} A} \, . \rule{0ex}{3ex}
\end{array} \end{displaymath}
Consequently,
\begin{equation}
B^{i}_{\; j A} = C^{i}_{\; j KL} s^{|KL|}_{\hspace{3ex} A},
\end{equation}
so in any active regular basis one has
\begin{equation} \begin{array}{rcl}
B^{i}_{\; j \alpha} & = & C^{i}_{\; j \alpha 5} + C^{i}_{\; j \mu \nu}
s^{|\mu \nu|}_{\hspace{3ex} \alpha} \\ B^{i}_{\; j 5} & = &
C^{i}_{\; j \mu \nu} s^{|\mu \nu|}_{\hspace{3ex} 5}. \rule{0ex}{3ex}
\end{array} \end{equation}

The latter formulae elucidate the meaning of the bivector gauge fields.
Within the traditional gauge field theory scheme, the parallel transport of
nonspacetime vectors is independent of torsion in the sense that there is
no direct relation between the latter and the corresponding gauge fields
associated with the covariant derivative. According to the scheme we are
now discussing, the parallel transport of nonspacetime vectors {\em is}
torsion-dependent, which manifests itself in an additional rotation of
transported vectors compared to the case where torsion is zero. Let me also
note that the scheme with ordinary gauge fields can be viewed as a particular
case of the one we are now considering, which corresponds to the situation
where the bivector gauge fields $C^{i}_{\; j \mu \nu}$ in any regular
five-vector basis are all identically zero.

As in the case of four-vector and five-vector fields, the bivector derivative
operator for the fields of nonspacetime vectors can be split into two parts:
\begin{equation}
{\sf D}_{\cal A} = {\sf D}_{\cal A^{E}} + {\sf D}_{\cal A^{Z}}.
\end{equation}
The first operator in the right-hand side can be regarded as a function
of the four-vector $\bf A$ that corresponds to the $\cal E$-component of
the bivector $\AAA$ (or as a function of the corresponding five-vector from
$\cal Z$), and it is a simple matter to show that when regarded this way,
it has all the properties of an ordinary covariant derivative, which
permits one to denote this operator as $\dotnabla_{\bf A}$. It is
easy to see that in any four-vector basis the connection coefficients
associated with $\dotnabla$, defined according to formula (37) of part V,
equal $C^{i}_{\; j \mu 5}$ provided that the latter are evaluated for the
corresponding active regular five-vector basis.

In a similar manner, the second operator in the right-hand side of formula
(126) can be viewed as a function of the four-vector bivector $\bf B$ that
corresponds to $\AAA^{\cal Z}$, and by analogy with the case of four- and
five-vectors, I will denote it as $\widehat{\bf M}_{\bf B}$. Naturally, in
the case of nonspacetime vectors the components of $\widehat{\bf M}$ will no
longer equal $(M_{\mu \nu})^{\alpha}_{\, \beta}$ or $(M_{\mu \nu})^{A}_{\,
B}$, but instead, in any four-vector basis one will have
\begin{displaymath}
(\widehat{\bf M}_{\alpha \beta})^{i}_{\; j} = C^{i}_{\; j \alpha \beta},
\end{displaymath}
where the bivector gauge fields in the right-hand side are to be evaluated
in the corresponding regular five-vector basis. The latter fact reflects the
fundamental difference between the case of four- and five-vectors and the
case of nonspacetime vectors in relation to the bivector derivative: whereas
for the former the operator $\widehat{\bf M}$ is fixed and its components
are constructed from the Lorentz-invariant quantities $g_{\alpha \beta}$ and
$\delta^{\alpha}_{\, \beta}$, for the latter the operator $\widehat{\bf M}$
can be as arbitrary as is allowed by the constraints imposed on $\sf D$ and
its components represent an independent element of the geometry associated
with the considered type of nonspacetime vectors, just as within the
traditional scheme this is done by ordinary gauge fields. Such a state of
affairs has a certain logic to it. Since the components of the operator
$\widehat{\bf M}$ for four-vector fields are fixed, the additional rotation
of such vectors in the process of their parallel transport compared to the
case where torsion is zero but the Riemannian geometry is the same, is
determined only by the quantities $s^{\mu \nu}_{\hspace{1ex} A}$, and having
found the latter this way, one can then make a similar comparison for the
transport of considered nonspacetime vectors and determine the combinations
$C^{i}_{\; j \mu \nu} s^{|\mu \nu|}_{\hspace{2.5ex} A}$, from which, knowing
the torsion, one can find the quantities $C^{i}_{\; j \mu \nu}$ themselves.

Thus, in the case of nonspacetime vectors, too, the bivector derivative
can be presented as in formula (25), only now both $\dotnabla$ and
$\widehat{\bf M}$ are independent of metric. Let me also stress that the
derivative $\dotnabla$ should not be mixed up with the ``differential part''
of the operator $\plaision$, which earlier I have denoted as $\nabla$
and which is related to $\plaision$ according to formula (12) of part V. That
these two derivatives are different is already seen from the fact that the
connection coefficients for $\nabla$ equal $B^{i}_{\; j \alpha} =
C^{i}_{\; j \alpha 5} + C^{i}_{\; j \mu \nu} s^{|\mu \nu|}_{\hspace{3ex}
\alpha}$, and not $C^{i}_{\; j \alpha 5}$ as they do for $\dotnabla$.

Let us now consider the analog of the field strength tensor. From
everything that has been said in section C it is apparent that the
latter can be defined by the equation
\begin{equation}
< \FS, \AAA \otimes \BB > \; = {\sf D}_{\cal A} {\sf D}_{\cal B} -
{\sf D}_{\cal B} {\sf D}_{\cal A} - {\sf D}_{\cal [ \, A,B \, ]},
\end{equation}
where the operators on both sides act on the fields of the considered
nonspacetime vectors. Consequently, the components of $\FS$ have the
following symmetry properties:
\begin{equation} \begin{array}{l}
{\sf F}^{i}_{\; j KLMN} = - \, {\sf F}^{i}_{\; j LKMN} = - \,
{\sf F}^{i}_{\; j KLNM} \\ {\sf F}^{i}_{\; j KLMN} = - \,
{\sf F}^{i}_{\; j MNKL}, \rule{0ex}{3ex}
\end{array} \end{equation}
and one can easily derive for them the following expression in terms of the
corresponding bivector gauge fields:
\begin{equation} \begin{array}{l}
{\sf F}^{i}_{\; j KLMN} = {\sf D}_{KL} C^{i}_{\; j MN} - {\sf D}_{MN}
C^{i}_{\; j KL} \\ \hspace{12ex} + \; C^{i}_{\; s KL} C^{s}_{\; j MN} -
C^{i}_{\; s MN} C^{s}_{\; j KL} \rule{0ex}{3ex} \\ \hspace{12ex} - \;
C^{i}_{\; j ST} Q^{|ST|}_{\hspace{2.5ex} KL \, MN} \, , \rule{0ex}{3ex}
\end{array} \end{equation}
where $Q^{ST}_{\hspace{1.5ex} KL \, MN}$ are the commutation constants for
basis bivector fields, which are defined in the following evident way:
\begin{displaymath}
[ \, {\bf e}_{K} \wedge {\bf e}_{L}, {\bf e}_{M} \wedge {\bf e}_{N} \, ] =
{\bf e}_{S} \wedge {\bf e}_{T} \, Q^{|ST|}_{\hspace{2.5ex} KL \, MN}.
\end{displaymath}
By using definition (50) one can easily prove that for any active regular
coordinate basis one has $Q^{ST}_{\hspace{1.5ex} \kappa 5 \, \mu 5} = 0$,
$Q^{\sigma 5}_{\hspace{1.5ex} \kappa 5 \, \mu \nu} = 0$, and $Q^{\sigma
\tau}_{\hspace{1.5ex} \kappa 5 \, \mu \nu} = \delta^{\sigma}_{\, \mu}
G^{\tau}_{\; \nu \kappa} - \delta^{\sigma}_{\, \nu} G^{\tau}_{\; \mu \kappa}
- \delta^{\tau}_{\, \mu} G^{\sigma}_{\; \nu \kappa} + \delta^{\tau}_{\, \nu}
G^{\sigma}_{\; \mu \kappa}$, where $G^{\sigma}_{\; \mu \kappa}$ are the
connection coefficients for five-vector fields, associated with the
derivative $\dotnabla$, so in this basis one has
\begin{equation} \begin{array}{l}
{\sf F}^{i}_{\; j \, \kappa 5 \mu 5} = \partial_{\kappa} C^{i}_{\; j \mu 5}
- \partial_{\mu} C^{i}_{\; j \kappa 5} \\ \hspace{15ex} + \;
C^{i}_{\; s \kappa 5} C^{s}_{\; j \mu 5} -
C^{i}_{\; s \mu 5} C^{s}_{\; j \kappa 5} \rule{0ex}{3ex}
\end{array} \end{equation}
and
\begin{equation} \begin{array}{l}
{\sf F}^{i}_{\; j \, \kappa 5 \mu \nu} = \partial_{\kappa}
C^{i}_{\; j \mu \nu} + C^{i}_{\; s \kappa 5} C^{s}_{\; j \mu \nu} \\
\hspace{3ex} - \; C^{i}_{\; s \mu \nu} C^{s}_{\; j \kappa 5} -
C^{i}_{\; j \omega \nu} G^{\omega}_{\; \mu \kappa} -
C^{i}_{\; j \mu \omega} G^{\omega}_{\; \nu \kappa}. \rule{0ex}{3ex}
\end{array} \end{equation}
Likewise, one can prove that for the indicated basis the commutation
constants $Q^{ST}_{\hspace{1.5ex} \kappa \lambda \, \mu \nu}$ are given
by equation (51), so one has
\begin{equation} \begin{array}{l}
{\sf F}^{i}_{\; j \, \kappa \lambda \mu \nu} =  C^{i}_{\; s \kappa \lambda}
C^{s}_{\; j \mu \nu} - C^{i}_{\; s \mu \nu} C^{s}_{\; j \kappa \lambda} \\
\hspace{13ex} - \; g_{\kappa \mu} C^{i}_{\; j \lambda \nu} +
g_{\kappa \nu} C^{i}_{\; j \lambda \mu} \rule{0ex}{3ex} \\
\hspace{17ex} + \; g_{\lambda \mu} C^{i}_{\; j \kappa \nu} -
g_{\lambda \nu} C^{i}_{\; j \kappa \mu} \, . \rule{0ex}{3ex}
\end{array} \end{equation}

As in the case of bivector analogs of the Riemann tensor, one can express
the components of $\FS$ in terms of the corresponding four-vector quantities.
For that one should first introduce the ordinary gauge fields, $A^{i}_{\; j
\alpha}$, associated with the derivative $\dotnabla$ defined above. Denoting
the components of the four-vector field strength tensor constructed out of
them as $F^{i}_{\; j \alpha \beta}$, one can rewrire equation (130) as
\begin{equation}
{\sf F}^{i}_{\; j \, \kappa 5 \mu 5} =  F^{i}_{\; j \kappa \mu},
\end{equation}
which is the analog of formula (31). One can then introduce a four-vector
2-form whose values are tensors of rank $(1,1)$ over $\VV$ and whose
components in any four-vector basis coincide with the quantities
$C^{i}_{\; j \mu \nu}$ evaluated for the corresponding regular five-vector
basis. Denoting the components of this 2-form as $E^{i}_{\; j \mu \nu}$,
one can rewrite equation (131) as
\begin{equation} \begin{array}{l}
{\sf F}^{i}_{\; j \, \kappa 5 \mu \nu} = \partial_{\kappa}
E^{i}_{\; j \mu \nu} + A^{i}_{\; s \kappa} E^{s}_{\; j \mu \nu} -
E^{i}_{\; s \mu \nu} A^{s}_{\; j \kappa} \\ \hspace{7ex} - \;
E^{i}_{\; j \omega \nu} \Gamma^{\omega}_{\; \mu \kappa} -
E^{i}_{\; j \mu \omega} \Gamma^{\omega}_{\; \nu \kappa} =
E^{i}_{\; j \mu \nu ; \hspace{0.1ex} \kappa} \rule{0ex}{3ex} \, ,
\end{array} \end{equation}
where the semicolon denotes the covariant differentiation corresponding to
$\dotnabla$. Finally, equation (132) will acquire the form
\begin{equation} \begin{array}{l}
{\sf F}^{i}_{\; j \, \kappa \lambda \mu \nu} =  E^{i}_{\; s \kappa \lambda}
E^{s}_{\; j \mu \nu} - E^{i}_{\; s \mu \nu} E^{s}_{\; j \kappa \lambda} \\
\hspace{13ex} - \; g_{\kappa \mu} E^{i}_{\; j \lambda \nu} + g_{\kappa \nu}
E^{i}_{\; j \lambda \mu} \rule{0ex}{3ex} \\ \hspace{13ex} + \;
g_{\lambda \mu} E^{i}_{\; j \kappa \nu} - g_{\lambda \nu}
E^{i}_{\; j \kappa \mu} \, . \rule{0ex}{3ex}
\end{array} \end{equation}
It is evident that formulae (133)--(135) will hold in {\em any} active
regular basis.

As $\RS$, $\FS$ satisfies a differential identity, which can be easily
obtained from the general identity (61) and which has the same form as
equation (62):
\begin{equation} \begin{array}{l}
{\sf D}_{\cal A} < \FS , \BB \wedge \CC > - < \FS ,
[ \, \AAA , \BB \, ] \wedge \CC > \\
\hspace{4ex} + \mbox{ cyclic permutations }
= \; {\sf D}_{\bf \Delta (\cal A,B,C )} \, ,
\end{array} \end{equation}
where $\FS$ is regarded as a tensor. It is interesting to see how from
this latter identity one can derive identity (53) of part V for the
five-vector field strength tensor $\bf F$. To this end let us first rewrite
equation (136) for the case where $\FS$ is regarded as an operator:
\begin{displaymath} \begin{array}{l}
[ \, {\sf D}_{\cal A} , < \FS , \BB \wedge \CC > \, ] \; - < \FS , [ \,
\AAA , \BB \, ] \wedge \CC > \\
\hspace{6ex}
+ \; {\sf D}_{\cal [A,[B,C]]} \, + \,
\mbox{cyclic permutations} \; = \; {\bf 0},
\end{array} \end{displaymath}
where I have also written out explicitly the quantity $\bf \Delta (\AAA,
\BB,\CC)$. By making simple transformations one can cast the latter equation
into the form
\begin{equation} \begin{array}{l}
[ \, {\sf D}_{\cal A}, ( < \FS , \BB \wedge \CC > + {\sf D}_{\cal [B,C]} \,
) \, ] \\ \hspace{13ex} + \mbox{ cyclic permutations} = {\bf 0}.
\end{array} \end{equation}
By using the definitions of the tensors $\FS$ and $\bf F$, it is not
difficult to prove that for any two five-vector fields $\bf v$ and $\bf w$,
\begin{displaymath} \begin{array}{l}
\bf < \FS , \, \sigma(v) \wedge \sigma (w) > + \; {\sf D}_{[ \, \sigma (v),
\, \sigma(w) \, ]} \\ \bf \hspace{17ex} = \; < F , v \wedge w > + \; \,
\plaision_{[v,w]} \, , \rule{0ex}{3ex}
\end{array} \end{displaymath}
where $\sigma$ is the same as in equation (123). Substituting this expression
into the left-hand side of equation (137) at $\AAA = \sigma(\bf u)$, $\BB =
\sigma(\bf v)$, and $\CC = \sigma(\bf w)$, one finds that
\begin{displaymath} \begin{array}{l}
\bf [ \; \plaision_{u}, ( < F, v \wedge w > + \; \plaision_{[\, v, \, w \,]}
\, ) \, ] \; + \; \mbox{\rm cyclic perm.} \\ \bf \hspace{6ex} = \; [ \;
\plaision_{u}, < F, v \wedge w > \, ] \; +  < F, u \wedge [ \, v,w \, ] >
\rule{0ex}{2.5ex} \\ \bf \hspace{18.5ex} + \; \, \plaision_{[ \, u, \,
[ \, v, \, w \, ]]} \; + \; \mbox{\rm cyclic perm.} \rule{0ex}{2.5ex} \\
\bf \hspace{6ex} = \; \plaision_{([ \, u, \, [ \, v, \, w \, ]] + [ \, v, \,
[\, w, \, u \, ]] + [ \, w, \, [ \, u, \, v \, ]])} \rule{0ex}{2.5ex} \\ \bf
\hspace{18.5ex} + <dF,u \wedge v \wedge w> \; = \; {\bf 0},\rule{0ex}{2.5ex}
\end{array} \end{displaymath}
and since the commutators of five-vector fields satisfy the Jacobi
identity, from the latter equation it follows that $\bf dF = 0$.

\vspace{3ex}

To give the reader an idea of the consequences the considered generalization of the traditional gauge field theory framework leads to, let us see how the formalism developed above applies to classical electrodynamics. In this case the fields $A_{\alpha}$ and $E_{\mu \nu}$ will apparently lose their idices ${i}$ and $j$, and formulae (133)--(135) will acquire the following simpler form:
\begin{equation} \begin{array}{l}
{\sf F}_{\mu 5 \alpha 5} = F_{\mu \alpha}, \; {\sf F}_{\mu 5 \alpha
\beta} = - {\sf F}_{\alpha \beta \mu 5} = \partial_{\mu} E_{\alpha \beta},
\\
{\sf F}_{\mu \nu \alpha \beta} = - \; g_{\mu \alpha} E_{\nu \beta} +
g_{\nu \alpha} E_{\mu \beta} \rule{0ex}{3ex} \\ 
\hspace{18ex} + \; g_{\mu \beta} E_{\nu \alpha} - g_{\nu
\beta} E_{\mu \alpha}. \rule{0ex}{3ex}
\end{array} \end{equation}
From the point of view of brother physicist, who typically does not care much for the fancy mathematics that underlies a new concept or theory and is interested only in the result, the discussed generalization comes to that in addition to the usual electromagnetic interaction one introduces a new interaction mediated by an antisymmetric tensor field. This is by no means a new idea in physics, and the Lagrangian density for the pair of fields $A_{\alpha}$ and $E_{\mu \nu}$ at an appropriate normalization of the latter is usually taken to be 
\begin{equation}
- {\scriptstyle \frac{1}{4}}
F_{\alpha \beta} F^{\alpha \beta} + \; {\scriptstyle \frac{1}{4}} \,
\partial_{\mu} E_{\alpha \beta} \, \partial^{\mu} \! E^{\alpha \beta} -
(\partial^{\mu} \! E_{\mu \alpha})^{2} ,
\end{equation}
see e.g.\ ref.[1]. If one does keep in mind the underlying mathematical concept, one would expect that the Lagrangian density in formula (139) can be expressed in terms of the components of the five-tensor $\FS$ as some bilinear combination of the latter. It is not difficult to check that from the quantities ${\sf F}_{ABCD}$ one can construct only two independent true scalars, for instance, ${\sf F}^{ABCD} {\sf F}_{ABCD}$ and ${\sf F}^{AC}_{\hspace{2ex} AD} \, {\sf F}^{BD}_{\hspace{2ex} BC}$. Consequently, the Lagrangian density can be expressed in the following form:
\begin{displaymath}
a \cdot {\sf F}^{ABCD} {\sf F}_{ABCD} + b \cdot {\sf F}^{AC}_{\hspace{2ex}AD}
\, {\sf F}^{BD}_{\hspace{2ex} BC}.
\end{displaymath}
Substituting into this expression the values of the components given by formulae (138) and considering that in an active regular basis $h^{55} = \kappa^{-2}$, one finds that the above sum equals
\begin{displaymath} \begin{array}{c}
(4a-b) \cdot \kappa^{-4} \cdot (F^{\alpha \beta} F_{\alpha \beta}) \\ + \;
4a \cdot \kappa^{-2} \cdot (\partial^{\mu} E^{\alpha \beta}) (\partial_{\mu}
E_{\alpha \beta}) - 2b \cdot \kappa^{-2} \cdot
(\partial^{\mu} E_{\mu \alpha})^{2} \rule{0ex}{3ex} \\ + \; 4b \cdot \kappa^{-2} \cdot
(F^{\alpha \beta} E_{\alpha \beta}) + (8a-4b) \cdot (E^{\alpha \beta}
E_{\alpha \beta}) \rule{0ex}{3ex}.
\end{array} \end{displaymath}
Requiring that the coefficient in front of the combination $F^{\alpha \beta} \! F_{\alpha \beta}$ equal $-1/4$, one obtains
\begin{displaymath} \begin{array}{c}
{-\scriptstyle \frac{1}{4}} F^{\alpha \beta} \! F_{\alpha \beta} +
(-{\scriptstyle \frac{1}{4}} \kappa^{2} + b \kappa^{-2}) \cdot (\partial^{\mu}
E^{\alpha \beta}) (\partial_{\mu} E_{\alpha \beta}) \\ \hspace{3ex} - \, 2b
\kappa^{-2} \cdot (\partial^{\mu} E_{\mu \alpha})^{2} + 4b \kappa^{-2} \cdot
(F^{\alpha \beta} E_{\alpha \beta}) \rule{0ex}{3ex} \\ - \; ({\scriptstyle \frac{1}{2}} \kappa^{4}
+ 2b) \cdot (E^{\alpha \beta} E_{\alpha \beta}) \rule{0ex}{3ex}.
\end{array} \end{displaymath}
It is evident that the coefficient $b$ should be of dimension
{\em length}$^{-4}$, so one can put $b = \kappa^{4} d$, where $d$ is
simply a number, and obtain
\begin{displaymath} \begin{array}{c}
{-\scriptstyle \frac{1}{4}} F^{\alpha \beta} \! F_{\alpha \beta} +
(-{\scriptstyle \frac{1}{4}} + d) \cdot \kappa^{2} \cdot (\partial^{\mu}
E^{\alpha \beta}) (\partial_{\mu} E_{\alpha \beta}) \\ \hspace{3ex} - \,
2d \cdot \kappa^{2} \cdot (\partial^{\mu} E_{\mu \alpha})^{2} + 4d \cdot
\kappa^{2} \cdot (F^{\alpha \beta} E_{\alpha \beta}) \rule{0ex}{3ex} \\ - \; 
({\scriptstyle \frac{1}{2}} + 2d) \cdot \kappa^{4} \cdot (E^{\alpha \beta} E_{\alpha \beta})
\rule{0ex}{3ex}.
\end{array} \end{displaymath}
Redefining the antisymmetric tensor field according to the rule
$E_{\alpha \beta} \rightarrow (\varepsilon \kappa)^{-1} E_{\alpha \beta}$,
where $\varepsilon$ is an unknown nonzero number, and requiring the
coefficient in the second term to equal $1/4$, one obtains
\begin{displaymath} \begin{array}{c}
{-\scriptstyle \frac{1}{4}} F^{\alpha \beta} \! F_{\alpha \beta} +
{\scriptstyle \frac{1}{4}} \, \partial^{\mu} E^{\alpha \beta} \partial_{\mu}
E_{\alpha \beta} - {\scriptstyle \frac{1}{2}} (\varepsilon^{2}+1) \,
(\partial^{\mu} E_{\mu \alpha})^{2} \\ + \, \kappa \,
(\varepsilon^{2}+1) \cdot F^{\alpha \beta} \! E_{\alpha \beta}
- {\scriptstyle \frac{1}{2}} \kappa^{2} (\varepsilon^{2}+2)
\cdot E^{\alpha \beta} \! E_{\alpha \beta} \rule{0ex}{3ex}.
\end{array} \end{displaymath}
Finally, requiring the coefficient in the third term to be $-1$ one
finds that $\varepsilon^{2} = 1$ and obtains
\begin{equation} \begin{array}{l}
{-\scriptstyle \frac{1}{4}} F^{\alpha \beta} \! F_{\alpha \beta} +
{\scriptstyle \frac{1}{4}} \, \partial^{\mu} E^{\alpha \beta} \partial_{\mu}
E_{\alpha \beta} - (\partial^{\mu} E_{\mu \alpha})^{2} \\ \hspace{13ex} 
+ \; 2 \kappa \, F^{\alpha \beta} \! E_{\alpha \beta} - {\scriptstyle \frac{3}{2}}
\kappa^{2} \, E^{\alpha \beta} \! E_{\alpha \beta} \rule{0ex}{3ex}.
\end{array} \end{equation}

As one can see, in addition to the standard Lagrangian densities for the
electromagnetic and antisymmetric tensor fields, one obtains an interaction
term between $F_{\alpha \beta}$ and $E_{\alpha \beta}$ and a term that has
the form of a mass term for the field $E_{\alpha \beta}$. Owing to the first
of these terms, Maxwell's equation and the equation for $E_{\alpha \beta}$
cease being independent from each other and in vacuum acquire the following form:
\begin{equation}
\partial^{\alpha} \! F_{\alpha \beta} = 4 \kappa \, \partial^{\alpha}
\! E_{\alpha \beta} ,
\end{equation}
\begin{equation} \begin{array}{c}
\partial^{2} \! E_{\alpha \beta} +
2 \, \partial^{\lambda} ( \partial_{\alpha} \! E_{\beta \lambda} -
\partial_{\beta} \! E_{\alpha \lambda} )  + 6 \kappa^{2} E_{\alpha \beta} \\ 
= 4 \kappa \, F_{\alpha \beta} . \rule{0ex}{3ex}
\end{array} \end{equation}
The general solution of these equations can be presented as follows:
\begin{displaymath} \begin{array}{l}
F_{\alpha \beta} = F^{(1)}_{\alpha \beta} + F^{(2)}_{\alpha \beta} \\
E_{\alpha \beta} = E^{(1)}_{\alpha \beta} + E^{(2)}_{\alpha \beta} + E^{(3)}_{\alpha \beta} .  \rule{0ex}{3ex}
\end{array} \end{displaymath}
Here $F^{(1)}_{\alpha \beta}$ is a solution of Maxwell's equation in vacuum and 
\begin{displaymath}
E^{(1)}_{\alpha \beta} = {\scriptstyle \frac{2}{3}} \, \kappa^{-1} F^{(1)}_{\alpha \beta} .
\end{displaymath}
$F^{(2)}_{\alpha \beta}$ is the electromagnetic field created by the source $C_{\beta} \equiv \partial^{\alpha} \! E_{\alpha \beta}$ that satisfies the equation
\begin{displaymath} 
\partial^{2} C_{\beta} + 10 \kappa^{2} \, C_{\beta} = 0 .
\end{displaymath}
The field $F^{(2)}_{\alpha \beta}$ has a nonzero longitudinal component and {\it nonzero mass} $m_{\rm ph} \equiv \sqrt{10} \kappa$, and one has
\begin{displaymath}
E^{(2)}_{\alpha \beta} = {\scriptstyle \frac{1}{4}} \, \kappa^{-1} F^{(2)}_{\alpha \beta} .
\end{displaymath}
$E^{(3)}_{\alpha \beta}$ is a solution of the Kaluza-Klein equation
\begin{displaymath} 
\partial^{2} E^{(3)}_{\alpha \beta} + 6 \kappa^{2} \, E^{(3)}_{\alpha \beta} = 0
\end{displaymath}
and such that $\partial^{\alpha} \! E^{(3)}_{\alpha \beta} = 0$. It is not accompanied by an electromagnetic field.

In the presence of matter the field equations acquire the following form:
\begin{equation}
\partial^{\alpha} \! F_{\alpha \beta} = 4 \kappa \, \partial^{\alpha}
\! E_{\alpha \beta} + j_{\beta} ,
\end{equation}
\begin{equation} \begin{array}{c}
\partial^{2} \! E_{\alpha \beta} +
2 \, \partial^{\lambda} ( \partial_{\alpha} \! E_{\beta \lambda} -
\partial_{\beta} \! E_{\alpha \lambda} )  + 6 \kappa^{2} E_{\alpha \beta} \\ 
= 4 \kappa \, F_{\alpha \beta} + \kappa^{-1} j_{\alpha \beta} , \rule{0ex}{3ex}
\end{array} \end{equation}
where $j_{\alpha}$ and $j_{\alpha \beta}$ can be obtained by varying the part of the action that describes the interaction of matter with bivector gauge fields with respect to $A_{\alpha}$ and $E_{\alpha \beta}$. The construction of a consistent model that would describe the interaction of a classical charged point particle with the field $E_{\alpha \beta}$ is discussed in ref.[2].

\vspace{3ex}

When considering the bivector analog of the field strength tensor and
other similar quantities, it is convenient to introduce some new notations
and terminology. First of all, one may denote
\begin{displaymath}
{\bf e}_{AB} \equiv {\bf e}_{A} \wedge {\bf e}_{B} \; \; \mbox{ and } \; \;
\widetilde{\bf o}^{AB} \equiv \widetilde{\bf o}^{A} \wedge
\widetilde{\bf o}^{B}.
\end{displaymath}
One will then have
\begin{displaymath} \begin{array}{rcl}
< \widetilde{\bf o}^{AB}, {\bf e}_{CD} > & = &
< \widetilde{\bf o}^{A} \wedge \widetilde{\bf o}^{B}, {\bf e}_{C} \wedge
{\bf e}_{D} > \\ & = & \delta^{A}_{C} \delta^{B}_{D} - \delta^{A}_{D}
\delta^{B}_{C} \; \; \equiv \; \; \delta^{AB}_{CD}. \rule{0ex}{3ex}
\end{array} \end{displaymath}
Secondly, for any two five-vector bivectors $\AAA$ and $\BB$ one can define
\begin{equation}
\AAA \wedge \BB \equiv \AAA \otimes \BB - \BB \otimes \AAA.
\end{equation}
Thus, according to these notations the wedge product of two bivectors
is constructed as if $\AAA$ and $\BB$ were tangent vectors, i.e.\ by
antisymmetrizing their tensor product, and not as it is usually done in
exterior calculus, where, for example,
\begin{displaymath}
\bf (a \wedge b) \wedge (c \wedge d) = a \wedge b \wedge c \wedge d.
\end{displaymath}
In a similar manner, one may take that
\begin{equation}
\widetilde{\bf o}^{AB} \wedge \widetilde{\bf o}^{CD} =
\widetilde{\bf o}^{AB} \otimes \widetilde{\bf o}^{CD} -
\widetilde{\bf o}^{CD} \otimes \widetilde{\bf o}^{AB},
\end{equation}
and postulate that
\begin{displaymath} \begin{array}{l}
< \widetilde{\bf o}^{AB} \wedge \widetilde{\bf o}^{CD}, {\bf e}_{KL} \wedge
{\bf e}_{MN} > \\ \hspace{7ex} = \; <\widetilde{\bf o}^{AB}, {\bf e}_{KL}> \,
< \widetilde{\bf o}^{CD}, {\bf e}_{MN} > \\ \hspace{12ex} - \; < \widetilde
{\bf o}^{AB},{\bf e}_{MN} > \, < \widetilde{\bf o}^{CD}, {\bf e}_{KL} >.
\end{array} \end{displaymath}

In all these formulae five-vector bivectors and five-vector 2-forms are
treated as if they were tangent vectors and 1-forms. In accordance with this,
one can call quantity (145) a ``bivector'' and quantity (146) a ``2-form''.
To distinguish such ``bivectors made of bivectors'' and ``2-forms made of
2-forms'' from the bivectors and 2-forms made of tangent vectors and 1-forms
one may add the adjective ``bivector'' to the corresponding noun. In that
case, however, one will have such awkward combinations as ``bivector
bivector'' and such misleading terms as ``bivector 2-form''. To avoid this,
instead of the adjective ``bivector'' one can use the adjective ``adjoint'',
adopting it from group theory. Thus, a five-vector bivector can be called an
{\em adjoint vector}\hspace{0.2ex}; a five-vector 2-form can be called an
{\em adjoint} 1-{\em form}\hspace{0.1ex}; the wedge product (145) will be
an {\em adjoint bivector}\hspace{0.2ex}; and tensors $\RS$ and $\FS$ will
be operator-valued {\em adjoint} 2-{\em forms}.

One can go even further and introduce a new type of indices that would
replace the pairs of antisymmetrized five-vector indices that lable
the components of five-vector bivectors and five-vector 2-forms and the
corresponding basis elements. I will call such indices {\em adjoint} and
will denote them with capital Gothic letters. By definition, an adjoint
index runs through ten values, which, for example, can be chosen to be 01,
02, 03, 05, 12, 13, 15, 23, 25, and 35. Then, instead of
\begin{displaymath}
\AAA = {\cal A}^{|KL|} {\bf e}_{KL} \; \; \mbox{ and } \; \;
\BB = {\cal B}_{|KL|} \widetilde{\bf o}^{KL}
\end{displaymath}
one can write
\begin{displaymath}
\AAA = {\cal A}^{\Re} {\bf e}_{\Re} \; \; \mbox{ and } \; \;
\BB = {\cal B}_{\Re} \widetilde{\bf o}^{\Re},
\end{displaymath}
and instead of formula (29),
\begin{displaymath} \begin{array}{rcl}
\RS & = & {\bf E}_{\alpha} \otimes \widetilde{\bf O}^{\beta} \,
{\sf R}^{\alpha}_{\; \beta \, \Im \Re} \, \widetilde{\bf o}^{\Im} \otimes
\widetilde{\bf o}^{\Re} \\ & = & {\scriptstyle \frac{1}{2}} \,
{\bf E}_{\alpha} \otimes \widetilde{\bf O}^{\beta} \,
{\sf R}^{\alpha}_{\; \beta \, \Im \Re} \, (\widetilde{\bf o}^{\Im} \otimes
\widetilde{\bf o}^{\Re} - \widetilde{\bf o}^{\Re} \otimes \widetilde
{\bf o}^{\Im}) \\ & = & {\scriptstyle \frac{1}{2}} \, {\bf E}_{\alpha}
\otimes \widetilde{\bf O}^{\beta} \, {\sf R}^{\alpha}_{\; \beta \, \Im \Re}
\, \widetilde{\bf o}^{\Im} \wedge \widetilde{\bf o}^{\Re}.
\end{array} \end{displaymath}
Definition (127) can now be presented in the form
\begin{displaymath}
< \FS, \AAA \wedge \BB > \; = {\sf D}_{\cal A} {\sf D}_{\cal B} -
{\sf D}_{\cal B} {\sf D}_{\cal A} - {\sf D}_{\cal [ \, A,B \, ]},
\end{displaymath}
similar to the usual definition of the field strength tensor. Finally, one
can introduce a certain order on the set of values of adjoint indices, for
example, $01<02<03<05<12<13<15<23<25<35$, and present the above expression
for $\RS$ as
\begin{displaymath}
\RS = {\bf E}_{\alpha} \otimes \widetilde{\bf O}^{\beta} \,
{\sf R}^{\alpha}_{\; \beta \, |\Im \Re|} \widetilde{\bf o}^{\Im}
\wedge \widetilde{\bf o}^{\Re},
\end{displaymath}
where, as usual, the  bars around the indices mean that summation extends
only over $\Im < \Re$.

\vspace{3ex} \begin{flushleft}
I. \it Integration of adjoint forms
\end{flushleft}
In conclusion of this paper let us briefly discuss one more item that
concerns the five-tensors which according to the terminology introduced in
the previous section are adjoint forms. Up to now I have discussed only
the differential properties of such tensors and have said nothing about
their integration. Obviously, one can take that the integral, say, of the
adjoint 2-form $\FS$ over a given surface by definition equals the integral
of the corresponding four-vector 2-form $\bf F$ over the same surface. In
that case, however, the contribution to the integral will be given only by
the components of the type ${\sf F}_{\mu 5 \, \nu 5}$, while the rest of
$\FS$ will be of absolutely no account. Considering this, it will be more
correct to formulate the question about the integration of adjoint forms in
the following way: can one define the integral of an adjoint form {\em as
a whole}? It is evident that what one has in mind is a certain ``bivector''
generalization of ordinary exterior calculus, where the objects of
integration will be adjoint forms and where the infinitesimal elements
of integration volumes will be characterized by adjoint vectors or by
antisymmetrized tensor products of the latter. As in the case of ordinary
exterior calculus, the starting point for developing such a formalism is
the notion of a vector (in our case, of an adjoint one) tangent to a curve,
the definition of which I will now discuss.

It is evident that there is no distinguished way of putting into
correspondence to an ordinary parametrized curve a five-vector bivector
with a nonzero $\cal Z$-component, and therefore one has to find a more
specific class of curves that possess some additional structure. The
simplest and the most obvious solution are the curves endowed with a rule
for transporting four-vectors along them, provided that this transport
is linear and conserves the scalar product $g$. By using these rules of
transport, at any point $Q$ of any such curve $\cal C$ one can evaluate
the derivative of any sufficiently smooth four-vector field $\bf W$ defined
in the vicinity of $Q$. By analogy with the derivative of a scalar field
along a parametrized curve, let us denote this derivative as $D_{\cal C}
{\bf W}|_{Q}$. Making use of the mentioned properties of the transport in
question, it is not difficult to prove that at each point of $\cal C$ there
exists a five-vector bivector $\AAA$ such that for any sufficiently smooth
four-vector field $\bf W$
\begin{equation}
D_{\cal C} {\bf W} = {\sf D}_{\cal A} {\bf W}.
\end{equation}
It can be shown that the four-vector corresponding to the $\cal E$-component
of $\AAA$ coincides with the four-vector $\bf U$ tangent to the curve at that
point and that the $\cal Z$-component of $\AAA$ is a homogeneous function of
$\bf U$ in the sense that when the parametrization of the curve is changed,
it changes in the same proportion as $\bf U$. Equation (147) offers one a way
of putting into correspondence to each curve endowed with a transport rule
for four-vectors, at each its point a certain five-vector bivector, which I
will call the {\em bivector} (or {\em adjoint vector}) {\em tangent to the
curve} at that point.

Let us now consider integrals along such curves. Defining them similarly to
the integrals along ordinary parametrized curves, one obtains that any such
integral can be presented as
\begin{displaymath}
\int^{\lambda_{b}}_{\lambda_{a}} d \lambda \; \phi(\lambda),
\end{displaymath}
where $\lambda_{a}$ and $\lambda_{b}$ are the end-point values of the curve
parameter and $\phi(\lambda)$ is a certain numerical function, which may also
depend on the adjoint vector tangent to the curve at the integration point.
In the following I will be interested only in the integrals whose value
is independent of the curve parametrization (at the same transport rule).
Such integrals will be called invariant. As in ordinary exterior calculus,
one can prove that for a given integral to be invariant it is sufficient
that $\phi$ be a homogeneous function of the tangent bivector. As in part IV,
let us confine ourselves to the case of integrals whose integrand depends
on the tangent bivector {\em linearly}.

Already at this stage it is best to ask the following question: where may
the transport rules discussed above come from in the real world? To this
natural question there exists an equally natural answer: as rules for
transporting four-vectors along any given curve one should take the rules
of parallel transport fixed by the five-vector affine connection, which one
assumes to be defined throughout space-time. From what has been said above it
then follows that for any parametrized curve the adjoint tangent vector will
be $\AAA = \sigma ({\bf u})$, where $\bf u$ is the homogeneous tangent
five-vector for this curve at the considered point. As it has already been
said, the $\cal E$-component of $\AAA$ will be the five-vector bivector
corresponding to ${\bf u}^{\cal Z}$ and the $\cal Z$-component of
$\AAA$ will be completely determined by the four-vector torsion 1-form
$\widetilde{\bf S}$ introduced in section A. In particular, when the latter
is zero, $\AAA^{\cal Z}$ is zero, too, and the considered curve becomes
equivalent to an ordinary parametrized curve.

Let us now consider $m$-dimensional surfaces ($m = 2$, 3 or 4) endowed with
transport rules for four-vectors. Let us take one such surface, introduce on
it a certain parametrization $\lambda^{(1)}, \ldots, \lambda^{(m)}$, and by
using the method described above, define at each its point a set of $m$
adjoint vectors $(\AAA^{(1)}, \ldots, \AAA^{(m)})$, each of which is tangent
to the corresponding inner coordinate line at that point. If, as in the case
of curves, one assumes that the rules of transport along the considered
surface are determined by five-vector affine connection, one will have
$\AAA^{(k)} = \sigma ({\bf u}^{(k)})$ for all $k$ from 1 to $m$, where
${\bf u}^{(k)}$ is the homogeneous tangent five-vector to the coordinate
line $\lambda^{(k)}$. An integral over such a surface will have the form
\begin{equation}
\int_{\Lambda} d \lambda^{(1)} \ldots d \lambda^{(m)}
\phi(\lambda^{(1)}, \ldots, \lambda^{(m)}),
\end{equation}
where $\Lambda$ is the range of variation of inner coordinates and $\phi
(\lambda^{(1)}, \ldots, \lambda^{(m)})$ is some numerical function that
may also depend on the adjoint vectors tangent to coordinate lines at
the integration point. As in part IV, let us confine ourselves to the
case where $\phi$ is a linear function of each of the tangent bivectors
$\AAA^{(k)}$. Since each of the latter, in its turn, is a linear function
of the corresponding homogeneous tangent five-vector, integral (148) will
actually be an integral of the type considered in section B of part IV,
and therefore it will not depend on the surface parametrization only if its
integrand is completely antisymmetric in all the tangent bivectors and each
of the latter is taken to be $\sigma ({\bf u}^{(k) \, \cal Z})$, not $\sigma
({\bf u}^{(k)})$. Thus, the integrand of any invariant integral of the
considered type should be a contraction of some adjoint $m$-form with
the adjoint multivector of rank $m$ of the form
\begin{equation}
\sigma ({\bf u}^{(1) \, \cal Z}) \wedge \ldots \wedge
\sigma ({\bf u}^{(m) \, \cal Z}).
\end{equation}
In the following such integrals will be referred to as {\em integrals of the
first kind}. From what has been said above it follows that any such integral
can also be regarded as an integral over the same surface of a certain
five-vector $m$-form, which is obtained from the adjoint $m$-form mentioned
above by substituting into the latter as its arguments $m$ samples of the
1-form $\widetilde{\bf s}$ introduced in section A. This duality between
integrals of adjoint forms and integrals of ordinary five-vector forms
proves to be very useful.

Comparing the results we have just obtained with those obtained for integrals
along curves, we see that for the latter the condition of invariance is
less stringent: for them it is enough that the integrand be homogeneous
in ${\bf u}^{\cal Z}$, whereas for the integrals over volumes of greater
dimension it is necessary that the integrand be a {\em linear} function of
the $\cal Z$-components of all the tangent five-vectors. This weaking of the
invariance condition for integrals along curves is evidently a consequence
of the latter being one-dimensional and results in that invariant integrals
are not only those whose integrand is a contraction of some adjoint 1-form
with the bivector $\sigma ({\bf u}^{\cal Z})$, but also those whose integrand
is a contraction of an adjoint 1-form with the bivector $\sigma ({\bf u})$,
and therefore is not linear in ${\bf u}^{\cal Z}$. For uniformity, only the
integrals of the former type will be considered as integrals of the
first kind, while the integrals whose integrand is proportional to
${\bf u}^{\cal E}$ will be regarded as nonlinear integrals, which I will
not consider here.

Thus, for any $m$-dimensional volume (now $m = 1, 2 ,3 ,4$) an integral of
the first kind has the form
\begin{equation}
\int_{\Lambda} \! \! d \lambda^{(1)} \! \! \ldots d \lambda^{(m)} \! \!
< \! \widetilde{\NN}, \sigma ({\bf u}^{(1) \, \cal Z}) \wedge \ldots
\wedge \sigma ({\bf u}^{(m) \, \cal Z}) \! > ,
\end{equation}
where $\widetilde{\NN}$ is some adjoint $m$-form, which, naturally, may
depend on torsion itself. As one can see, in this case the adjoint
multivector that characterizes the infinitesimal element of the integration
volume is independent of $\sigma ({\bf n})$ (according to the definition
of the homogeneous tangent five-vector, one has $\sigma({\bf u}^{(k) \,
\cal E}) = |g({\bf u}^{(k)},{\bf u}^{(k)})|^{1/2} \cdot \sigma ({\bf n})$
for every $k$) and therefore is determined only by that part of four-vector
torsion which corresponds to the derivative $\nabla_{\bf u} \equiv
\plaision_{({\bf u}^{\cal Z})}$, and which in the following will be referred
to as $\nabla$-{\em torsion}. In order to construct an integral that would
explicitly depend on $\sigma ({\bf n})$, let us recall the second type of
integrals of ordinary five-vector forms, where the infinitesimal element of
an $m$-dimensional integration volume is characterized not by the multivector
${\bf u}^{(1) \, \cal Z} \wedge \ldots \wedge {\bf u}^{(m) \, \cal Z}$ but
by the multivector ${\bf u}^{(1)} \wedge \ldots \wedge {\bf u}^{(m)} \wedge
{\bf 1}$. It is evident that for this type of integrals of five-vector forms
the corresponding integrals of adjoint forms have the following form:
\begin{equation}
\int_{\Lambda} \! \! d \lambda^{(1)} \! \! \ldots d \lambda^{(m)} \! \!
< \! \widetilde{\NN}, \sigma ({\bf u}^{(1)}) \wedge
\ldots \wedge \sigma ({\bf u}^{(m)}) \wedge \sigma ({\bf 1}) \! > ,
\end{equation}
where $\widetilde{\NN}$ is now a certain adjoint $(m+1)$-form. Integrals of
this type will be referred to as {\em integrals of the second kind}. As in
the case of integrals of the first kind, one can regard integral (151) as
an integral over the same surface of an ordinary five-vector $(m+1)$-form,
which can be obtained from $\widetilde{\NN}$ by substituting into the
latter as its arguments $m+1$ samples of the 1-form $\widetilde{\bf s}$.

As one can see from the formulae obtained, in contrast to the case of
ordinary five-vector forms, for which the difference between integrals
of the first and second kind is more a formality and comes only to the
difference in the rank of the integrated form, for adjoint forms the
difference between integrals (150) and (151) is more significant: in the
first case the adjoint multivector that characterizes the element of
the integration volume depends only on $\nabla$-torsion, whereas in the
second case this multivector also explicitly depends on the quantity
$\sigma ({\bf 1})$, which is the value of some independent bivector field.
Another significant difference between integrals (150) and (151) is the
following: since for any nonzero ${\bf u} \in {\cal Z}$ the bivector
$\sigma ({\bf u})$ is never zero, multivector (149) does not vanish at
{\em any} five-vector torsion, whereas for the adjoint multivector in
integral (151) to be nonvanishing one must have $\sigma ({\bf 1}) \neq 0$.
Therefore, when speaking of integrals of the second kind one should assume
the latter condition to be obeyed everywhere within the integration volume.

In conclusion let me say a few words about the situation with the generalized
Stokes theorem, which in the case of ordinary forms enables one to transform
integrals over a closed surface into integrals over the volume the latter
encloses. Owing to the mentioned correspondence between integrals of adjoint
forms and integrals of ordinary five-vector forms, a similar transformation
can also be performed with integrals (150) and (151). However, since in
either case the multivector that characterizes the element of the integration
volume is torsion-dependent, the volume integral one obtains will in general
not be an integral of an adjoint form. As an example, let us consider an
integral of the type (150) over some closed three-dimensional surface
$\partial V$ that limits a four-dimensional volume $V$, and let us write
this integral down in components relative to an active regular basis
associated with some coordinate system $x^{\alpha}$. One has
\begin{displaymath} \begin{array}{l}
\displaystyle \int_{\partial V} \! \! \widetilde{\NN} \propto \displaystyle
\int_{\partial V} \! \! d \lambda^{(1)} d \lambda^{(2)} d \lambda^{(3)} \,
{\cal N}_{\Re_{1} \Re_{2} \Re_{3}} \, s^{\Re_{1}}_{\alpha_{1}}
s^{\Re_{2}}_{\alpha_{2}} s^{\Re_{3}}_{\alpha_{3}} \rule{0ex}{3ex} \\
\hspace{17ex} \times \; ({\bf u}^{(1)})^{\alpha_{1}}
({\bf u}^{(2)})^{\alpha_{2}} ({\bf u}^{(3)})^{\alpha_{3}} \rule{0ex}{3ex} \\
\hspace{7ex} = \displaystyle \int_{V} d^{4} \! x \; \partial_{\, [ \, 0} \;
( \, s^{\Re_{1}}_{1} s^{\Re_{2}}_{2} s^{\Re_{3}}_{3 \, ]} \;
{\cal N}_{\Re_{1} \Re_{2} \Re_{3}} \, ) \\ \hspace{7ex} = \displaystyle
\int_{V} d^{4} \! x \; \partial_{\, [ \, 0} \; ( \, s^{\Re_{1}}_{1}
s^{\Re_{2}}_{2} s^{\Re_{3}}_{3 \, ]} \, ) \cdot
{\cal N}_{\Re_{1} \Re_{2} \Re_{3}} \rule{0ex}{3.5ex} \\ \hspace{11ex}
\displaystyle  + \; \int_{V} d^{4} \! x \; s^{\Re_{1}}_{[ \, 1}
s^{\Re_{2}}_{\rule{0ex}{1.3ex} 2} s^{\Re_{3}}_{\rule{0ex}{1.3ex} 3} \cdot
\partial_{\rule{0ex}{1.5ex} 0 \, ]} \, {\cal N}_{\Re_{1} \Re_{2} \Re_{3}} \\
\hspace{7ex} = \displaystyle \int_{V} d^{4} \! x \;
{\cal N}_{\Re_{1} \Re_{2} \Re_{3}} \cdot \partial_{\, [ \, 0} \; ( \,
s^{\Re_{1}}_{1} s^{\Re_{2}}_{2} s^{\Re_{3}}_{3 \, ]} \, ) \\ \hspace{9ex}
\displaystyle  + \; \int_{V} d^{4} \! x \; {\sf D}_{[ \, \Re} \,
{\cal N}_{\Re_{1} \Re_{2} \Re_{3} \, ]} \cdot s^{\Re}_{0} \,
s^{\Re_{1}}_{1} s^{\Re_{2}}_{2} s^{\Re_{3}}_{3}.
\end{array} \hspace{-1ex} \end{displaymath}
In the general case, the first integral in the right-hand side is not zero
and its integrand cannot be presented as a contraction of some adjoint 4-form
with a multivector of the type (149) at $m=4$. Consequently, the sum of the
two terms in the right-hand side, too, will not be an integral of an adjoint
form over $V$. The latter fact, however, does not prevent one from using the
generalized Stokes theorem when integrating adjoint forms.

\vspace{6ex} \begin{flushleft}
\bf Acknowledgement
\end{flushleft}
I would like to thank V. D. Laptev for supporting this work. I am grateful
to V. A. Kuzmin for his interest and to V. A. Rubakov for a very helpful
discussion and advice. I am indebted to A. M. Semikhatov of the Lebedev
Physical Institute for a very stimulating and pleasant discussion and to
S. F. Prokushkin of the same institute for consulting me on the Yang-Mills
theories of the de Sitter group. I would also like to thank L. A. Alania,
S. V. Aleshin, and A. A. Irmatov of the Mechanics and Mathematics Department
of the Moscow State University for their help and advice.

\vspace{6ex} \begin{flushleft}
\bf Reference
\end{flushleft} \begin{enumerate}
 \item L. V. Avdeev and M. V. Chizhov, Phys. Lett. B {\bf 321} (1994) 212.
 \item A. Krasulin, Bivector gauge fields from classical electrodynamics, In: Dvoeglazov, V.V. (ed.): Photon and Poincare group, 1999, p.218-230.
\end{enumerate}

\end{document}